\documentclass[showpacs,twocolumn,pra,superscriptaddress,notitlepage]{revtex4-1}
\usepackage{amsthm}  
\newtheorem{assumption}{Assumption}  
\usepackage[dvips]{graphicx}
\usepackage{amsmath,amssymb,amsthm,mathrsfs,amsfonts,dsfont,amscd,keyval}
\usepackage{mathtools,mathrsfs}

\usepackage{yfonts} 
\usepackage{ytableau}
\usepackage{subfigure, epsfig}
\usepackage{braket}
\usepackage{bm} 
\usepackage{fancyhdr}
\usepackage{enumerate}
\usepackage{color}
\usepackage{hyperref}
\usepackage[normalem]{ulem}
\usepackage{tabularx}
\usepackage{times}
\hypersetup{colorlinks=true, linkcolor=blue, citecolor=blue, urlcolor=black}
\usepackage{multirow}
\usepackage{float}
\usepackage{tensor}
\usepackage{multirow}
\usepackage{physics}
\usepackage{soul}

\renewcommand\vec{\mathbf}

\newcommand{\ie}{\textit{i.e.}}
\newcommand{\eg}{\textit{e.g.}}

\theoremstyle{definition}
\newtheorem{definition}{Definition}[section]

\newtheorem{proposition}[definition]{Proposition}

\theoremstyle{plain}
\newtheorem{theorem}[definition]{Theorem}
\newtheorem{lemma}[definition]{Lemma}

\theoremstyle{remark}
\newtheorem*{remark}{Remark}
\newcommand{\comments}[1]{}

\makeatletter
\def\l@subsubsection#1#2{} 
\makeatother

\begin{document}

\let\oldaddcontentsline\addcontentsline

\renewcommand{\addcontentsline}[3]{}

\title{An Analytic Theory of Quantum Imaginary Time Evolution}

\author{Min Chen}
\affiliation{Department of Computer Science, University of Pittsburgh, Pittsburgh, PA 15260, USA}

\author{Bingzhi Zhang}
\affiliation{Ming Hsieh Department of Electrical and Computer Engineering, University of Southern California, Los Angeles, CA 90089, USA}
\affiliation{Department of Physics and Astronomy, University of Southern California, Los Angeles, CA 90089, USA}

\author{Quntao Zhuang}
\affiliation{Ming Hsieh Department of Electrical and Computer Engineering, University of Southern California, Los Angeles, CA 90089, USA}
\affiliation{Department of Physics and Astronomy, University of Southern California, Los Angeles, CA 90089, USA}

\author{Junyu Liu}
\affiliation{Department of Computer Science, University of Pittsburgh, Pittsburgh, PA 15260, USA}

\date{Dated: November 11, 2024}

\date{\today}

\begin{abstract}
Quantum imaginary time evolution (QITE) algorithm is one of the most promising variational quantum algorithms (VQAs), bridging the current era of Noisy Intermediate-Scale Quantum devices and the future of fully fault-tolerant quantum computing. 
Although practical demonstrations of QITE and its potential advantages over the general VQA trained with vanilla gradient descent (GD) in certain tasks have been reported, a first-principle, theoretical understanding of QITE remains limited. Here, we aim to develop an analytic theory for the  dynamics of QITE. First, we show that QITE can be interpreted as a form of a general VQA trained with Quantum Natural Gradient Descent (QNGD), where the inverse quantum Fisher information matrix serves as the learning-rate tensor.  This equivalence is established not only at the level of gradient update rules, but also through the action principle: the variational principle can be directly connected to the geometric geodesic distance in the quantum Fisher information metric, up to an integration constant. Second, for wide quantum neural networks, we employ the quantum neural tangent kernel framework to construct an analytic model for QITE. We prove that QITE always converges faster than GD-based VQA, though this advantage is suppressed by the exponential growth of Hilbert space dimension. This helps explain certain experimental results in quantum computational chemistry. Our theory encompasses linear, quadratic, and more general loss functions. We validate the analytic results through numerical simulations. Our findings establish a theoretical foundation for QITE dynamics and provide analytic insights for the first-principle design of variational quantum algorithms.

\end{abstract}

\maketitle

\section{Introduction}























\begin{figure*}[ht]
\includegraphics[width=0.8\textwidth]{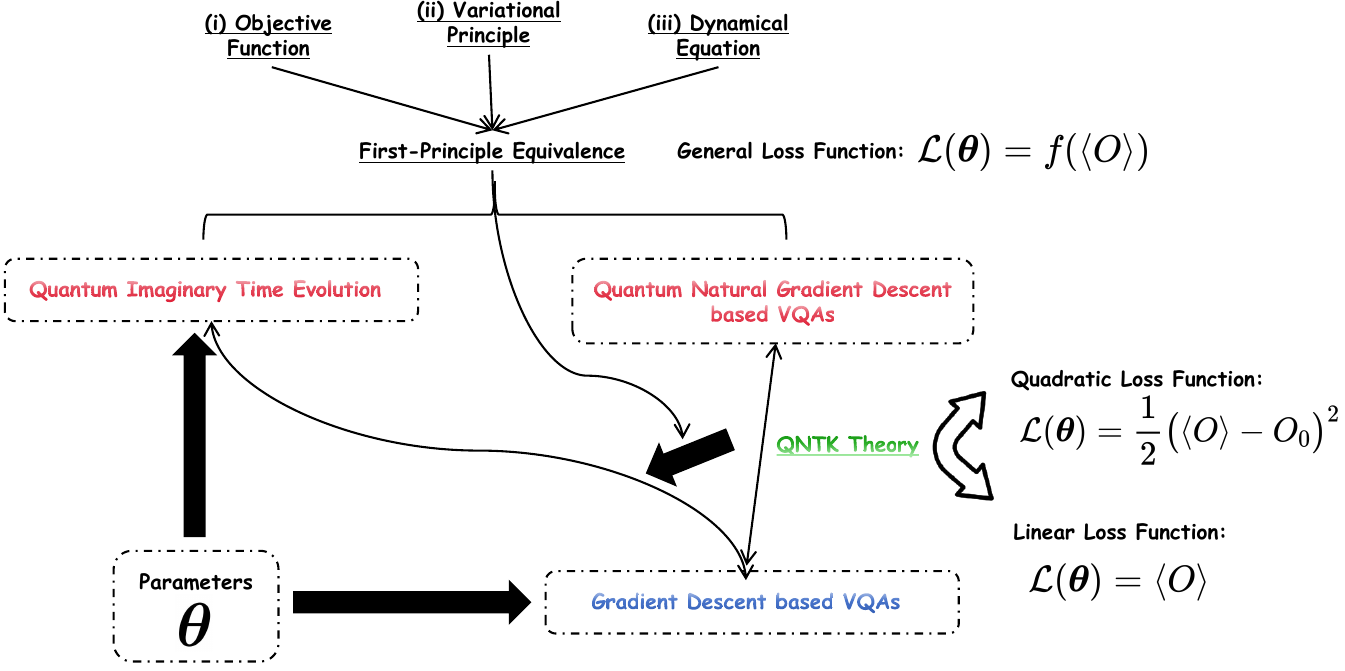}
\caption{Overview of Main Results. Firstly, we uncover and establish a first-principle equivalence between QITE and QNGD-based VQAs by deriving that the objective functions, the variational principles and the dynamical equations are identical up to an integration constant with general loss function in continuous-time limit.
Secondly, we focus on quadratic and linear loss, and leverage QNTK theory to derive a first-principle model to characterize the training dynamics of QITE in the regimes of interest.}
\label{Overview_framework}
\end{figure*}




Current quantum devices \cite{preskill2018quantum,preskill2025beyond} operate with a limited number of qubits and are subject to significant noise. While challenges such as restricted qubit coherence remain, variational quantum algorithms~(VQAs) have emerged as promising approaches 
~\cite{liu2022application,lubasch2020variational,amaro2022case,singh2023benchmarking,peruzzo2014variational,wittek2014quantum,abbas2021power}. In particular, quantum imaginary time evolution~(QITE) algorithm~\cite{motta2020determining,1812.08767,mcardle2019variational} stands out by enabling efficient ground state convergence and state preparation through emulating non-unitary imaginary time dynamics in quantum chemistry and condensed matter systems. Beyond that, it can also be applied to general learning and optimization problems by following its variational principle~\cite{mcardle2019variational} such as the ``McLachlan variational principle'' which determines the optimal evolution by minimizing the instantaneous difference between the exact and variational time derivatives, often leading to faster convergence. These properties have led to improved performance in diverse applications, including state preparation~\cite{mcardle2019variational,tsuchimochi2023improved} and simulations of quantum many-body systems~\cite{sun2021quantum,beach2019making}. 

While there has been evidence that QITE exhibits experimental advantages~\cite{guo2024experimental}, the lack of analytic studies hinders a deeper understanding of its training dynamics and constrains the potential of  designing new quantum algorithms~\cite{schuld2022quantum}. Notably, while substantial analytic progress has been made for general VQAs or quantum neural networks (QNNs) trained with vanilla gradient descent~(GD), including studies on trainability~\cite{mcclean2018barren,pesah2021absence,cerezo2021cost}, expressivity~\cite{holmes2022connecting,sim2019expressibility,nakaji2021expressibility,du2022efficient,sharma2022reformulation}, generalization~\cite{abbas2021power,caro2022generalization,huang2021information,huang2021power}, and convergence~\cite{PRXQuantum.3.030323,liu2023analytic,zhang2024dynamical}, a comparable systematic analytic framework for QITE is still lacking. This motivates us to develop a systematic analytic framework for QITE, aiming to establish a first-principle understanding of its training dynamics. 



In this work, we begin by developing the analytic theory via interpreting QITE as a general VQA trained with Quantum Natural Gradient Descent (QNGD)~\cite{stokes2020quantum} (QNGD-based VQAs) where the inverse quantum Fisher information matrix serves as the learning-rate tensor. This interpretation is based on the action principle where we relate the variational principle of QITE with the geometric geodesic distance in the quantum Fisher information metric. 
Our contribution proceeds along two complementary directions: (i) we show the equivalence between the objective of QITE and that of QNGD-based VQAs, and (ii) we formulate an action principle for QNGD-based VQAs and prove the direct equivalence with the variational principle of QITE, both up to an integration constant under the continuous-time limit. This also explains why these two seemingly different processes yield identical parameter update rules~\cite{stokes2020quantum}. To pursue generality, we go beyond the scheme with linear loss function by formulating the analysis with general loss functions, along with the extension of the variational principles. 



We further develop the analytic theory for the dynamics of QITE from the natural geometric perspective of QNGD, allowing us to directly compare the dynamics with general VQAs trained with GD (GD-based VQAs), focusing on a wide QNN. The results are based on the Quantum Neural Tangent Kernel (QNTK) frameworks~\cite{PRXQuantum.3.030323,PhysRevLett.130.150601,Liu_2024,wang2022symmetric,Yu_2024,zhang2024dynamical}. We calculate the dynamics with respect to residual training error for sufficiently random quantum circuits modeled as unitary $k$-designs~\cite{roberts2017chaos,cotler2017chaos,liu2018spectral,liu2020scrambling}. 
For sufficiently large number of variational parameters modelled as overparameterized regime~\cite{PRXQuantum.3.030323}, analytic solutions characterizing the convergence relation between QITE and GD-based VQAs in terms of the residual training error are established~\cite{PRXQuantum.3.030323}.
Notably, prior studies~\cite{PRXQuantum.3.030323,liu2023analytic,zhang2024dynamical} have predominantly tackled quadratic loss without showing any convergence speed-up in linear loss, where we argue that it is limited as QITE is often applied to problems such as ground state energy estimation~\cite{aspuru2005simulated,kandala2017hardware,kandala2019error,lehtovaara2007solution} which naturally involve linear objectives. Therefore, we target both the quadratic and linear loss here. 
Specifically, our analysis with quadratic loss function focuses on lazy training regime~\cite{liu2023analytic,PRXQuantum.3.030323}, where the QNTK \(K\) becomes a constant at late time. Meanwhile, for the ground state search problem naturally formulated with a linear loss,  we show that the dynamics of both \(K\) and the residual training error \(\epsilon\) decay exponentially with their dynamics driven by the relative dQNTK \(\lambda\). 
We demonstrate that the out theories offer a geometric perspective to analytically explain its convergence advantages in the studied regimes, and show that QITE with  either type of loss function can present convergence advantages than GD-based VQAs, though these advantages are suppressed by the exponential growth of Hilbert space dimension with number of qubits. These theories help explain certain experimental results in quantum computational chemistry~\cite{guo2024experimental}. We verify these results with numerical studies.

In summary, the workflow of this study is outlined in Fig.~\ref{Overview_framework}. We first introduce the analytic theories, including the action-principle equivalence and the analytic model, with detailed derivations provided in the \textit{Appendix}. We then show numerical simulations for validating the convergence alignments, and outline the methods we adopt. Finally, we discuss broader implications and potential future directions related with this work.

\section{Results}


We consider a universal VQA or QNN that prepares a normalized state 
$|\psi(\bm\theta)\rangle$ with circuit parameters $\bm\theta$. 
The learning objective  is a differentiable function of an observable expectation, 
\begin{equation}
    \mathcal{L}(\bm\theta) \;=\; f\!\left(\langle O\rangle\right),
\quad 
\langle O\rangle := \langle \psi(\bm\theta)|\,O\,|\psi(\bm\theta)\rangle,
\end{equation}
where $O$ is a Hermitian observable and $f:\mathbb{R}\to\mathbb{R}$ is differentiable. 
This covers common energy-based VQAs by taking $O=H$ and $f(x)=x$ or any monotone $f$.
Below we define (i) \emph{quantum imaginary time evolution} (\textit{QITE}) and (ii) \textit{the general VQAs trained with QNGD} (termed \textit{QNGD-based VQAs}) such that we can establish the equivalence relation later.

\textbf{QITE.} QITE is widely adopted for approximating the non-unitary imaginary time dynamics. 
\begin{equation}
    \frac{d|\psi(\tau)\rangle}{d\tau} = -O|\psi(\tau)\rangle,
\end{equation}
where \(\tau = it\) defines the imaginary time, and the \(|\psi(\tau)\rangle\) defines the evolved quantum state. In practice, one restricts the dynamics to a parametrized variational manifold $|\psi(\bm \theta)\rangle$, and the goal is to determine $\bm \theta(\tau)$ that best approximates the true imaginary time evolution. A principled way is given by McLachlan’s variational principle, which is of our focus with the reasons illustrated in \textit{Appendix \ref{sec:Variational Formulation of QITE}}. It states that the exact imaginary time derivative should be projected onto the tangent space of the variational manifold by requiring that the residual norm follows:
\begin{equation}
    \delta \left\| \left( \frac{\partial}{\partial \tau} + O - E_\tau \right) | \psi( \bm \theta(\tau))\rangle \right\| = 0.
\end{equation}
This leads to equations that determines the infinitesimal parameter update $\dot{\bm \theta}$ such that the variational trajectory remains as close as possible to the exact flow. In a discrete time setting, instead of matching derivatives, one considers a finite step $\Delta \tau$ of imaginary time evolution. The evolved state can be written as $e^{-O \Delta \tau} \psi(\bm \theta)$, which in general does not lie within the variational manifold. Therefore, the projected QITE addresses this by choosing the updated variational state $\psi(\bm \theta + \Delta \bm \theta)$ that maximizes its fidelity with the evolved state, namely the approximation objective of QITE is defined below:


\begin{equation}
  \Delta \bm \theta_{\text{QITE}}
  = \underset{\Delta \bm\theta}{\arg\max}\;
  \left| \bra{\psi(\bm\theta)}\, e^{-\Delta\tau\, O}\, \ket{\psi(\bm\theta+\Delta\bm\theta)} \right|^{2}.
\label{eq:QITE_update_rule}
\end{equation}

In the small step limit $\Delta \tau \to 0$, this discrete time fidelity maximization is equivalent to McLachlan’s residual minimization, so that projected QITE can be interpreted as a finite-step realization of McLachlan’s variational principle.


\textbf{QNGD-based VQAs:} QNGD-based VQAs seek an update $\bm{\Delta\bm \theta}$ such that the loss 
$\mathcal{L}(\bm{\bm \theta} + \bm{\Delta\bm \theta})$ is minimized 
where $\bm{\bm \theta}$ denotes the QNN parameters, while constraining the induced change in quantum state fidelity. To facilitate a unified geometric understanding of QITE, we reformulate QNGD-based VQAs in the setting of general differentiable loss functions \(f(\langle O \rangle)\), where \(f\) is an arbitrary differentiable function. Let \(|\psi(\bm \theta)\rangle\) denotes the quantum state prepared by a parameterized circuit, then instead of constraining \(\|\bm \Delta\bm \theta\|\) in Euclidean space, QNGD equips the geometric manifold with the fidelity distance between \(|\psi(\bm \theta)\rangle\) and \(|\psi(\bm \theta + \bm \Delta\bm \theta)\rangle\) defined as~\cite{meyer2021fisher}:

\begin{equation}
\begin{aligned}
d_f(\psi(\bm \theta), \psi(\bm \theta + \bm \Delta\bm \theta)) 
&:= 1 - |\langle \psi(\bm \theta) | \psi(\bm \theta + \bm \Delta\bm \theta) \rangle|^2.
\end{aligned}
\end{equation}

In the limit of \(\bm \Delta\bm \theta \to 0\), the fidelity distance reduces to the squared line element \(\mathrm{d}s^2\) on the quantum state manifold~\cite{stokes2020quantum}:

\begin{equation}
\begin{aligned}
d_f \approx \mathrm{d}s^2 
&= \frac{1}{4} \sum_{\ell_1,\ell_2} \mathcal{F}_{\ell_1\ell_2}(\bm \theta) \Delta\bm \theta_{\ell_1} \Delta\bm \theta_{\ell_2}, \\
&=  \sum_{\ell_1,\ell_2} g_{\ell_1\ell_2}(\bm \theta) \Delta\bm \theta_{\ell_1} \Delta\bm \theta_{\ell_2},
\end{aligned}
\end{equation}
where \(\mathcal{F}(\bm \theta)\) is the quantum Fisher information matrix (QFIM). It is also connected to the Fubini–study metric tensor \(g\) by \(\mathcal{F}_{\ell_1 \ell_2}(\bm \theta) = 4 g_{\ell_1, \ell_2}(\bm \theta) = \, 4\Re \big[ G_{\ell_1 \ell_2}(\bm \theta) \big]\), with \(G\)  to be the quantum geometric tensor (QGT) ~\cite{kolodrubetz2017geometry} defined as follows:

\begin{equation}
\begin{aligned}
G_{\ell_1 \ell_2}(\bm \theta) 
&:= \left\langle \frac{\partial \psi(\bm \theta)}{\partial \bm \theta_{\ell_1}} \Bigg| \frac{\partial \psi(\bm \theta)}{\partial \bm \theta_{\ell_2}} \right\rangle \\
&\quad - \left\langle \frac{\partial \psi(\bm \theta)}{\partial \bm \theta_{\ell_1}} \Bigg| \psi(\bm \theta) \right\rangle 
          \left\langle \psi(\bm \theta) \Bigg| \frac{\partial \psi(\bm \theta)}{\partial \bm \theta_{\ell_2}} \right\rangle.
\label{quantum_geometric_tensor}
\end{aligned}
\end{equation}

Applying the Lagrangian formulation, the approximation objective of QNGD-based VQAs becomes:

\begin{equation}
\begin{aligned}
&\Delta \bm \theta_{\text{QNGD}} = 
\nonumber
\\
&\underset{\Delta \bm \theta}{\text{argmin}}
\left(
\; \nabla_{\bm \theta} \mathcal{L}(\bm \theta) \cdot \bm \Delta\bm \theta 
+ \frac{1}{4} \lambda \sum_{\ell_1, \ell_2} \mathcal{F}_{\ell_1 \ell_2}(\bm \theta) \Delta\bm \theta_{\ell_1} \Delta\bm \theta_{\ell_2}
\right),
\label{eq:QNGD_update_rule}
\end{aligned}
\end{equation}
where \(\lambda > 0\) is the Lagrange multiplier. Full derivations are provided in \textit{Appendix~\ref{A Review for Quantum Natural Gradient Descent}}.

\subsection{A First-principle Equivalence Between QITE and QNGD-based VQAs}

We first analyze the linear loss function defined as a linear observable expectation \(\mathcal{L}(\bm \theta) = \langle O \rangle\).

\begin{theorem}[The Objective Equivalence of QITE and QNGD-based VQAs in the Continuous-Time Limit] In the infinitesimal-step limit \(\eta \to 0\) with \(\mathcal{L}(\bm \theta) = \langle O \rangle\), the objective function of QNGD-based VQAs becomes equivalent to that of QITE.
\end{theorem}

\textit{Sketch of Proof.}
Following~\citet{stokes2020quantum}, we expand the fidelity between the imaginary-time evolved state \(e^{-O\Delta\tau} \psi(\bm \theta)\) and the variationally updated state \(\psi(\bm \theta + \Delta\bm \theta)\) to second order. Letting \(\bm \delta := \frac{d \bm \theta}{d \tau}\) and taking the continuous-time limit \(\Delta \tau \to 0\), then QITE's objective reduces to:

\begin{equation}
\begin{aligned}
\frac{\partial \langle \psi(\bm \theta) | O | \psi(\bm \theta) \rangle}{\partial \bm \theta_{\ell_1}} \delta_{\ell_1} 
+ \Re\left[G_{\ell_1 \ell_2}(\bm \theta)\right] \delta_{\ell_1} \delta_{\ell_2},
\end{aligned}
\end{equation}
which matches the objective of QNGD-based VQAs in Eq.~(\ref{eq:QNGD_update_rule}) for \(\lambda = 1\), thereby proving the equivalence up to an overall scaling factor. See detailed derivations in \textit{Appendix~\ref{appendix:equiv_objective}}.

Furthermore, to establish a first-principle equivalence, we now turn to the continuous-time variational formulation of QNGD-based VQAs and compare it with QITE with the McLachlan principle.  We first formalize the variational principle of QNGD-based VQAs with general loss function below:

\begin{theorem}[Variational Principle of QNGD-based VQAs with General Loss Function]
\label{thm:qngd_general_vp}
Let \( |\psi(\bm{\bm \theta})\rangle \) be a parameterized quantum state, and let the general loss be a differentiable function of the observable expectation defined as \( \mathcal{L}(\bm{\bm \theta}) = f(\langle O \rangle) \), where \( O \) is a Hermitian observable. Then the variational principle underlying QNGD-based VQAs is given by
\begin{equation}
\delta \left[ f'\big( \langle O \rangle_{\bm \theta} \big) \, \nabla \langle O \rangle_{\bm \theta}^\top \Delta {\bm \theta} + \frac{1}{2\eta} \Delta{\bm \theta}^\top \mathcal{F} \Delta{\bm \theta} \right] = 0,
\end{equation}
where $\nabla \langle O \rangle_{\bm \theta}$ is the gradient of the expectation value $\langle O \rangle$ with respect to the parameters $\bm{\theta}$, \( \eta > 0 \) is the learning rate, and \( \mathcal{F} \) is the QFIM, and $\delta$ indicates the first variation with respect to $\Delta \bm \theta$. This variational principle defines the variational functional for QNGD-based VQAs with general loss function:
\begin{equation}
\mathcal{J}_{\text{General}}[\Delta{\bm \theta}] = f'\big( \langle O \rangle_{\bm \theta} \big) \, \nabla \langle O \rangle_{\bm \theta}^\top \Delta \bm \theta + \frac{1}{2\eta} \Delta{\bm \theta}^\top \mathcal{F} \Delta{\bm \theta}.
\end{equation}
\end{theorem}
Detailed formalism can be found in \textit{Appendix \ref{sec:QNGD_variational_principle}}. To establish the equivalence step by step, we first give our finding with linear loss function:

\begin{theorem}[Variational Principle Equivalence between QNGD-based VQAs and QITE with Linear Loss Function]
\label{thm:qngd_qite_variational_equiv}
Let \( |\psi(\bm \theta)\rangle \) be a normalized parameterized quantum state, and let the linear loss function be an observable expectation \( \mathcal{L}(\bm \theta) = \langle O \rangle \). In the continuous-time limit \( \eta \to 0 \), the variational functional of QNGD-based VQAs \(\mathcal{J}_{\text{Linear}}\) and QITE \(\mathcal{D}_{\text{Linear}}\) become
\begin{equation}
\begin{aligned}
    &\mathcal{J}_{\text{Linear}}[\bm \theta(\tau)] =\\
&\int \left( \sum_{\ell_1} \text{Re}\left( \langle \partial_{\ell_1} \psi | O | \psi \rangle \right) \dot{\bm \theta}_{\ell_1} + \frac{1}{2} \sum_{\ell_1, \ell_2} F_{\ell_1\ell_2} \dot{\bm \theta}_{\ell_1} \dot{\bm \theta}_{\ell_2} \right) d\tau,
\end{aligned}
\end{equation}
\begin{equation}
\mathcal{D}_{\text{Linear}}[\bm \theta(\tau)] = \left\| \left( \partial_\tau + O - E_\tau \right) |\psi(\bm \theta(\tau))\rangle \right\|^2, E_\tau = \langle O \rangle_{\bm \theta(\tau)}.
\end{equation}
We interchangeably use 
$\partial_\tau \psi$ to denote
\(
\partial_\tau |\psi(\bm \theta(\tau))\rangle 
= \sum_\ell 
\frac{\partial |\psi(\bm \theta)\rangle}{\partial \bm \theta_\ell} \,
\frac{d\bm \theta_\ell}{d\tau}.
\) Then, up to an additive constant and an overall scaling factor that do not affect the variational dynamics,
\begin{equation}
\mathcal{D}_{\text{Linear}} \propto \mathcal{J}_{\text{Linear}} + \text{constant}.
\end{equation}
Therefore, QNGD-based VQAs and QITE induce the same dynamics in the linear loss setting.
\end{theorem}


\textit{Sketch of proof.}
Utilizing the derivative chain rule 
\(
\partial_\tau |\psi\rangle = \sum_{\ell_1} \partial_{\bm \theta_{\ell_1}} |\psi\rangle\, \dot{\bm \theta}_{\ell_1}
\), the first term in \(\mathcal{J}_{\text{Linear}}\) can be reduced to \(\text{Re}\left( \langle \partial_\tau \psi | O | \psi \rangle \right)\). While for the second term we use the normalization condition \(\partial_\tau \langle \psi | \psi \rangle = 0\) and have \(
\sum_{\ell_1, \ell_2} F_{\ell_1\ell_2} \dot{\bm \theta}_{\ell_1} \dot{\bm \theta}_{\ell_2} = \langle \partial_\tau \psi | \partial_\tau \psi \rangle\). Therefore we can rewrite \(\mathcal{J}_{\text{Linear}}\) as
\(
\mathcal{J}_{\text{Linear}} = \int \left( \mathrm{Re}\langle \partial_\tau \psi | O | \psi \rangle + \tfrac{1}{2} \langle \partial_\tau \psi | \partial_\tau \psi \rangle \right) d\tau.
\)

On the other hand, expanding the QITE functional
\(
\mathcal{D}_{\text{Linear}} = \| (\partial_\tau + O - E_\tau)|\psi\rangle \|^2
\)
and discarding constant terms (e.g., \( \langle O^2 \rangle - E_\tau^2 \)) yields the same integrand as \( \mathcal{J}_{\text{Linear}} \), up to an overall scaling factor. Hence, the two variational functionals differ only by a constant scaling and additive offset, proving that they yield the same variational dynamics. Details can be found in \textit{Appendix \ref{sec:qite_qngd_first}}.

Now we discuss the case with general loss function. We first need to extend the variational principle of QITE as formulated below:

\begin{theorem}[Generalized McLachlan Variational Principle for QITE with general loss function]
Given a general differentiable loss function \(\mathcal{L} = f\big( \langle \psi(\bm \theta) | O | \psi(\bm \theta) \rangle \big)\), the variational principle for QITE is given by:
\begin{equation}
    \delta \left\| \left( \frac{\partial}{\partial \tau} + f'\big(E_\tau\big)(O - E_\tau) \right) |\psi(\bm \theta(\tau))\rangle \right\| = 0,
\end{equation}
where \( E_\tau = \langle O \rangle =  \langle \psi(\bm \theta(\tau)) | O | \psi(\bm \theta(\tau)) \rangle \) denotes the expectation value of \( O \). 

Detailed formalism can be found in \textit{Appendix \ref{sec:general_mclachlan}}. 
\begin{remark}
    This formulation reduces to the standard McLachlan variational principle with linear loss function \(\mathcal{L}(\bm \theta) = \langle O \rangle\).
\end{remark}
\end{theorem}
Accordingly, we establish the equivalence below:
\begin{theorem}[Variational Principle Equivalence of QITE and QNGD-based VQAs with General Loss Function]
\label{thm:qite_qngd_general_equivalence}
Given a general differentiable loss function \( \mathcal{L} = f(\langle O\rangle) \), in the continuous-time limit \( \eta \to 0 \), the variational functionals of QNGD-based VQAs and QITE become

\begin{equation}
     \mathcal{J}_{\text{general}} = \int \left[ f'\big(\langle O \rangle\big) \, \text{Re}\left( \langle \partial_\tau \psi | O | \psi \rangle \right) + \frac{1}{2} \langle \partial_\tau \psi | \partial_\tau \psi \rangle \right] d\tau,
\end{equation}

\begin{equation}
    \mathcal{D}_{\text{general}} = \left\| \left( \partial_\tau + f'\big(\langle O \rangle\big)(O - \langle O \rangle) \right) |\psi\rangle \right\|^2.
\end{equation}

Then, up to an additive constant and an overall scaling factor that do not affect the variational dynamics,

\begin{equation}
\mathcal{D}_{\text{Linear}} \propto \mathcal{J}_{\text{Linear}} + \text{constant}.
\end{equation}

Therefore, QNGD-based VQAs and QITE induce the same dynamics in the general loss setting.
\end{theorem}

Detailed proofs can be found in \textit{Appendix \ref{sec:general_loss_variational}}.

\subsection{Dynamics of QITE with Quadratic Loss Function}
\label{sec:dynamics_quadratic}

The first-principle equivalence establishes a bridge between the parameter-space QNGD-based VQAs and the quantum state evolution of QITE with general loss function. As a result, we can analyze the training dynamics of QITE through the lens of QNGD using QNTK theory (See Section \ref{sec:Variational Principles and First-Principle Equivalence}). In the following, we go beyond the scope of GD-based VQAs typically analyzed in previous works~\cite{PRXQuantum.3.030323, PhysRevLett.130.150601, Liu_2024, wang2022symmetric, Yu_2024, zhang2024dynamical}, and adopt this QNGD-informed perspective to study QITE. We formalize our regimes of interest as assumptions below:


\begin{assumption}[Lazy training regime]
\label{assump:lazy}
The variational parameters remain close to their initialization during training, resulting in small updates to the quantum state~\cite{PRXQuantum.3.030323,PhysRevLett.130.150601}.
\end{assumption}

\begin{assumption}[Random ansatz structure]
\label{assump:random}
The parameterized unitary \(U(\boldsymbol{\bm \theta})\) is sufficiently random. Specifically, for each layer index \(\ell\), \(U_{\ell-}\) and \(U_{\ell+}\), defined in Eq.~\eqref{quantum_wavefunction} are independent and match the Haar distribution up to the second moment.
\end{assumption}

The variational quantum wavefunction~\cite{peruzzo2014variational,farhi2014quantum,mcclean2016theory,kandala2017hardware,mcardle2020quantum,cerezo2021variational} is defined as
\begin{equation}
\begin{aligned}
    \ket{\psi(\boldsymbol{\bm \theta})} &= U(\boldsymbol{\bm \theta})\ket{\psi_0} =  \prod_{\ell=1}^L W_\ell \, \mathrm{e}^{i\bm \theta_\ell X_\ell} \ket{\psi_0} \\
    &= \prod_{\ell=1}^L W_\ell V_\ell(\bm \theta_\ell) \ket{\psi_0} = U_{\ell^+} U_{\ell^-} \ket{\psi_0},\\
& \text{with } U_{\ell^-} = \prod_{k=1}^{\ell-1} W_k V_k(\bm \theta_k), U_{\ell^+} = \prod_{k=\ell}^L W_k V_k(\bm \theta_k),
\label{quantum_wavefunction}
\end{aligned}
\end{equation}
where each unitary gate is decomposed into a fixed gate \(W_\ell\) and a parametrized rotation \(V_\ell(\bm \theta_\ell) = \mathrm{e}^{i\bm \theta_\ell X_\ell}\) with \(X_\ell\) denoting a Hermitian generator. In this work, we focus on the common case where each \(X_\ell\) is a traceless operator, typically a single Pauli operator or a Pauli string of tensor products of Pauli operators adopted in many VQA ansätze. \(\ket{\psi_0}\) denotes the input state. The quadratic loss function is defined as

\begin{equation}
    \mathcal{L}(\bm \theta) = \frac{1}{2} \left( \langle O \rangle - O_0 \right)^2 \equiv \frac{1}{2} \epsilon^2,
\label{loss_quadratic}
\end{equation}
where \(O\) is the Hermitian observable and $\langle O \rangle$ is the expectation value defined in Eq.(1). \(O_0\) is the target value and  \(\epsilon \equiv \langle O \rangle - O_0\) defines the residual training error~\cite{PRXQuantum.3.030323}. 


When applying QITE, the \(\ell_1\)-th variational parameter is updated according to the difference equation below:

\begin{equation}
\begin{aligned}
    \delta \bm \theta_{\ell_1}(t) \equiv \bm \theta_{\ell_1}(t+1) - \bm \theta_{\ell_1}(t) 
    &= -\eta \sum_{\ell_2} g_{\ell_1 \ell_2}^{+} \frac{\partial \mathcal{L}}{\partial \bm \theta_{\ell_2}}, \\
    &= -\eta \epsilon(\boldsymbol{\bm \theta}) \sum_{\ell_2} g_{\ell_1 \ell_2}^{+}  \frac{\partial \epsilon}{\partial \bm \theta_{\ell_2}},
\end{aligned}
\end{equation}
where \(\eta\) is the learning rate. When \(\eta\) is small, by Taylor expansion to the first order in \(\eta\),
the time difference equation for \(\epsilon\) becomes: 

\begin{equation}
\begin{aligned}
    \epsilon(t+1) - \epsilon(t) \equiv \delta \epsilon 
    &\approx \sum_{\ell} \frac{\partial \epsilon}{\partial \bm \theta_{\ell}} \delta \bm \theta_{\ell} \\ 
    &= - \eta \epsilon(\boldsymbol{\bm \theta}) \sum_{\ell_1,\ell_2} \frac{\partial \epsilon }{ \partial \bm \theta_{\ell_1} } g_{\ell_1 \ell_2}^{+} \frac{\partial \epsilon}{\partial \bm \theta_{\ell_2} } \\
\end{aligned}
\label{QITE Taler Expansion}
\end{equation}

We define the QNTK \(K_{\text{QITE}}\) for QITE:

\begin{definition}[Quantum Neural Tangent Kernel (QNTK) for QITE]
The QNTK for QITE is defined as:
\begin{equation}
\begin{aligned}
    K_{\text{QITE}} &= \sum_{\ell_1,\ell_2} \frac{\partial \epsilon }{ \partial \bm \theta_{\ell_1} } \, g_{\ell_1 \ell_2}^{+} \, \frac{\partial \epsilon}{\partial \bm \theta_{\ell_2} } \\
    &= \sum_{\ell_1,\ell_2} g_{\ell_1 \ell_2}^{+} \, \frac{\partial \epsilon }{ \partial \bm \theta_{\ell_1} } \, \frac{\partial \epsilon}{\partial \bm \theta_{\ell_2} }
    \quad \text{(by the symmetry of } g^{+} \text{)}.
\end{aligned}
\label{K_QITE}
\end{equation}
It governs the decay rate of the residual optimization error \(\epsilon(t)\) by:
\begin{equation}
\epsilon(t+1) - \epsilon(t) \equiv \delta \epsilon 
\approx - \eta \epsilon(\boldsymbol{\bm \theta}) K_\text{QITE}.
\label{QITE Taylor Expansion}
\end{equation}
\end{definition}

We first model \(g\) below:

\begin{lemma}
    Under Assumption~\ref{assump:random}, the average of \(g_{\ell_1 \ell_2}\), \ie, \(\overline{g_{\ell_1 \ell_2}}\), turns out to be a constant dependent on the dimension of Hilbert Space (HS) :
    \begin{equation}
        \overline{g_{\ell_1 \ell_2}} = \frac{N}{N + 1} \delta_{\ell_1 \ell_2},
    \label{overline_g}
    \end{equation}
    where \(N = 2^n\) represents the dimension of HS, and \(n\) denotes the number of qubits. \(\delta_{\ell_1 \ell_2}\) represents the Kronecker delta:
    \begin{equation}
        \delta_{\ell_1 \ell_2} = 
        \begin{cases} 
            0 & \text{if } \ell_1 \neq \ell_2, \\
            1 & \text{if } \ell_1 = \ell_2.
        \end{cases}
    \end{equation}
    
    In the large-\(N\) limit, the fluctuations in \(g_{\ell_1 \ell_2}\) around \(\overline{g_{\ell_1 \ell_2}}\) where we adopt the variance of \(g_{\ell_1 \ell_2}\) around its expectation \(\overline{g_{\ell_1 \ell_2}}\) is given by
    \begin{equation}
        \Delta g_{\ell_1 \ell_2}^2 \coloneqq
        \mathbb{E}(g_{\ell_1 \ell_2}^2) - \overline{g_{\ell_1 \ell_2}}^2 \approx
        \begin{cases}
            \dfrac{2}{N^2}, & \text{if } \ell_1 = \ell_2 = \ell, \\[6pt]
            \dfrac{1}{2N}, & \text{if } \ell_1 \ne \ell_2.
        \end{cases}
    \end{equation}
\label{lemma:quadratic_g}
\end{lemma}

Numerical study for Lemma~\ref{lemma:quadratic_g} is provided in Section \ref{sec:quadratic_loss}, along with the derivation provided in \textit{Appendix \ref{Average Fisher information haar}} and \textit{Appendix \ref{sec:fluctuations_g}}. On average, in the large-\(N\) limit, \(g\) approaches an identity-like structure scaled by a dimension-dependent factor, with the fluctuations become negligible. Consequently, we approximate the average pseudoinverse \(\overline{g^+(\vec{\bm \theta})}\) using \((\overline{g})^+\), greatly simplifying the analysis. 

We now derive the results regarding  \(K_\text{QITE}\). Reviewing the definition of QNTK for GD-based VQAs:

\begin{equation}
    K_{\text{GD}} = \sum_{\ell} \frac{\partial \epsilon }{ \partial \bm \theta_{\ell} } \, \frac{\partial \epsilon}{\partial \bm \theta_{\ell} },
\label{eq:k_gd}
\end{equation}
thus we can summarize the findings below:

\begin{proposition}
Under Assumption~\ref{assump:random}, on average $K_\text{QITE}$ defined in Eq.~(\ref{K_QITE}) and \(K_\text{GD}\) defined in Eq.~(\ref{eq:k_gd}) satisfy:

\begin{equation}
\overline{K_{\text{QITE}}} \simeq \frac{N+1}{N} 
\overline{K_{\text{GD}}},    
\end{equation}
where $N$ is the dimension of the Hilbert space. When \(N\) goes to infinity, theoretically we recover $\lim_{N \to \infty} \overline{K_\text{QITE}} \approx \overline{K_\text{GD}}$.

Furthermore, for small \(\eta\), in the lazy training regime, on average the residual training error \(\epsilon\) follows:

\begin{equation}
    \epsilon(t) \approx \epsilon(0)\exp(-\eta \overline{K} t),
\label{eq:eps_quadratic}
\end{equation}
which implies that \(\epsilon_\text{QITE}\) and \(\epsilon_\text{GD}\) exhibit the following:

\begin{equation}
\epsilon_{\text{QITE}}(t) \approx \epsilon_{\text{GD}}(t) \cdot \exp\left( -\frac{\eta t}{N} \overline{K_{\text{GD}}} \right),    
\end{equation}
where \(\overline{K_{\text{GD}}} = \frac{L \Tr{O^2}}{N^2}\)~\cite{liu2023analytic} under the Assumption \ref{assump:lazy} and Assumption \ref{assump:random}. 
\label{Theorem_K_Quadratic}
\end{proposition}
The numerical study for Proposition~\ref{Theorem_K_Quadratic} is provided in Section \ref{sec:quadratic_loss}, with the derivation in \textit{Appendix \ref{sec:semi_proof_quadratic}}.


At the level of \(K\) itself, the difference is \(\mathcal{O}(1/N)\) and hence negligible in the large-\(N\) limit. However, since the training loss evolves exponentially, even a small discrepancy in \(K\) can lead to a significant error gap over time. We define the logarithmic residual error gap as
\begin{equation}
    \delta_{\log}(t)\;:=\;\ln \epsilon_{\text{GD}}(t) - \ln \epsilon_{\text{QITE}}(t) = \frac{\eta t}{N}\,\overline{K_\text{GD}} \geq 0,
\label{eq:logarithmic_residual_error_gap}
\end{equation}
and the relative error gap between \(\epsilon_{\text{GD}}(t)\) and \(\epsilon_{\text{QITE}}(t)\) as
\begin{equation}
    \delta_{\mathrm{rel}}(t) := \frac{\epsilon_\text{GD}(t) - \epsilon_\text{QITE}(t)}{\epsilon_\text{GD}(t)} = 1 - e^{-\delta_{\log}(t)}.
\end{equation}

\textbf{(i) Convergence dynamics with respect to time in quadratic loss function.}  
When \(t \ll N/(\eta K_\text{GD})\), \ie, \(\delta_{\log}(t) \ll 1\), we perform a Taylor expansion:
\begin{equation}
    e^{-\delta_{\log}(t)} = 1 - \delta_{\log}(t) + \mathcal{O}\left(\delta_{\log}^2(t) \right),
\end{equation}
leading to
\begin{equation}
    \delta_{\mathrm{rel}}(t) = \delta_{\log}(t) + \mathcal{O}(\delta_{\log}^2(t)).
\end{equation}

Using the XXZ scaling~\cite{zhang2024dynamical},
\begin{equation}
    \overline{K_\text{GD}} \sim \mathcal{O}\left(\frac{L n}{N}\right),
\end{equation}
we obtain
\begin{equation}
    \delta_{\mathrm{rel}}(t) \approx \delta_{\log}(t) = \Theta\left(\frac{L n\, \eta t}{N^2} \right).
\end{equation}
Thus, in this time scale, the relative error gap scales as \(\bm \theta(Ln\,\eta t/N^2)\), indicating that the two evolutions remain indistinguishable up to a vanishing error.

However, when \(t = \Theta\bigl(N/(\eta K_\text{GD})\bigr) = \Theta(N^2 / \eta L n)\), we have \(\delta_{\log}(t) = \Theta(1)\), and hence
\begin{equation}
    \epsilon_\text{QITE}(t) = \epsilon_\text{GD}(t) \cdot e^{-\delta_{\log}(t)} \leq e^{-c} \cdot \epsilon_\text{GD}(t),
\end{equation}
where \(c = \Theta(1)\) is a system-size independent constant. This gives a non-vanishing relative error gap:
\begin{equation}
    \delta_{\mathrm{rel}}(t) = 1 - e^{-\delta_{\log}(t)} \geq 1 - e^{-c} = \Theta(1).
\end{equation}

\textbf{(ii) Asymptotic consistency.}  
Assume \(t = \mathcal{O}(N^k)\), \(\eta = \mathcal{O}(N^m)\), and \(L = \mathcal{O}(N^\ell)\), with constants \(k, m, \ell \geq 0\). Then
\begin{equation}
    \delta_{\log}(t) = \mathcal{O}\left(N^{k + m + \ell - 2} \log N\right).
\end{equation}
In the limit \(N \to \infty\), if \(k + m + \ell < 2\), then \(\delta_{\log}(t) \to 0\), implying
\begin{equation}
    \frac{\epsilon_{\mathrm{QITE}}(t)}{\epsilon_{\mathrm{GD}}(t)} = e^{-\delta_{\log}(t)} \to 1.
\end{equation}
Hence, QITE and GD-based VQAs become asymptotically indistinguishable when the total scaling budget is sub-quadratic. The extra \(\log N\) factor is subleading and does not affect the threshold.

\medskip



\subsection{Dynamics of QITE with Linear Loss Function}
\label{sec:linear_qite}

QITE is often applied in ground state search with a linear loss function defined below:

\begin{equation}
    \mathcal{L}(\bm \theta) = \langle O \rangle = \langle \psi_0 | U^\dagger(\boldsymbol{\bm \theta}) O U(\boldsymbol{\bm \theta}) | \psi_0 \rangle.
\label{eq:loss_linear}
\end{equation}
Its convergence with the linear loss function can be described using its residual training error 
\(\epsilon = \langle O \rangle - O_{\text{min}}\). Therefore the difference equation for \(\ell_1\)-th \(\bm \theta \) is formulated as below:

\begin{equation}
\begin{aligned}
    \delta \bm \theta_{\ell_1}(t) \equiv \bm \theta_{\ell_1}(t+1) - \bm \theta_{\ell_1}(t) &= -\eta \sum_{\ell_2} g_{\ell_1 \ell_2}^{+} \frac{\partial \mathcal{L}}{\partial \bm \theta_{\ell_2}} \\
    &= -\eta \sum_{\ell_2} g_{\ell_1 \ell_2}^{+}  \frac{\partial \epsilon}{\partial \bm \theta_{\ell_2}}
\end{aligned}
\end{equation}

With the linear loss, both \(K_\text{QITE}\) and \(\epsilon_\text{QITE}\) exhibit non-linear dynamics~\cite{zhang2024dynamical}. Thereby we go beyond the linear order expansion in Eq.~(\ref{QITE Taler Expansion}) and introduce a higher order correction to the Taylor expansion of the time difference equation for \(\epsilon\):

\begin{equation}
\begin{aligned}
     &\epsilon(t+1) - \epsilon(t) \equiv \delta \epsilon (t) \\
     &\simeq \sum_{\ell_1} \frac{\partial \epsilon }{ \partial \bm \theta_{\ell_1} } \delta \bm \theta_{\ell_1} (t) + \frac{1}{2} \sum_{\ell_1, \ell_2} \frac{\partial^2 \epsilon}{\partial \bm \theta_{\ell_1} \partial \bm \theta_{\ell_2}} \delta \bm \theta_{\ell_1} \delta \bm \theta_{\ell_2} \\
     &= -\eta \sum_{\ell_1,\ell_2} g_{\ell_1 \ell_2}^{+} \frac{\partial \epsilon }{ \partial \bm \theta_{\ell_1} }  \frac{\partial \epsilon}{\partial \bm \theta_{\ell_2} } \\
     &+ \frac{1}{2} \eta^2 \sum_{\ell_1,\ell_2, \ell_3,\ell_4} g_{\ell_1 \ell_3}^{+} g_{\ell_2 \ell_4}^{+} \frac{\partial^2 \epsilon}{\partial \bm \theta_{\ell_1} \partial \bm \theta_{\ell_2}}  \frac{\partial \epsilon }{ \partial \bm \theta_{\ell_3} }  \frac{\partial \epsilon}{\partial \bm \theta_{\ell_4} }.
\end{aligned}
\end{equation}
Here we define the \textit{quantum meta-kernel} (\textit{dQNTK}) for QITE:

\begin{definition}[Quantum Meta-Kernel for QITE]
The quantum meta-kernel $\mu_\text{QITE}$ associated with QITE is defined as:
\begin{equation}
    \mu_\text{QITE} = \sum_{\ell_1,\ell_2, \ell_3,\ell_4} g_{\ell_1 \ell_3}^{+} g_{\ell_2 \ell_4}^{+} \frac{\partial^2 \epsilon}{\partial \bm \theta_{\ell_1} \partial \bm \theta_{\ell_2}}  \frac{\partial \epsilon }{ \partial \bm \theta_{\ell_3} }  \frac{\partial \epsilon}{\partial \bm \theta_{\ell_4} }.
\end{equation}
\end{definition}
Then, \(\delta \epsilon (t)\) reduces to:
\begin{equation}
    \delta \epsilon (t) \simeq -\eta K_\text{QITE} + \frac{1}{2}\eta^2\mu_\text{QITE}.
\label{eq:delta_eps_second_order}
\end{equation}
We also define the \textit{relative dQNTK} for QITE as follows: 

\begin{definition}[Relative quantum meta-kernel (dQNTK) for QITE]
The relative dQNTK for QITE is defined as the ratio between $\mu_\text{QITE}(t)$ and $K_\text{QITE}(t)$, given by:
\begin{equation}
\lambda_\text{QITE}(t) = \frac{\mu_\text{QITE}(t)}{K_\text{QITE}(t)}.
\end{equation}
\end{definition}

This is similar to the way of defining \textit{relative dQNTK} for GD-based VQAs \(\lambda_\text{GD}(t)\)~\citep{zhang2024dynamical}:

\begin{equation}
    \lambda_\text{GD}(t) = \frac{\mu_\text{GD}(t)}{K_\text{GD}(t)},
\end{equation}
where 

\begin{equation}
    \mu_\text{GD} = \sum_{\ell_1,\ell_2} \frac{\partial^2 \epsilon}{\partial \bm \theta_{\ell_1} \partial \bm \theta_{\ell_2}}  \frac{\partial \epsilon }{ \partial \bm \theta_{\ell_1} }  \frac{\partial \epsilon}{\partial \bm \theta_{\ell_2} }.
\end{equation}




Accordingly, we derive a first-order difference equation for \(K_\text{QITE}\) with linear loss function (See \textit{Appendix \ref{sec:time_diff_K_QITE}} for detailed derivations):

\begin{equation}
\begin{aligned}
    \delta K_\text{QITE} &= -2\eta \mu_\text{QITE}  + \mathcal{O}(\eta^2).\\
\label{time_diff_K}
\end{aligned}
\end{equation}

Combining the results from Eq.~(\ref{time_diff_K}) and Eq.~(\ref{eq:delta_eps_second_order}), we can solve this as~\cite{zhang2024dynamical}:

\begin{equation}
    2\lambda_\text{QITE} \epsilon_\text{QITE}(t) = K_\text{QITE}(t) \propto e^{-2\eta \lambda_\text{QITE} t},
\end{equation}
where we can see that the dynamics of both \(K_{\text{QITE}}\) and \(\lambda_{\text{QITE}}\) are driven by \(\lambda_{\text{QITE}}\). Therefore we now formalize the findings:

\begin{proposition}
Under Assumption~\ref{assump:random}, QITE and GD-based VQAs dynamics with linear loss function exhibit the following analytic structure:

First, on average \(\lambda_\text{QITE}\) and \(\lambda_\text{GD}(t)\) satisfies:
\begin{equation}
\overline{\lambda_{\text{QITE}} (t)} \simeq \frac{N+1}{N} \overline{\lambda_{\text{GD}}(t)},
\end{equation}
where \(N\) is the dimension of HS. Based on this, the relation between \(K_\text{QITE}\) and \(K_\text{GD}\) follows:

\begin{equation}
K_{\text{QITE}}(t) \approx K_{\text{GD}} (t) \cdot \exp\left( -\frac{2 \eta t}{N} \overline{\lambda_{\text{GD}} (t)} \right).
\end{equation}

Consequently, the residual training error of QITE satisfies the following approximate expression:
\begin{equation}
\epsilon_{\text{QITE}}(t) \approx \frac{N}{2(N+1)\overline{\lambda_{\text{GD}}} } K_{\text{GD}} (t) \cdot \exp\left( -\frac{2 \eta t}{N} \overline{\lambda_{\text{GD}} (t)} \right)
\end{equation}

\label{Theorem_K_Linear}
\end{proposition}

The numerical study for Proposition~\ref{Theorem_K_Linear} is provided in Section \ref{sec:numeric_linear}, with the derivation in \textit{Appendix \ref{sec:semi_proof_linear}}.

A direct substitution of the XXZ results into Eq.~(\ref{eq:logarithmic_residual_error_gap}) gives
\begin{equation}
\begin{aligned}
    \delta_{\log}(t) &:= \ln \epsilon_{\text{GD}}(t) - \ln \epsilon_{\text{QITE}}(t) \\
    &= \ln\left( \frac{1}{2\lambda_{\text{GD}}(t)} K_{\text{GD}}(t) \right) \\
    &\quad - \ln\left( \frac{N}{2(N+1) \lambda_{\text{GD}}(t)} K_{\text{GD}}(t) \cdot e^{ -\frac{2\eta t}{N} \lambda_{\text{GD}}(t) } \right) \\
    &= \ln\left( \frac{N+1}{N} \right) + \frac{2\eta t}{N} \lambda_{\text{GD}}(t).
\end{aligned}
\label{eq:linear_log_gap_xxz}
\end{equation}

Analyzing the first term in the large-$N$ limit:
\[
\ln\left(\frac{N+1}{N}\right) = \ln\left(1 + \frac{1}{N}\right) = \frac{1}{N} - \frac{1}{2N^2} + \mathcal{O}\left(\frac{1}{N^3}\right),
\]
which is of order $\mathcal{O}(1/N)$ and hence negligible as $N \to \infty$.

From the XXZ scaling results~\cite{zhang2024dynamical}, we have
\begin{equation}
    \lambda_{\text{GD}}(t) \simeq \overline{\lambda_0} \sim \mathcal{O}(L/N).
\end{equation}

Substituting into Eq.~\eqref{eq:linear_log_gap_xxz} and neglecting the constant offset, we obtain
\begin{equation}
    \delta_{\log}(t) \approx \frac{2\eta L t}{N^2}.
\label{eq:delta_log_xxz}
\end{equation}

\textbf{(i) Convergence dynamics with respect to time in linear loss function.}  
When \( t \ll N^2 / (\eta L) \), the error gap remains small:
\begin{equation}
    \delta_{\mathrm{rel}}(t) = \delta_{\log}(t) + \mathcal{O}(\delta_{\log}^2(t)) \approx \frac{2\eta L t}{N^2}.
\end{equation}
Therefore, in this regime, GD-based VQAs and QITE are indistinguishable up to a vanishingly small relative error.

However, for \( t = \Theta(N^2 / \eta L) \), we have \( \delta_{\log}(t) = \Theta(1) \), leading to an exponential gap:
\begin{equation}
\begin{aligned}
   \epsilon_{\mathrm{QITE}}(t) &= \epsilon_{\mathrm{GD}}(t)\, e^{-\delta_{\log}(t)} \leq e^{-c} \epsilon_{\mathrm{GD}}(t), \\
   \delta_{\mathrm{rel}}(t) &\geq 1 - e^{-c} = \Theta(1).
\end{aligned}
\end{equation}
Thus, QITE removes an $\mathcal{O}(1)$ fraction of the error of GD-based VQAs in this regime.

\textbf{(ii) Asymptotic consistency.}  
Assume \( t = \mathcal{O}(N^{k}) \), \( \eta = \mathcal{O}(N^{m}) \), and \( L = \mathcal{O}(N^\ell) \), with constants \(k, m, \ell \geq 0\). Then
\[
    \delta_{\log}(t) = \frac{2\eta L t}{N^2} = \mathcal{O}(N^{k + m + \ell - 2}).
\]
Hence in $N \to \infty$:
\[
    k + m + \ell < 2 \;\;\Longrightarrow\;\;
    \delta_{\log}(t) \to 0 
    \;\;\Longrightarrow\;\;
    \epsilon_{\mathrm{QITE}}(t) / \epsilon_{\mathrm{GD}}(t) \to 1,
\]
\ie, QITE and GD-based VQAs become asymptotically indistinguishable.


\begin{figure*}
    \centering
    \includegraphics[width=0.85\textwidth]{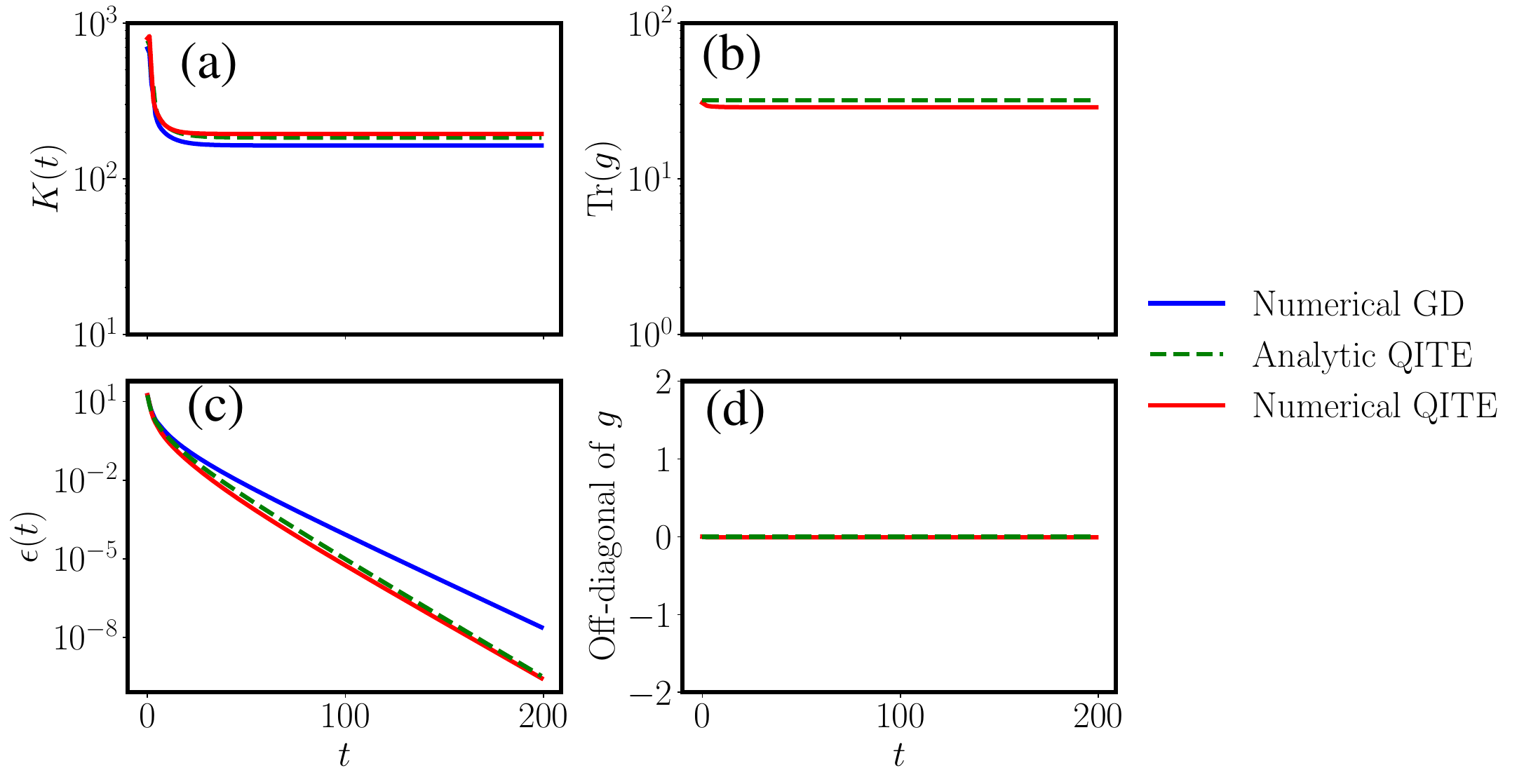}    
    \vspace{-10pt}
    \caption{Training Dynamics of GD-based VQAs and QITE with quadratic loss function. Here, in the example of XXZ model, we respectively investigate the QNTK \(K(t)\), the residual error \(\epsilon(t)\), the average trace and the off-diagonal terms of the Fubini-study metric tensor \(g\). Each numerical curves are plotted by averaging over 50 times, indicating 50 initializations. We adopt HEA ansatz with 6 layers, and set the number of qubits \(n = 3\). The learning rate for optimization is \(\eta = 0.001\) with 200 steps. Red curves (denoted as ``Numerical QITE") represent ensemble average results of QITE. Blue curves (denoted as ``Numerical GD") represent the ensemble average numerical results of GD-based VQAs. Green dashed curves represent the analytic prediction of the dynamics of QITE. We also plot the gray lines in the plot of average trace of \(g\), indicating 50 random samples.}
    \label{fig:Quadratic_results}
\end{figure*}

\begin{figure*}
    \centering
\includegraphics[width=0.85\textwidth]{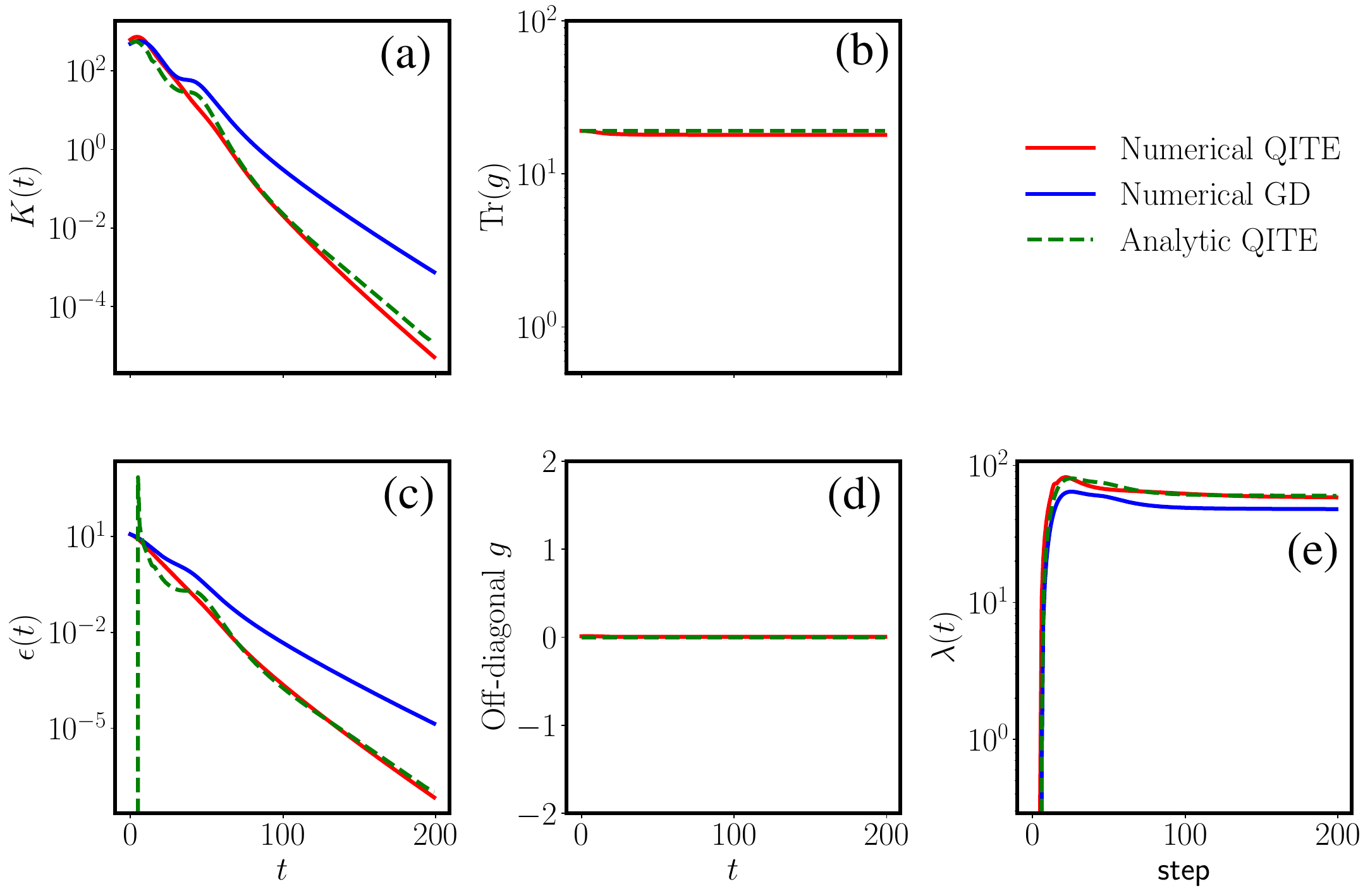}    
    \vspace{-10pt}
    \caption{Training Dynamics of QNNs in GD-based VQAs and QITE with linear loss function. Comparing with Fig \ref{fig:Quadratic_results}, we additionally investigate the relative dQNTK value \(\lambda (t)\) which drives the dynamics of \(K(t)\) and \(\epsilon(t)\). Each numerical curves are plotted by averaging over 50 times, indicating 50 initializations. The settings is identical to Fig.~\ref{fig:Quadratic_results}.}
    \label{Linear_results}
\end{figure*}

\subsection{Numerical Studies}

\subsubsection{Numerical Studies with Quadratic Loss Function}
\label{sec:quadratic_loss}

We numerically examine the training dynamics of GD-based VQAs and QITE using the XXZ model with a quadratic loss. As shown in Fig.~\ref{fig:Quadratic_results}, the QNTK value $K(t)$ exhibits distinct behaviors under the two VQAs. Both \(K_\text{QITE}\) and \(K_\text{GD}\) values remain constant throughout the optimization, and \(K_\text{QITE}\) is larger than \(K_\text{GD}\). Besides, the relation is analytically predictable by following the formula derived in Theorem~\ref{Theorem_K_Quadratic}. Correspondingly, the training error $\epsilon(t)$ of QITE demonstrates a steeper descent compared to GD-based VQAs. Importantly, the QITE error dynamics also closely matches the analytic prediction given by $\epsilon_{\text{GD}}(t) \cdot \exp(-\frac{\eta t}{N} K_{\text{GD}})$, validating our theoretical approximation in Theorem~\ref{Theorem_K_Quadratic}.

According to our theory, the discrepancy in $\epsilon(t)$ trajectories arises from the properties of the underlying Fubini–study metric tensor $g(t)$, particularly its trace component. To validate this, we conduct a numerical analysis of both the average trace $\operatorname{Tr}(g)$ and the off-diagonal elements over the course of training. As shown in Fig.~\ref{fig:Quadratic_results}, the trace of QITE remains relatively stable and closely follows the analytic prediction. In contrast, the significant suppression of the off-diagonal terms indicates that the QITE dynamics are primarily driven by local parameter updates (the diagonal terms), with non-local correlations having only a negligible impact.

\subsubsection{Numerical Studies with Linear Loss Function}
\label{sec:numeric_linear}

We further investigate the training dynamics with the linear loss function defined in Eq.~(\ref{eq:loss_linear}). As shown in Fig.~\ref{Linear_results},  the relative dQNTK \(\lambda(t)\) converges to a stable value at late time. Notably, the converged value of \(\lambda_{\text{QITE}}\) is consistently larger than that of \(\lambda_{\text{GD}}\), and their ratio follows the analytic prediction in Theorem~\ref{Theorem_K_Linear}. Correspondingly, the QNTK \(K(t)\) no longer remains constant as in the quadratic loss case. Both \(K_{\text{GD}}(t)\) and \(K_{\text{QITE}}(t)\) exhibit exponential decay, with \(K_{\text{QITE}}(t)\) decaying at a faster rate. This relative difference is again analytically predictable, as the ratio \(K_{\text{QITE}}(t)/K_{\text{GD}}(t)\) follows the exponential form derived from the difference in \(\lambda\) values. As a consequence, the training error \(\epsilon(t)\) of QITE and GD-based VQAs exhibits a similar exponential decay behavior to that of \(K(t)\). These dynamics are well captured by the analytic expression as described in Theorem~\ref{Theorem_K_Linear}.

We also examine the numerical properties of \(g(t)\) with linear loss function. Similar to the case with quadratic loss function, the trace \(\operatorname{Tr}(g)\) of QITE remains constant over time. Additionally, the off-diagonal components of \(g\) are strongly suppressed. These structural trends in \(g(t)\) directly govern the observed behavior of \(\lambda(t)\), \(K(t)\), and \(\epsilon(t)\), and are in excellent agreement with our theoretical analysis.


\section{Methods}

\subsection{Tasks, Ansatz, and Observable}

\textbf{Tasks.} 
The first task is to establish a first-principle equivalence between QITE and QNGD-based VQAs through the variational principle, encompassing linear, quadratic and more general loss functions up to an integration constant. We then focus on the cases of quadratic and linear loss functions. The tasks lie in two assumptions, formulated as Assumption \ref{assump:lazy} and Assumption \ref{assump:random}. Specifically, the task requires developing a closed-form model that quantitatively connects the training dynamics of classical GD-based VQAs and QNGD-based VQAs, thereby characterizing the behavior of QITE. This enables a systematic comparison of the convergence behavior, curvature scaling, and training error dynamics across these two paradigms.

\textbf{Ansatz.}
We employ a hardware-efficient ansatz (HEA)~\cite{kandala2017hardware} architecture to parameterize the quantum state \( |\psi(\bm{\bm \theta})\rangle \) used throughout our experiments. The circuit is composed of \(D\) alternating layers of local rotations and entangling operations applied to \(n\) qubits. Specifically, in each layer, we apply a sequence of single-qubit gates \( R_Y(\bm \theta^{(d)}_{i,1}) \) and \( R_Z(\bm \theta^{(d)}_{i,2}) \) to every qubit \(i\), where \(d\) indexes the circuit depth. These rotations are followed by entangling operations implemented as a brickwall-patterned array of CNOT gates acting on nearest neighbors. Each pair of \( R_Y \) and \( R_Z \) rotations contributes two trainable parameters per qubit per layer. Thus, for a circuit with depth \(D\), the total number of trainable parameters is \( L = 2nD \). The variational parameters \( \bm{\bm \theta} = (\bm \theta_1, \dots, \bm \theta_L) \) are initialized independently at random from a uniform distribution. While the HEA ansatz is not exactly Haar-distributed, it has been observed~\cite{mcclean2018barren, cerezo2021cost} that such circuits with sufficient depth can approximate Haar randomness and form approximate unitary 2-designs, especially when randomized layer permutations or Pauli basis choices are employed. Consequently, we adopt HEA to simulate random circuit behavior in our numerical experiments.

\textbf{Observable.}
We consider optimization tasks involving a general Hermitian observable \( O \), acting on an \(n\)-qubit system. To simplify the analysis, we often assume \( O \) to be traceless, \ie, \( \operatorname{Tr} (O) = 0 \), as this removes constant energy offsets that do not affect optimization dynamics. A typical traceless observable can be expanded in the Pauli basis as
\begin{equation}
    O = \sum_{i=1}^{N} c_i \, \hat{P}_i,
\end{equation}
where \( \hat{P}_i \in \{\hat{\sigma}^x, \hat{\sigma}^y, \hat{\sigma}^z\}^{\otimes n} \setminus \{\mathbb{I}^{\otimes n}\} \) are nontrivial Pauli strings and \( c_i \in \mathbb{R} \) are real coefficients. For concrete examples and exact expressions, we consider structured XXZ Hamiltonians below:
\begin{equation}
    O_{\text{XXZ}} = - \sum_{i=1}^{n}
    \left[ \hat{\sigma}^x_i \hat{\sigma}^x_{i+1} +
           \hat{\sigma}^y_i \hat{\sigma}^y_{i+1} +
           J\left( \hat{\sigma}^z_i \hat{\sigma}^z_{i+1} + \hat{\sigma}^z_i \right) \right],
\end{equation}
where \(J\) is a tunable interaction strength.

\subsection{Haar Random Ensemble As A Statistical Assumption}

For analytic tractability, we follow previous studies~\cite{mcclean2018barren, liu2023analytic, cerezo2021cost} and adopt the \emph{Haar random ensemble} to model the typical parameterized unitaries. Under random initialization, the variational circuit is modeled as being drawn from the Haar measure on the unitary group \( \mathcal{U}(N) \). This modeling relies on the assumption that the circuit ensemble forms a \emph{unitary \(k\)-design}~\cite{roberts2017chaos, cotler2017chaos, liu2018spectral, liu2020scrambling}. Formally, an ensemble \(\mathcal{E} = \{U_i\}\) is said to form a unitary \(k\)-design if for any degree-\(k\) polynomial \(P(U, U^\dagger)\) in the matrix elements of \(U\), we have
\begin{equation}
    \mathbb{E}_{U \sim \mathcal{E}}[P(U, U^\dagger)] = \mathbb{E}_{U \sim \text{Haar}}[P(U, U^\dagger)].
\end{equation}
This condition ensures that the ensemble \(\mathcal{E}\) statistically mimics the Haar measure up to the \(k\)-th moment, allowing us to analytically compute quantities like

\begin{equation}
    \overline{ \frac{\partial \epsilon}{\partial \bm \theta_\ell} } = 0,
\qquad
\overline{ g_{\ell_1 \ell_2} } = \frac{N}{N+1} \delta_{\ell_1 \ell_2},
\end{equation}
and other key quantities used in our analysis. For example, a unitary 2-design suffices to match the second-order statistical properties of the Haar distribution, such as the expected value of gradients and metric tensors.

\subsection{Variational Principles and First-Principle Equivalence}
\label{sec:Variational Principles and First-Principle Equivalence}

\textbf{Variational principles.}
Generally, a variational principle denotes that the evolution of a system can be characterized as the stationary point of a \textit{functional}, \ie, a mapping that assigns to each trial function a numerical quantity. Instead of solving the governing equations directly, one introduces a parametrized family of trial functions and determines the parameters by requiring that a chosen functional is optimized with respect to the variational parameters. 
The choice of functional encodes the physical or mathematical structure of the problem. Typically, it could be an \emph{action functional}, defined as the time integral of the Lagrangian, whose stationarity yields the Euler--Lagrange equations. Similarly, in optimization and machine learning scheme, analogous energy-like or loss functionals define objective landscapes to be minimized. 
Formally, the variational principle enforces a variational functional $\mathcal{F}[\psi]$  satisfying
\begin{equation}
\delta \mathcal{F}[\psi] = 0,
\end{equation}
subject to admissible variations of $\psi$. Though different problems instantiate different functionals, the unifying principle is that the governing equations are recovered by requiring that $\mathcal{F}$ is stationary under variations within the chosen family of trial functions. 
This formulation emphasizes that the essence of a variational method lies not in dynamical details, but in the specification of the functional whose extremum characterizes the desired variational solution~\cite{kibble2004classical,arnol2013mathematical,cerezo2021variational}.

\textbf{First-principle equivalence.}
Two algorithms $\mathcal{A}$ and $\mathcal{B}$ can be regarded as \emph{first-principle equivalent} if they induce the same continuous time flow on the variational manifold, up to a relabeling of coordinates or a rescaling of time. 
In this definition, \textit{``first-principle''} emphasizes that the equivalence is established at the level of the underlying variational functional rather than at the level of discretized updates or empirical performance. 
In particular, if the Euler–Lagrange equations derived from the variational principles underlying $\mathcal{A}$ and $\mathcal{B}$ coincide, then the induced state trajectories are identical, thereby justifying the notion of first-principle equivalence. More concretely, consider two algorithms $\mathcal{A}$ and $\mathcal{B}$, each formulated through a variational principle:
\begin{equation}
    \delta \mathcal{F}_{\mathcal{A}}[\psi] = 0, 
    \qquad 
    \delta \mathcal{F}_{\mathcal{B}}[\psi] = 0,
\end{equation}
where $\mathcal{F}_{\mathcal{A}}$ and $\mathcal{F}_{\mathcal{B}}$ are the respective variational functionals. 
If the two functionals coincide, \ie, \ $\mathcal{F}_{\mathcal{A}}[\psi] \equiv \mathcal{F}_{\mathcal{B}}[\psi]$ up to an integration constant, then the corresponding Euler--Lagrange equations are identical, leading to the same continuous-time dynamics on the variational manifold. Thus, although $\mathcal{A}$ and $\mathcal{B}$ may arise from different algorithmic constructions, \eg, GD-based VQAs versus QITE, their underlying dynamics are identical and independent of implementation details. Therefore, it is adequate to analyze QITE from the perspective of QNGD in the continuous-time limit because the variational flow of QNGD-based VQAs faithfully reproduces that of QITE, allowing us to apply the QNTK framework and thereby providing a principled and tractable approach to its analysis. \\

\section{Discussion}

This work goes beyond existing analytic theories of quantum learning tasks driven by GD in general QNNs. By building a first-principle equivalence, this work explains why QITE and QNGD-based VQAs are equivalent up to a constant scaling factor within continuous time limit, enabling QITE to be interpreted through QNGD's natural geometric framework. Building on this connection, we derive the corresponding training dynamics and compare them with those of GD-based VQAs. Our analysis reveals regimes where QITE offers a convergence advantage, and further suggests that well-developed analytical frameworks, such as the QNTK theory, can be applied to advanced VQAs. This could potentially contribute to the identification and design of quantum algorithms that outperform their classical counterparts.

The theory developed in this work has several potential applications. One direction is to analyze QITEs with specific structures, \eg, symmetry~\cite{wang2022symmetric,larocca2022diagnosing}, that may enhance the performance of QITE. Such potential can be harnessed to guide the design of new quantum algorithms. Another direction is to extend the analysis beyond the lazy training regime or the specific problem considered here, exploring how the dynamics may change and whether the convergence advantage persists. In addition, the notations and formulations developed in this work could be adopted for other analytical studies, such as investigations of expressivity.

\section*{Acknowledgment}

MC and JL is supported in part by the University of Pittsburgh, School of Computing and Information, Department of Computer Science, Pitt Cyber, Pitt Momentum fund, PQI Community Collaboration Awards, John C. Mascaro Faculty Scholar in Sustainability, NASA under award number 80NSSC25M7057, and Fluor Marine Propulsion LLC (U.S. Naval Nuclear Laboratory) under award number 140449-R08. This research used resources of the Oak Ridge Leadership Computing Facility, which is a DOE Office of Science User Facility supported under Contract DE-AC05-00OR22725. QZ and BZ acknowledge support from NSF (OMA-2326746,2350153, CCF-2240641), ONR (N00014-23-1-2296), DARPA (HR00112490453, HR0011-24-9-0362, D24AC00153-02) and AFOSR MURI FA9550-24-1-0349.

\bibliography{egbib}
\onecolumngrid
\newpage
\let\addcontentsline\oldaddcontentsline
\tableofcontents

\section{Summary of Notations}

Notations are elaborated in Table \ref{Notation}.

\begin{table}[hpt!]
\centering
\caption{Notations}
\begin{tabular}{l|p{12cm}}
\hline
\textbf{Symbol} & \textbf{Description} \\
\hline
\(D\) & Depth of QNN, \ie, number of layers in the parameterized quantum circuit (PQC). \\
\(n\) & Number of qubits. \\
\(L\) & Number of variational parameters; for hardware-efficient ansatz, \(L = 2nD\). \\
\(N\) & Hilbert space dimension, \(N = 2^n\). \\
\(\bm \theta\) & Vector of variational parameters, \(\bm \theta = (\bm \theta_1, \dots, \bm \theta_L)\). \\
\(|\psi(\bm \theta)\rangle\) & Normalized parameterized quantum state prepared by PQC. \\
\(|\psi_0\rangle\) & Input (initial) quantum state. \\
\(U(\bm \theta)\) & Parameterized unitary of the ansatz circuit. \\
\(U_\ell^-, U_\ell^+\) & Partial products of the ansatz unitary before and after parameter \(\bm \theta_\ell\), as in Eq.~(16). \\
\(V_\ell(\bm \theta_\ell)\) & Parameterized single-qubit rotation \(e^{i\bm \theta_\ell X_\ell}\) with Hermitian generator \(X_\ell\). \\
\(X_\ell\) & Hermitian generator of parameter \(\bm \theta_\ell\), often a single Pauli or tensor product of Paulis. \\
\hline
\(O\) & Hermitian observable; in this work taken as the system Hamiltonian \(H\). \\
\(H\) & System Hamiltonian; for the XXZ model, see \(O_{\mathrm{XXZ}}\). \\
\(O_0\) & Target value of observable \(O\). \\
\(O_{\min}\) & Minimum eigenvalue of \(O\) (ground state energy). \\
\(E_\tau\) & Expectation value of \(O\) at imaginary time \(\tau\), \(E_\tau = \langle \psi(\bm \theta(\tau))| O | \psi(\bm \theta(\tau)) \rangle\). \\
\(\langle O \rangle\) & Shorthand for \(E_\tau\). \\
\(\mathcal{L}(\bm \theta)\) & Loss function.\\
\(\epsilon\) & Residual training error. \\
\hline
\(\eta\) & Learning rate. \\
\(\tau\) & Imaginary time variable; related to gradient descent steps via \(\eta \to 0\) continuous-time limit. \\
\(\partial_\tau\) & Imaginary-time derivative; interchangeably written as \(\frac{\partial}{\partial \tau}\). \\
\(\delta_{\ell_1\ell_2}\) & Kronecker delta. \\
\hline
\(g_{\ell_1\ell_2}(\bm \theta)\) & Fubini–Study metric tensor (real part of the quantum geometric tensor). \\
\(g^+_{\ell_1\ell_2}(\bm \theta)\) & Pseudoinverse of \(g_{\ell_1\ell_2}(\bm \theta)\). \\
\(F_{\ell_1\ell_2}(\bm \theta)\) & Quantum Fisher information matrix (QFIM), \(F = 4g\). \\
\(K_{\mathrm{GD}}\) & Quantum neural tangent kernel (QNTK) for gradient descent: \(K_{\mathrm{GD}} = \sum_{\ell} (\partial_\ell \epsilon)^2\). \\
\(K_{\mathrm{QITE}}\) & QNTK for QITE-based optimization: \(K_{\mathrm{QITE}} = \sum_{\ell_1,\ell_2} g^+_{\ell_1\ell_2} \, \partial_{\ell_1} \epsilon \, \partial_{\ell_2} \epsilon\). \\
\(\mu_{\mathrm{GD}}\) & Quantum meta-kernel (dQNTK) for GD: \(\mu_{\mathrm{GD}} = \sum_{\ell_1,\ell_2} \partial^2_{\ell_1\ell_2} \epsilon \, \partial_{\ell_1}\epsilon \, \partial_{\ell_2}\epsilon\). \\
\(\mu_{\mathrm{QITE}}\) & Quantum meta-kernel for QITE: \(\mu_{\mathrm{QITE}} = \sum_{\ell_1,\ell_2,\ell_3,\ell_4} g^+_{\ell_1\ell_3} g^+_{\ell_2\ell_4} \, \partial^2_{\ell_1\ell_2} \epsilon \, \partial_{\ell_3}\epsilon \, \partial_{\ell_4}\epsilon\). \\
\(\lambda_{\mathrm{GD}}\) & Relative dQNTK for GD: \(\lambda_{\mathrm{GD}} = \mu_{\mathrm{GD}} / K_{\mathrm{GD}}\). \\
\(\lambda_{\mathrm{QITE}}\) & Relative dQNTK for QITE: \(\lambda_{\mathrm{QITE}} = \mu_{\mathrm{QITE}} / K_{\mathrm{QITE}}\). \\
\hline
\(\delta_{\mathrm{log}}(t)\) & Logarithmic residual error gap: \(\delta_{\mathrm{log}}(t) = \ln \epsilon_{\mathrm{GD}}(t) - \ln \epsilon_{\mathrm{QITE}}(t)\). \\
\(\delta_{\mathrm{rel}}(t)\) & Relative error gap: \(\delta_{\mathrm{rel}}(t) = 1 - e^{-\delta_{\mathrm{log}}(t)}\). \\
\hline
\(O_{\mathrm{XXZ}}\) & XXZ Hamiltonian: \(O_{\mathrm{XXZ}} = -\sum_{i=1}^n \left[ \sigma^x_i\sigma^x_{i+1} + \sigma^y_i\sigma^y_{i+1} + J \left( \sigma^z_i\sigma^z_{i+1} + \sigma^z_i \right) \right]\). \\
\(J\) & Coupling parameter in the XXZ model; controls anisotropy between \(z\)-axis and \(x/y\)-axis interactions. \\
\(\sigma^x, \sigma^y, \sigma^z\) & Pauli operators. \\
\(\hat{P}_i\) & Generic Pauli string in the expansion \(O = \sum_i c_i \hat{P}_i\). \\
\(c_i\) & Real coefficient of Pauli string \(\hat{P}_i\) in observable \(O\). \\
\hline
\end{tabular}
\label{Notation}
\end{table}

For clarity, we consider the observable \(O\) to be the system Hamiltonian \(H\) throughout this work. Besides, we interchangeably use \(\frac{\partial}{\partial \tau}\) and \(\partial_\tau\). We also interchangeably use \(\langle O \rangle\) and \(E_\tau\).

\section{Overview of QITE}
\label{app:overview_qite}

In this section, we briefly review quantum simulation tasks~\cite{haegeman2011time, li2017efficient, yuan2019theory}, with a particular focus on both real-time and imaginary-time evolution. We then describe the variational principle underlying QITE, restricted to the pure state scenario~\cite{mcardle2019variational}. We denote the Hermitian observable generating the dynamics as \(O\), which replaces the standard Hamiltonian symbol \(H\). This notation aligns with our later discussion, where the observable also defines the loss function.

\subsection{Real Time Evolution}

In real time, the evolution of a quantum state is governed by the Schrödinger equation:
\begin{equation}
    \frac{d|\psi(t)\rangle}{dt} = -iO|\psi(t)\rangle,
\end{equation}
where the reduced Planck constant \(\hbar\) is absorbed into the definition of \(O\).

The corresponding unitary evolution operator is:
\begin{equation}
    U(t) = e^{-iOt}.
\end{equation}

The normalized state at time \(t\) is then given by:
\begin{equation}
    |\psi(t)\rangle = A(t) U(t) |\psi(0)\rangle,
\end{equation}
where the normalization factor is
\begin{equation}
    A(t) = \frac{1}{\sqrt{\langle\psi(0)| e^{-2iOt} |\psi(0)\rangle}}.
\end{equation}

\subsection{Imaginary Time Evolution}

Imaginary time evolution replaces \(t\) with \(\tau = it\), resulting in a non-unitary flow. The imaginary-time Schrödinger equation becomes:
\begin{equation}
    \frac{d|\psi(\tau)\rangle}{d\tau} = -O|\psi(\tau)\rangle.
\end{equation}

Its solution up to normalization is:
\begin{equation}
    |\psi(\tau)\rangle = A(\tau) e^{-O\tau} |\psi(0)\rangle,
\end{equation}
where the normalization factor is
\begin{equation}
    A(\tau) = \frac{1}{\sqrt{\langle\psi(0)| e^{-2O\tau} |\psi(0)\rangle}}.
\end{equation}

\subsection{Differentiating the Normalized State}

Differentiating the normalized QITE state gives:
\begin{equation}
    \frac{d}{d\tau} |\psi(\tau)\rangle
    = \frac{dA}{d\tau} e^{-O\tau} |\psi(0)\rangle + A(\tau) \frac{d}{d\tau} e^{-O\tau} |\psi(0)\rangle.
    \label{eq:diff_normalized_qite}
\end{equation}

Using the normalization condition \(\langle \psi(\tau) | \psi(\tau) \rangle = 1\), and the fact that \(A(\tau) \in \mathbb{R}\), \(O^\dagger = O\), and \([e^{-O\tau}, O] = 0\), we compute the derivative of \(A(\tau)\) as follows:
\begin{equation}
\begin{aligned}
    \frac{dA}{d\tau} &= \frac{d}{d\tau} \left( \langle \psi(0) | e^{-2O\tau} | \psi(0) \rangle \right)^{-1/2} \\
    &= -\frac{1}{2} A^3(\tau) \langle \psi(0) | (-2O) e^{-2O\tau} | \psi(0) \rangle \\
    &= A^3(\tau) \langle \psi(0) | O e^{-2O\tau} | \psi(0) \rangle \\
    &= A(\tau) \langle \psi(\tau) | O | \psi(\tau) \rangle \\
    &= A(\tau) E_\tau,
\end{aligned}
\end{equation}
where \(E_\tau := \langle \psi(\tau) | O | \psi(\tau) \rangle\) denotes the instantaneous energy.

Substituting into Eq.~\eqref{eq:diff_normalized_qite}, we obtain the Wick-rotated Schrödinger equation:
\begin{equation}
\begin{aligned}
    \frac{d}{d\tau} |\psi(\tau)\rangle
    &= A(\tau) E_\tau e^{-O\tau} |\psi(0)\rangle - A(\tau) O e^{-O\tau} |\psi(0)\rangle \\
    &= (E_\tau - O) |\psi(\tau)\rangle.
\end{aligned}
\end{equation}

\subsection{Variational Formulation of QITE}
\label{sec:Variational Formulation of QITE}

On a variational manifold~\cite{verstraete2008matrix, ashida2018variational, yuan2019theory}, we consider a normalized parameterized trial state \(|\psi(\vec{\bm \theta}(\tau))\rangle\) with real parameters \(\vec{\bm \theta} \in \mathbb{R}^L\). The QITE evolution equation is then approximated as:
\begin{equation}
    \sum_{j} \frac{\partial |\psi(\vec{\bm \theta}(\tau))\rangle}{\partial \bm \theta_j} \, \dot{\bm \theta}_j \approx (E_\tau - O) |\psi(\vec{\bm \theta}(\tau))\rangle,
\end{equation}
where \(\dot{\bm \theta}_j := \frac{d\bm \theta_j}{d\tau}\) denotes the imaginary-time derivative of the parameters.  

This variational formulation enables efficient simulation of imaginary-time dynamics within a tractable subspace. As discussed in~\cite{yuan2019theory}, there exist three variational principles (Dirac–Frenkel, McLachlan, and time-dependent variational principle) that are equivalent when parameters are complex. However, since we restrict to real parameters, only McLachlan’s variational principle is applicable.



\textbf{McLachlan’s Variational Principle~\cite{mclachlan1964variational, mcardle2019variational, yuan2019theory}.}  
This principle offers a natural way to project non-unitary quantum dynamics, such as imaginary time evolution, onto a variational ansatz. Instead of requiring the trial state to follow the exact equation of motion, McLachlan’s approach minimizes the distance between the true derivative of the state and its projection within the variational manifold. Specifically, the evolution path is chosen such that the deviation
\(\left( \frac{\partial}{\partial \tau} + O - E_\tau \right) | \psi( \bm \theta(\tau))\rangle\)
remains orthogonal to the tangent space of allowed variations. The condition is formally expressed as:
\begin{equation}
    \delta \left\| \left( \frac{\partial}{\partial \tau} + O - E_\tau \right) | \psi( \bm \theta(\tau))\rangle \right\| = 0,
\label{eq:McLachlan’s_Variational_Principle}
\end{equation}
which ensures that the evolution follows the most faithful trajectory allowed by the variational parameters. This principle is particularly suitable when parameters are constrained to be real, as in many practical ansatz constructions.

\section{Reformulate QNGD-based VQAs and Build A First-principle Equivalence with QITE}

\citet{stokes2020quantum} presents that the optimizer QNGD induces same parameter update rule with QITE from a phenomenological observation, yet the reason behind remains unclear. We provide a detailed derivations to address this gap.

\subsection{Objective Function of QNGD-based VQAs with General Loss Function}
\label{A Review for Quantum Natural Gradient Descent}

QNGD-based VQAs update their variational parameters by determining an optimal update direction \(\bm \Delta\bm \theta\) within a local neighborhood of parameters \(\bm \theta\), while accounting for the underlying geometry of the quantum state manifold, which is described by the Quantum Fisher Information Matrix (QFIM)~\cite{meyer2021fisher, stokes2020quantum}. To generalize the optimization framework, we introduce a general loss function of the form:
\begin{equation}
    \mathcal{L}(\bm \theta) = f(\langle O \rangle),
\end{equation}
where \(\langle O \rangle = \langle \psi(\bm \theta) | O | \psi(\bm \theta) \rangle\) denotes the expected value of a given observable \(O\), and \(f\) is a differentiable scalar-valued function. A commonly used example is the quadratic loss function:
\begin{equation}
    \mathcal{L}(\bm \theta) = \frac{1}{2} \left( \langle O \rangle - O_0 \right)^2,
\end{equation}
where \(O_0\) is the target observable value.

In gradient descent, optimization is performed under Euclidean geometry, where the update direction is constrained by an \(\ell_2\) norm:
\begin{equation}
    \|\bm \Delta\bm \theta\| \leq \varepsilon, \quad \text{with } \bm \Delta\bm \theta = \epsilon \bm \nu,
\end{equation}
where \(\bm \nu\) is an arbitrary unit vector and \(\epsilon > 0\) is a small scalar step size. However, in quantum variational algorithms, such parameter displacements may not reflect the true distance between quantum states. Instead, QNGD regularizes updates using the fidelity distance between quantum states:
\begin{equation}
    d_f\left(|\psi(\bm \theta)\rangle, |\psi(\bm \theta + \bm \Delta\bm \theta)\rangle\right) = 1 - |\langle \psi(\bm \theta) | \psi(\bm \theta + \bm \Delta\bm \theta) \rangle|^2.
\end{equation}
In the infinitesimal limit \(\bm \Delta\bm \theta \rightarrow \bm 0\), the fidelity distance becomes the squared line element on the Hilbert space manifold:
\begin{equation}
    d_f \approx \mathrm{d}s^2 = \sum_{\ell_1, \ell_2} g_{\ell_1 \ell_2}(\bm \theta) \Delta\bm \theta_{\ell_1} \Delta\bm \theta_{\ell_2} = \frac{1}{4} \sum_{\ell_1, \ell_2} \mathcal{F}_{\ell_1 \ell_2}(\bm \theta) \Delta\bm \theta_{\ell_1} \Delta\bm \theta_{\ell_2},
\end{equation}
where \(g_{\ell_1 \ell_2}(\bm \theta)\) is the Fubini–study metric tensor, defined as:
\begin{equation}
    g_{\ell_1\ell_2}(\bm \theta) = \Re \left[ \left\langle \frac{\partial \psi(\bm \theta)}{\partial \bm \theta_{\ell_1}} \Bigg| \frac{\partial \psi(\bm \theta)}{\partial \bm \theta_{\ell_2}} \right\rangle 
    - \left\langle \frac{\partial \psi(\bm \theta)}{\partial \bm \theta_{\ell_1}} \Bigg| \psi(\bm \theta) \right\rangle \left\langle \psi(\bm \theta) \Bigg| \frac{\partial \psi(\bm \theta)}{\partial \bm \theta_{\ell_2}} \right\rangle \right],
\label{g_appendix}
\end{equation}
and \(\mathcal{F}(\bm \theta) = 4g(\bm \theta)\) is the QFIM given by:
\begin{equation}
    \mathcal{F}_{\ell_1\ell_2}(\bm \theta) = 4 \Re \left[ \left\langle \frac{\partial \psi(\bm \theta)}{\partial \bm \theta_{\ell_1}} \Bigg| \frac{\partial \psi(\bm \theta)}{\partial \bm \theta_{\ell_2}} \right\rangle 
    - \left\langle \frac{\partial \psi(\bm \theta)}{\partial \bm \theta_{\ell_1}} \Bigg| \psi(\bm \theta) \right\rangle \left\langle \psi(\bm \theta) \Bigg| \frac{\partial \psi(\bm \theta)}{\partial \bm \theta_{\ell_2}} \right\rangle \right].
\end{equation}

With fidelity-based regularization, the update direction is formulated as the solution to a constrained optimization problem:
\begin{equation}
    \bm \Delta\bm \theta^* = \arg \min_{\bm \Delta\bm \theta \;\text{s.t.}\; d_f(\psi(\bm \theta), \psi(\bm \theta + \bm \Delta\bm \theta)) = c} \; \mathcal{L}(\bm \theta + \bm \Delta\bm \theta),
\end{equation}
where \(c > 0\) is a fixed fidelity threshold. Reformulating this as a Lagrangian and applying a first-order Taylor expansion of \(\mathcal{L}\), we obtain:
\begin{equation}
\begin{aligned}
    \bm \Delta\bm \theta^* 
    &= \arg \min_{\bm \Delta\bm \theta} \left[ \mathcal{L}(\bm \theta) + \nabla_{\bm \theta} \mathcal{L}(\bm \theta) \cdot \bm \Delta\bm \theta + \frac{\lambda}{4} \sum_{\ell_1, \ell_2} \mathcal{F}_{\ell_1 \ell_2}(\bm \theta) \Delta\bm \theta_{\ell_1} \Delta\bm \theta_{\ell_2} - \lambda c \right] \\
    &\equiv \arg \min_{\bm \Delta\bm \theta} \nabla_{\bm \theta} \mathcal{L}(\bm \theta) \cdot \bm \Delta\bm \theta + \frac{\lambda}{4} \sum_{\ell_1, \ell_2} \mathcal{F}_{\ell_1 \ell_2}(\bm \theta) \Delta\bm \theta_{\ell_1} \Delta\bm \theta_{\ell_2},
\end{aligned}
\end{equation}
where \(\lambda\) is the Lagrange multiplier associated with the fidelity constraint. Substituting the definition of \(\mathcal{F}_{\ell_1\ell_2}\), we obtain the explicit objective of QNGD-based VQAs:
\begin{equation}
\begin{aligned}
    \bm \Delta\bm \theta^* = \arg \min_{\bm \Delta\bm \theta} &\; \nabla_{\bm \theta} \mathcal{L}(\bm \theta) \cdot \bm \Delta\bm \theta \\
    &+ \lambda \sum_{\ell_1, \ell_2} \Re \left[ \left\langle \frac{\partial \psi(\bm \theta)}{\partial \bm \theta_{\ell_1}} \Bigg| \frac{\partial \psi(\bm \theta)}{\partial \bm \theta_{\ell_2}} \right\rangle 
    - \left\langle \frac{\partial \psi(\bm \theta)}{\partial \bm \theta_{\ell_1}} \Bigg| \psi(\bm \theta) \right\rangle \left\langle \psi(\bm \theta) \Bigg| \frac{\partial \psi(\bm \theta)}{\partial \bm \theta_{\ell_2}} \right\rangle \right] \Delta\bm \theta_{\ell_1} \Delta\bm \theta_{\ell_2}.
\end{aligned}
\label{eq:qngd-objective}
\end{equation}

\subsection{Equivalence between QITE and QNGD-based VQAs In the Objective Function}
\label{appendix:equiv_objective}

In this section, we demonstrate that projected QITE shares an equivalent objective function with QNGD-based VQAs in the continuous-time limit. Specifically, projected QITE seeks to variationally approximate the imaginary-time evolved state \( e^{-O \Delta \tau} \psi(\bm \theta(\tau)) \) using a parametrized ansatz \( \psi(\bm \theta + \Delta \bm \theta) \), by maximizing their fidelity:
\begin{equation}
    \arg\max_{\Delta \bm \theta \in \mathbb{R}^{d}} \left| \left\langle e^{-O\Delta \tau} \psi_{\bm \theta}, \psi_{\bm \theta+\Delta\bm \theta} \right\rangle \right|^{2}
    \equiv 
    \arg\min_{\Delta \bm \theta \in \mathbb{R}^{d}} \left| 1 - \left\langle e^{-O\Delta \tau} \psi_{\bm \theta}, \psi_{\bm \theta+\Delta\bm \theta} \right\rangle \right|^{2}.
\end{equation}

Assuming small \(\Delta \tau\) and \(\Delta \bm \theta\), we follow the expansion technique introduced in~\citet{stokes2020quantum}. Letting \(\bar{\psi}_{\bm \theta} := e^{-O \Delta \tau} \psi_{\bm \theta}\), we perform a second-order Taylor expansion and obtain:
\begin{equation}
\begin{aligned}
    \arg \min_{\Delta \bm \theta \in \mathbb{R}^d} \left( 1 - \left| \left\langle \bar{\psi}_{\bm \theta}, \psi_{\bm \theta + \Delta \bm \theta} \right\rangle \right|^2 \right) 
    &= \arg \min_{\Delta \bm \theta \in \mathbb{R}^d} ( 1 - \left| \left\langle \bar{\psi}_{\bm \theta}, \psi(\bm \theta) \right\rangle \right|^2 \\
    &\quad + \left[ \left\langle \frac{\partial \psi(\bm \theta)}{\partial \bm \theta_{\ell_1}}, O \psi(\bm \theta) \right\rangle + \left\langle O \psi(\bm \theta), \frac{\partial \psi(\bm \theta)}{\partial \bm \theta_{\ell_1}} \right\rangle \right] \Delta \bm \theta_{\ell_1} \Delta \tau \\
    &\quad + \Re\left[ G_{\ell_1 \ell_2}(\bm \theta) \right] \Delta \bm \theta_{\ell_1} \Delta \bm \theta_{\ell_2}), 
\end{aligned}
\end{equation}
where \(G_{\ell_1 \ell_2}(\bm \theta)\) is defined in Eq.~(\ref{quantum_geometric_tensor}). Discarding constant terms and reorganizing the expression, we have:
\begin{equation}
\begin{aligned}
    \arg \min_{\Delta \bm \theta \in \mathbb{R}^d} &\left[ \left\langle \frac{\partial \psi(\bm \theta)}{\partial \bm \theta_{\ell_1}}, O \psi(\bm \theta) \right\rangle + \left\langle O \psi(\bm \theta), \frac{\partial \psi(\bm \theta)}{\partial \bm \theta_{\ell_1}} \right\rangle \right] \Delta \bm \theta_{\ell_1} \Delta \tau \\
    &+ \Re\left[ G_{\ell_1 \ell_2}(\bm \theta) \right] \Delta \bm \theta_{\ell_1} \Delta \bm \theta_{\ell_2} \\
    = \arg \min_{\Delta \bm \theta \in \mathbb{R}^d} (\Delta \tau)^2 &\left\{ \left[ \left\langle \frac{\partial \psi(\bm \theta)}{\partial \bm \theta_{\ell_1}}, O \psi(\bm \theta) \right\rangle + \left\langle O \psi(\bm \theta), \frac{\partial \psi(\bm \theta)}{\partial \bm \theta_{\ell_1}} \right\rangle \right] \frac{\Delta \bm \theta_{\ell_1}}{\Delta \tau} \right. \\
    &\left. + \Re\left[ G_{\ell_1 \ell_2}(\bm \theta) \right] \frac{\Delta \bm \theta_{\ell_1}}{\Delta \tau} \frac{\Delta \bm \theta_{\ell_2}}{\Delta \tau} \right\}.
\end{aligned}
\end{equation}

To take the continuous-time limit, we define the instantaneous update direction \(\bm \delta := \frac{d \bm \theta}{d \tau}\). Substituting \(\Delta \bm \theta = \bm \delta \cdot \Delta \tau\) into the expression above and letting \(\Delta \tau \to 0\), the objective becomes:
\begin{equation}
    \arg \min_{\bm \delta \in \mathbb{R}^d} \frac{\partial \langle \psi(\bm \theta(\tau)) | O | \psi(\bm \theta(\tau)) \rangle}{\partial \bm \theta_{\ell_1}} \delta_{\ell_1} 
    + \sum_{\ell_1, \ell_2 \in [d]} \Re\left[ G_{\ell_1 \ell_2}(\bm \theta) \right] \delta_{\ell_1} \delta_{\ell_2}.
\end{equation}

This final expression is identical in form to the objective of QNGD-based VQAs in Eq.~(\ref{eq:qngd-objective}) when the loss function is chosen as a linear expectation value \(\mathcal{L}(\bm \theta) = \langle O \rangle\), and the regularization parameter \(\lambda\) is set to 1. This equivalence reveals a deep connection between projected QITE and QNGD-based VQAs in the continuous-time regime, unifying them under a shared optimization principle rooted in the geometry of quantum state space.

\subsection{Variational Principle Formulation of QNGD-based VQAs with General Loss Function}
\label{sec:QNGD_variational_principle}

To facilitate a first-principle comparison with QITE, we formalize the variational principle of QNGD-based VQAs. Using a first-order Taylor approximation, the change in the loss function \(\mathcal{L}\) under a small update \( \bm{\Delta\bm \theta} \) is:
\begin{equation}
    \Delta\mathcal{L} \approx \nabla \mathcal{L}^{\top} \bm{\Delta\bm \theta},
\end{equation}
where \(\nabla \mathcal{L}\) represents the gradient. To respect the geometric structure of the quantum state manifold, QNGD-based VQAs introduce a curvature penalty via the Fubini–study metric~\cite{bukov2019geometric, kolodrubetz2017geometry, shapere1989geometric, stokes2020quantum}:
\begin{equation}
    \|\bm{\Delta\bm \theta}\|_{\mathcal{F}}^2 = \bm{\Delta\bm \theta}^{\top} \mathcal{F} \bm{\Delta\bm \theta},
\end{equation}

To unify optimization and geometry, we define a variational functional:
\begin{equation}
\begin{aligned}
\mathcal{S}[\bm{\Delta\bm \theta}]
&= \nabla \mathcal{L}^{\top} \bm{\Delta\bm \theta} + \frac{1}{2\eta} \bm{\Delta\bm \theta}^{\top} \mathcal{F} \bm{\Delta\bm \theta},
\end{aligned}
\end{equation}
where \( \eta > 0 \) is a regularization parameter controlling step size. The first term captures loss descent, while the second penalizes large displacements in the quantum state space.

The variational principle of QNGD-based VQAs then requires the first variation of this functional to vanish:
\begin{equation}
\delta \left[ \nabla \mathcal{L}^{\top} \bm{\Delta\bm \theta} + \frac{1}{2\eta} \bm{\Delta\bm \theta}^{\top} \mathcal{F} \bm{\Delta\bm \theta} \right] = 0,
\label{variational_principle_qngd_general}
\end{equation}

Now, Consider a general differentiable loss function \( \mathcal{L}(\bm \theta) = f(\langle O \rangle) \), where \( O \) is a Hermitian observable,
\begin{equation}
\Delta\mathcal{L} = f'\big( \langle O \rangle_{\bm \theta} \big) \, \Delta \langle O \rangle_{\bm \theta},
\end{equation}
and the first-order change of the expectation value is
\begin{equation}
\Delta \langle O \rangle_{\bm \theta} \approx \nabla \langle O \rangle_{\bm \theta}^\top \Delta \bm \theta,
\end{equation}
which yields the variational principle of QNGD-based VQAs with general loss function:
\begin{equation}
\delta \left[ f'\big( \langle O \rangle_{\bm \theta} \big) \, \nabla \langle O \rangle_{\bm \theta}^\top \Delta \bm \theta + \frac{1}{2\eta} \Delta\bm \theta^\top \mathcal{F} \Delta\bm \theta \right] = \delta[\mathcal{J}_{\text{General}}] = 0,
\end{equation}
where we define \(\mathcal{J}_{\text{General}}\) as the variational functional of QNGD-based VQAs with general loss function.

\subsection{Equivalence between QITE and QNGD-based VQAs In the Variational Principle with Linear Loss Function}
\label{sec:qite_qngd_first}

We firstly connect the two variational principles with linear loss function. As a review, the variational principle of QNGD-based VQAs focuses on parameter changes \(\Delta\bm \theta\), while QITE's variational principle directly constrains the quantum state’s time evolution \(\partial_\tau |\psi(\bm \theta(\tau))\rangle\). The connection between the two can be established by the effect of parameter changes on quantum state evolution:

A parameter change \(\Delta\bm \theta\) leads to a change in the quantum state:
\begin{equation}
    |\psi(\bm \theta + \Delta\bm \theta)\rangle \approx |\psi(\bm \theta)\rangle + \sum_{\ell_1} \partial_{\bm \theta_{\ell_1}} |\psi(\bm \theta)\rangle \Delta\bm \theta_{\ell_1}.
\end{equation}

In the continuous-time limit \(\eta \to 0\), the rate of change of parameters \(\dot{\bm \theta} = \frac{d\bm \theta}{d\tau}\) corresponds to the time derivative of the quantum state:
\begin{equation}
\partial_\tau |\psi(\bm \theta(\tau))\rangle = \sum_{\ell_1} \partial_{\bm \theta_{\ell_1}} |\psi\rangle \dot{\bm \theta}_{\ell_1}.
\label{eq:time_derivative_of_the_quantum_state}
\end{equation}

Accordingly, we can convert the variational principle of QNGD-based VQAs into the form including quantum state evolution. Reviewing Eq.~(\ref{variational_principle_qngd_general}), when with linear loss function:

\begin{equation}
\mathcal{L}(\bm \theta) = \langle O \rangle = \langle \psi(\bm \theta) | O | \psi(\bm \theta) \rangle,
\end{equation}
the gradient is computed as:
\begin{equation}
\nabla_{\ell_1} \mathcal{L} = \frac{\partial \mathcal{L}}{\partial \bm \theta_{\ell_1}} = \langle \partial_{\ell_1} \psi | O | \psi \rangle + \langle \psi | O | \partial_{\ell_1} \psi \rangle.
\end{equation}

Exploiting Hermitian symmetry, we have:
\begin{equation}
\begin{aligned}
\nabla_{\ell_1} \mathcal{L} &= \langle \partial_{\ell_1} \psi | O | \psi \rangle + \left( \langle \partial_{\ell_1} \psi | O | \psi \rangle \right)^* \\
&= 2\,\text{Re}\left( \langle \partial_{\ell_1} \psi | O | \psi \rangle \right).
\end{aligned}
\end{equation}

The factor of 2 can often be absorbed into the learning rate, leading to the standard gradient expression:
\begin{equation}
\nabla_{\ell_1} \mathcal{L} = \text{Re}\left( \langle \partial_{\ell_1} \psi | O | \psi \rangle \right).
\label{gradient_linear}
\end{equation}

Using Eq.~(\ref{gradient_linear}) and \(\Delta\bm \theta = \dot{\bm \theta} \eta\), the variational principle under the continuous-time limit becomes:
\begin{equation}
\delta \left[ \eta \sum_{\ell_1} \text{Re}\left( \langle \partial_{\ell_1} \psi | O | \psi \rangle \right) \dot{\bm \theta}_{\ell_1} + \frac{\eta}{2} \sum_{\ell_1, \ell_2} F_{\ell_1\ell_2} \dot{\bm \theta}_{\ell_1} \dot{\bm \theta}_{\ell_2} \right] = \delta \left[ \mathcal{S}_{\text{inst}} \right] = 0,
\end{equation}
where \(\mathcal{S}_{\text{inst}}\) is the instantaneous variational functional. 

As \(\eta \to 0\), the sum over time steps converges to an integral over \(\tau\) with \(\eta \to d\tau\). The total variation is then:
\begin{equation}
\delta \int \left( \sum_{\ell_1} \text{Re}\left( \langle \partial_{\ell_1} \psi | O | \psi \rangle \right) \dot{\bm \theta}_{\ell_1} + \frac{1}{2} \sum_{\ell_1, \ell_2} F_{\ell_1\ell_2} \dot{\bm \theta}_{\ell_1} \dot{\bm \theta}_{\ell_2} \right) d\tau = \delta \left[ \mathcal{J}_{Linear} \right] = 0,
\end{equation}
where we define \(\mathcal{J}_{Linear}\) as the continuous-time action for QNGD-based VQAs with linear loss function.

Substituting Eq.~\eqref{eq:time_derivative_of_the_quantum_state} into the first term:
\begin{equation}
\begin{aligned}
\sum_{\ell_1} \text{Re}\left( \langle \partial_{\ell_1} \psi | O | \psi \rangle \right) \dot{\bm \theta}_{\ell_1} &= \text{Re}\left( \sum_{\ell_1} \langle \partial_{\ell_1} \psi | O | \psi \rangle \dot{\bm \theta}_{\ell_1} \right) \\
&= \text{Re}\left( \langle \partial_\tau \psi | O | \psi \rangle \right).
\end{aligned}
\end{equation}

\begin{remark}
\textit{This transformation projects the parameter gradient onto the quantum state's evolution direction.}
\end{remark}

Substituting Eq.~\eqref{eq:time_derivative_of_the_quantum_state} into the second term:
\begin{equation}
\begin{aligned}
\sum_{\ell_1, \ell_2} F_{\ell_1\ell_2} \dot{\bm \theta}_{\ell_1} \dot{\bm \theta}_{\ell_2} &= \text{Re} \left( \sum_{\ell_1, \ell_2} \left[ \langle \partial_{\ell_1} \psi | \partial_{\ell_2} \psi \rangle - \langle \partial_{\ell_1} \psi | \psi \rangle \langle \psi | \partial_{\ell_2} \psi \rangle \right] \dot{\bm \theta}_{\ell_1} \dot{\bm \theta}_{\ell_2} \right) \\
&= \text{Re} \left( \langle \partial_\tau \psi | \partial_\tau \psi \rangle - \langle \partial_\tau \psi | \psi \rangle \langle \psi | \partial_\tau \psi \rangle \right).
\end{aligned}
\end{equation}

From the normalization condition \(\partial_\tau \langle \psi | \psi \rangle = 0\), we derive:
\begin{equation}
\langle \partial_\tau \psi | \psi \rangle + \langle \psi | \partial_\tau \psi \rangle = 0, \ie, \text{Re}\left( \langle \psi | \partial_\tau \psi \rangle \right) = 0.
\end{equation}

This simplifies the metric to:
\begin{equation}
\sum_{\ell_1, \ell_2} F_{\ell_1\ell_2} \dot{\bm \theta}_{\ell_1} \dot{\bm \theta}_{\ell_2} = \langle \partial_\tau \psi | \partial_\tau \psi \rangle.
\end{equation}

Therefore, the variational principle is reformulated as:
\begin{equation}
\delta \int \left( \text{Re}\left( \langle \partial_\tau \psi | O | \psi \rangle \right) + \frac{1}{2} \langle \partial_\tau \psi | \partial_\tau \psi \rangle \right) d\tau = \delta \left[ \mathcal{J}_{Linear} \right] = 0,
\label{eq:reformulate_QNG}
\end{equation}

Meanwhile, according to \textit{Appendix \ref{app:overview_qite}}, QITE’s variational principle is associated with the following instantaneous variational functional :
\begin{equation}
\left\| \left( \frac{\partial}{\partial \tau} + O - E_\tau \right) |\psi(\bm \theta(\tau))\rangle \right\|^2,
\label{eq:QITE_D}
\end{equation}

Expanding this gives:

\begin{equation}
\begin{aligned}
&\left\| \left( \partial_\tau + O - E_\tau \right) |\psi\rangle \right\|^2 \\
&= \langle \partial_\tau \psi | \partial_\tau \psi \rangle + \langle \psi | (O - E_\tau)^2 | \psi \rangle \\
&\quad + 2\,\text{Re}\left( \langle \partial_\tau \psi | (O - E_\tau) | \psi \rangle \right) \\
&= \langle \partial_\tau \psi | \partial_\tau \psi \rangle + \langle \psi | O^2 | \psi \rangle - E_\tau^2 + 2\,\text{Re}\left( \langle \partial_\tau \psi | O | \psi \rangle \right)
\label{eq:expand_mclachlan_}
\end{aligned}
\end{equation}

Ignoring constant terms (\ie, \(E_\tau^2\) and \(\langle O^2 \rangle\) if \(O\) is fixed) and dividing by an overall constant factor (which does not affect the variational dynamics), the variational principle can be reduced to:
\begin{equation}
\delta \int \left( \text{Re}\left( \langle \partial_\tau \psi | O | \psi \rangle \right) + \frac{1}{2} \langle \partial_\tau \psi | \partial_\tau \psi \rangle \right) d\tau = \delta \left[ \mathcal{D}_{Linear} \right] = 0,
\end{equation}
which is exactly the same as QNGD-based VQAs’ continuous-time variational principle with linear loss function in Eq.~(\ref{eq:reformulate_QNG}).

In sum, when \(\mathcal{L}(\bm \theta) = \langle O \rangle_{\bm \theta}\), \(E_\tau = \mathcal{L}(\bm \theta(\tau))\), and if \(O\) is a stationary operator, higher-order terms \(\langle O^2 \rangle - E_\tau^2\) can be treated as constants in the variational principle (or canceled by normalization), and thus do not affect the resulting dynamics. Thus, the two variational functionals are related as:
\begin{equation}
\mathcal{D}_{Linear} \propto \mathcal{J}_{Linear} + \text{constant},
\end{equation}
\ie, the two variational functionals are equivalent up to a constant and a scaling factor, hence their variational principles are equivalent and lead to the same dynamics.

\begin{remark}
\textit{The QNGD-based VQAs’ variational principle implicitly optimizes the quantum state evolution path through a balance between geometric penalty in the parameter space and the rate of observable expectation decay, while QITE’s variational principle explicitly constrains quantum state evolution to approximate imaginary time dynamics. When given a linear loss function, \ie, an observable expectation, both variational principles reduce to the same problem in the continuous-time limit.}
\end{remark}

\subsection{Extend the Variational Principle of QITE}
\label{sec:general_mclachlan}

Similar to the above section, we expand \(\left\| \left( \frac{\partial}{\partial \tau} + O - E_\tau \right) |\psi(\bm \theta(\tau))\rangle \right\|^2\), which gives:
\begin{equation}
\begin{aligned}
    &\left\| \left( \frac{\partial}{\partial \tau} + O - E_\tau \right) |\psi(\bm \theta(\tau))\rangle \right\|^2 \\
    &= \left( \left( \frac{\partial}{\partial \tau} + O - E_\tau \right) |\psi(\bm \theta(\tau))\rangle \right)^{\dagger} \left( \frac{\partial}{\partial \tau} + O - E_\tau \right) |\psi(\bm \theta(\tau))\rangle \\
    &= \sum_{\ell_1, \ell_2} \frac{\partial \langle \psi(\bm \theta(\tau)) |}{\partial \bm \theta_{\ell_1}} \frac{\partial |\psi(\bm \theta(\tau))\rangle}{\partial \bm \theta_{\ell_2}} \dot{\bm \theta}_{\ell_1} \dot{\bm \theta}_{\ell_2} + \sum_{\ell_1} \frac{\partial \langle \psi(\bm \theta(\tau)) |}{\partial \bm \theta_{\ell_1}} (O - E_\tau) |\psi(\bm \theta(\tau))\rangle \dot{\bm \theta}_{\ell_1} \\
    &+ \sum_{\ell_1} \langle \psi(\bm \theta(\tau)) | (O - E_\tau) \frac{\partial |\psi(\bm \theta(\tau))\rangle}{\partial \bm \theta_{\ell_1}} \dot{\bm \theta}_{\ell_1} + \langle \psi(\bm \theta(\tau)) | (O - E_\tau)^2 |\psi(\bm \theta(\tau))\rangle.
\end{aligned}
\end{equation}

Different loss function types only change \(\sum_{\ell_1} \frac{\partial \langle \psi(\bm \theta(\tau)) |}{\partial \bm \theta_{\ell_1}} (O - E_\tau) |\psi(\bm \theta(\tau))\rangle \dot{\bm \theta}_{\ell_1} + \sum_{\ell_1} \langle \psi(\bm \theta(\tau)) | (O - E_\tau) \frac{\partial |\psi(\bm \theta(\tau))\rangle}{\partial \bm \theta_{\ell_1}} \dot{\bm \theta}_{\ell_1}\), as analyzed in \textit{Appendix~\ref{sec:qite_qngd_first}}. Now we extend the principle to both quadratic loss function and general loss function.



\textbf{Quadratic Loss Extension.} For convenience, we do not consider an \(\frac{1}{x}\) scaling for the loss types beyond linear loss function. The quadratic loss function we consider here is \( \mathcal{L} = (\langle \psi | O | \psi \rangle)^2 \). Then the gradient becomes:

\begin{equation}
\begin{aligned}
    \frac{\partial  \langle \psi | O | \psi \rangle^2}{\partial \bm \theta_{\ell_1}} 
    &= \langle \psi | O | \psi \rangle \cdot \frac{\partial \langle \psi | O | \psi \rangle}{\partial \bm \theta_{\ell_1}} \\
    &= 2 E_\tau \left[ \frac{\partial \langle \psi |}{\partial \bm \theta_{\ell_1}} (O - E_\tau) |\psi\rangle 
+ \langle \psi | (O - E_\tau) \frac{\partial |\psi\rangle}{\partial \bm \theta_{\ell_1}} + \frac{\partial \langle \psi |}{\partial \bm \theta_{\ell_1}} E_\tau |\psi\rangle 
+ \langle \psi | E_\tau \frac{\partial |\psi\rangle}{\partial \bm \theta_{\ell_1}} \right] . \\
    &= 2 E_\tau \left[ \frac{\partial \langle \psi |}{\partial \bm \theta_{\ell_1}} (O - E_\tau) |\psi\rangle + \langle \psi | (O - E_\tau) \frac{\partial |\psi\rangle}{\partial \bm \theta_{\ell_1}} \right].
\end{aligned}
\end{equation}

where the last equality holds due to the normalization condition \(\langle \psi | \psi \rangle = 1\) and \(E_\tau\) is a scalar.

Substituting into the variational condition, we obtain:
\begin{equation}
    \delta \left\| \left( \frac{\partial}{\partial \tau} + E_\tau(O - E_\tau) \right) |\psi(\tau)\rangle \right\| = 0.
\end{equation}

\textbf{General Loss Extension.} Let \( \mathcal{L} = f(\langle \psi | O | \psi \rangle) \). The chain rule yields:
\begin{equation}
\begin{aligned}
    \frac{\partial \mathcal{L}}{\partial \bm \theta_{\ell_1}} &= f'(E_\tau) \cdot \frac{\partial \langle \psi | O | \psi \rangle}{\partial \bm \theta_{\ell_1}} \\
    &= f'(E_\tau) \cdot \left[ \frac{\partial \langle \psi |}{\partial \bm \theta_{\ell_1}} (O - E_\tau) |\psi\rangle + \langle \psi | (O - E_\tau) \frac{\partial |\psi\rangle}{\partial \bm \theta_{\ell_1}} \right].
\end{aligned}
\end{equation}

Thus, the generalized McLachlan variational principle becomes:
\begin{equation}
    \delta \left\| \left( \frac{\partial}{\partial \tau} + f'(E_\tau)(O - E_\tau) \right) |\psi(\tau)\rangle \right\| = 0.
\end{equation}

\begin{remark}
\textit{This generalized variational principle retains the fidelity to imaginary-time dynamics while enabling flexible loss definitions. When \( f(E_\tau) = E_\tau \), we recover the standard McLachlan variational formulation for QITE.}
\end{remark}

\subsection{Equivalence between QITE and QNGD-based VQAs In the Variational Principle with General Loss Function}
\label{sec:general_loss_variational} 

According to \textit{Appendix~\ref{sec:QNGD_variational_principle} and~\ref{sec:qite_qngd_first}}, for a general differentiable loss function \( f(\langle O \rangle) \), the continuous-time limit (\( \eta \to 0 \)) of QNGD-based VQAs yields the following variational functional:
\begin{equation}
     \mathcal{J}_{\text{general}} = \int \left[ f'\big(\langle O \rangle\big) \, \text{Re}\left( \langle \partial_\tau \psi | O | \psi \rangle \right) + \frac{1}{2} \langle \partial_\tau \psi | \partial_\tau \psi \rangle \right] d\tau,
\label{eq:general_qngd_action}
\end{equation}
where \( \langle O \rangle = \langle \psi(\bm \theta) | O | \psi(\bm \theta) \rangle \), and \( O \) denotes the observable.

From \textit{Appendix~\ref{sec:general_mclachlan}}, for the QITE with general loss function, the variational functional is given by:
\begin{equation}
    \mathcal{D}_{\text{general}} = \left\| \left( \partial_\tau + f'\big(\langle O \rangle\big)(O - \langle O \rangle) \right) |\psi\rangle \right\|^2.
\label{eq:general_qite_action}
\end{equation}

Expanding Eq.~\eqref{eq:general_qite_action}, we obtain:
\begin{align}
    \mathcal{D}_{\text{general}} &= \langle \partial_\tau \psi | \partial_\tau \psi \rangle 
    + f'\big(\langle O \rangle\big)^2 \langle (O - \langle O \rangle)^2 \rangle \nonumber \\
    &\quad + 2f'\big(\langle O \rangle\big)\, \text{Re}\left( \langle \partial_\tau \psi | O - \langle O \rangle | \psi \rangle \right).
\end{align}

Since \(\langle \psi | \psi \rangle = 1\), we similarly divide by an overall constant factor (which does not affect the variational dynamics), thus the variational functional reduces to:
\begin{equation}
    \mathcal{D}_{\text{general}} = \frac{1}{2} \langle \partial_\tau \psi | \partial_\tau \psi \rangle 
    + f'\big(\langle O \rangle\big) \, \text{Re}\left( \langle \partial_\tau \psi | O | \psi \rangle \right)
    + \text{constant},
\end{equation}
where the constant term \( f'(\langle O \rangle)^2 \langle (O - \langle O \rangle)^2 \rangle \) does not influence the dynamics.

Comparing with Eq.~\eqref{eq:general_qngd_action}, we observe:
\begin{equation}
    \mathcal{D}_{\text{general}} \propto \mathcal{J}_{\text{general}} + \text{const},
\end{equation}
which shows that minimizing \( \mathcal{D}_{\text{general}} \) is equivalent to minimizing \( \mathcal{J}_{\text{general}} \), up to an overall scaling and additive constant. Since these factors do not affect the optimization trajectory, the two variational principles are equivalent with general loss function.

\begin{remark}
    \textit{This equivalence reveals that under general differentiable loss functions \( f(\langle O \rangle) \), QNGD-based VQAs and QITE share the same variational principles. The scalar factor \( f'\big(\langle O \rangle\big) \) modulates the imaginary time evolution rate without altering the optimal descent direction. Hence, the equivalence between parameter-space natural gradient descent and quantum-state imaginary time evolution persists beyond the linear loss case.}
\end{remark}

\section{Proof for Proposition \ref{Theorem_K_Quadratic}}
\label{sec:semi_proof_quadratic}

We present a simplified, semi-rigorous proof for Proposition~\ref{Theorem_K_Quadratic}.

Firstly, we prove that

\begin{equation}
    \overline{K_{\mathrm{QITE}}} = \frac{N+1}{N}\, \overline{K_{\mathrm{GD}}}
\end{equation}

holds under Assumption \ref{assump:random}, without considering an explicit analysis of high-order fluctuations. We begin by expanding the expectation:
\begin{equation}
\begin{aligned}
\overline{K_{\mathrm{QITE}}}
&= \overline{ \sum_{\ell_1,\ell_2}
    \frac{\partial \epsilon}{\partial \bm \theta_{\ell_1}} \,
    g^+_{\ell_1\ell_2} \,
    \frac{\partial \epsilon}{\partial \bm \theta_{\ell_2}} }
= \sum_{\ell_1,\ell_2}
    \overline{
    \frac{\partial \epsilon}{\partial \bm \theta_{\ell_1}} \,
    g^+_{\ell_1\ell_2} \,
    \frac{\partial \epsilon}{\partial \bm \theta_{\ell_2}} }.
\end{aligned}    
\end{equation}

\begin{figure}[htbp]
    \centering
    \includegraphics[width=0.8\textwidth]{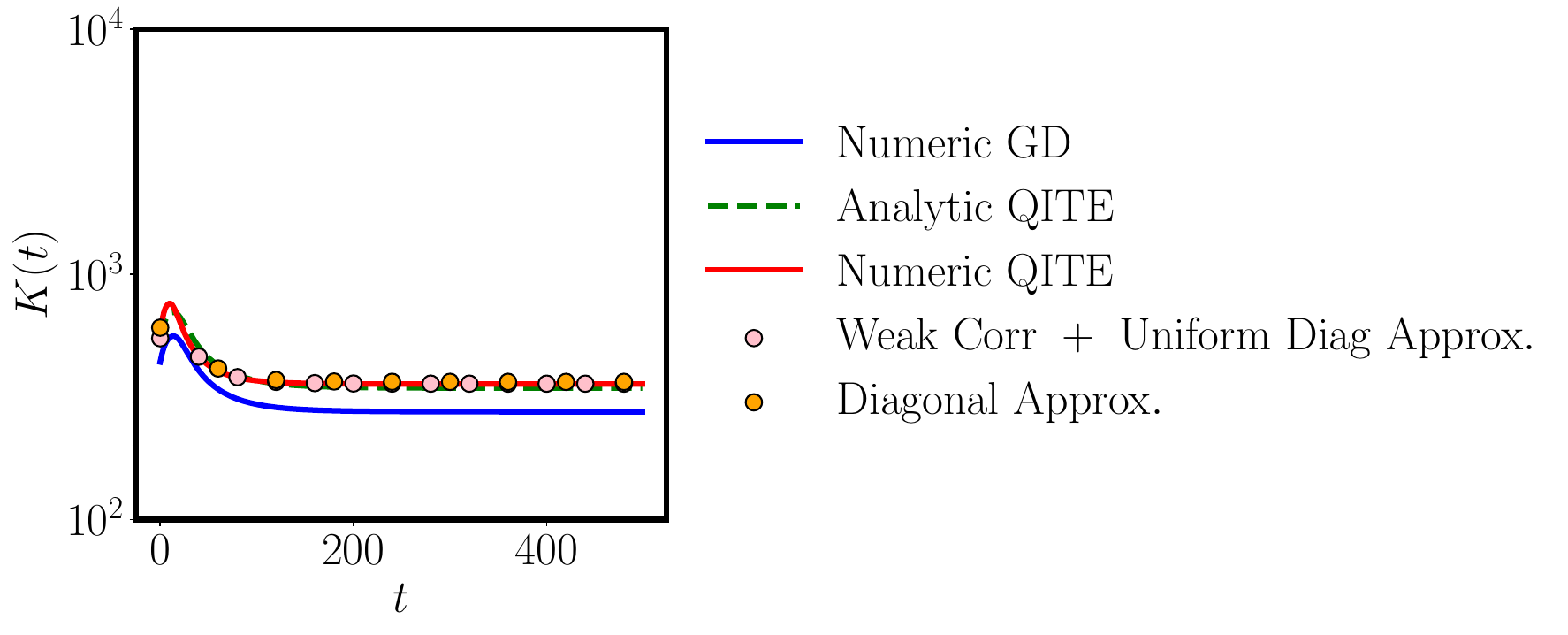}
    \caption{
       The diagonal approximation (yellow circles) is highlighted at selected steps to evaluate its agreement with the full \(K_\text{QITE}\). The agreement confirms that diagonal elements dominate the kernel structure, supporting the validity of the approximation in practice.
    }
    \label{fig:Kt_diagonal_approx_quadratic}
\end{figure}

From Lemma~\ref{lemma:quadratic_g}, the expected metric tensor is diagonal:

\begin{equation}
    \overline{g_{\ell_1\ell_2}} = \frac{N}{N+1}\, \delta_{\ell_1\ell_2},
\end{equation}

and the entrywise variance is given by:

\begin{equation}
    \mathrm{Var}(g_{\ell_1\ell_2}) =
\begin{cases}
\frac{2}{N^2}, & \text{if } \ell_1 = \ell_2, \\[2pt]
\frac{1}{2N}, & \text{if } \ell_1 \ne \ell_2.
\end{cases}
\end{equation}

Together with the results under Haar random ensemble:
\begin{equation}
\begin{aligned}
    &\overline{ \frac{\partial \epsilon}{\partial \bm \theta_\ell} } = 0,\\
    &\overline{ \frac{\partial \epsilon}{\partial \bm \theta_{\ell_1}} \frac{\partial \epsilon}{\partial \bm \theta_{\ell_2}} }
    = \delta_{\ell_1\ell_2} \cdot \overline{ \frac{\partial \epsilon}{\partial \bm \theta_\ell} \frac{\partial \epsilon}{\partial \bm \theta_\ell} },
\end{aligned}
\end{equation}
we can make a diagonal approximation~(justified by Fig.~\ref{fig:Kt_diagonal_approx_quadratic} that denotes ``\textit{Diagonal Approx.}'') as follows:

\begin{equation}
\begin{aligned}
\overline{K_{\mathrm{QITE}}}
&= \sum_{\ell_1, \ell_2} \overline{ \frac{\partial \epsilon}{\partial \bm \theta_{\ell_1}} \cdot g^+_{\ell_1\ell_2} \cdot \frac{\partial \epsilon}{\partial \bm \theta_{\ell_2}} } \\
&\overset{\text{(diagonal approx.)}}{\approx} \sum_{\ell} \overline{ \left( \frac{\partial \epsilon}{\partial \bm \theta_\ell} \right)^2 \cdot g^+_{\ell\ell} }.
\end{aligned}
\end{equation}

Since the fluctuations are small in the large-\(N\) limit, we can also approximate (see analysis in Section \ref{sec:dynamics_quadratic}):

\begin{equation}
    \overline{g^+_{\ell\ell}} \approx \left(\overline{g}^{-1}\right)_{\ell\ell} = \frac{N+1}{N}.
\label{eq:diag_inv}
\end{equation}

\begin{remark}
    Although \(\overline{A^+} \ne (\overline{A})^{-1}\) in general for a general matrix \(A\) , the deviation is suppressed by the small variance of \(g_{\ell_1\ell_2}\), therefore we assume the approximation holds for \(g\).
\end{remark}

Therefore, since we have $\overline{g^+_{\ell\ell}} = \frac{N+1}{N}$ (a constant independent of $\ell$), we make below approximation~(justified by Fig.~\ref{fig:Kt_diagonal_approx_quadratic} that denotes ``\textit{Weak Corr + Uniform Diag Approx.}''):

\begin{equation}
\begin{aligned}
        \overline{K_{\mathrm{QITE}}}
&\approx \frac{N+1}{N} \left( \sum_{\ell} \overline{ \frac{\partial \epsilon}{\partial \bm \theta_\ell} \frac{\partial \epsilon}{\partial \bm \theta_\ell} }  \right)
\end{aligned}
\end{equation}

where we assume weak correlations between \(g^+\) and the \(\left( \frac{\partial \epsilon}{\partial \bm \theta_\ell} \right)^2\) while uniform diagonals.

Thus we reach:

\begin{equation}
\begin{aligned}
\overline{K_{\mathrm{QITE}}} \approx \frac{N+1}{N} \, \overline{K_{\mathrm{GD}}}
.
\end{aligned}
\end{equation}

Now, we derive the relation regarding \(\epsilon\). Using Eq.~(\ref{eq:eps_quadratic}), we express the residual training error for both GD and QITE:

\begin{equation}
\begin{aligned}
        \epsilon_{\text{GD}}(t)
        &= \epsilon(0) \exp\left( -\eta \, \overline{K_{\mathrm{GD}}} \, t \right), \label{eq:eps_GD} \\[4pt]
    \epsilon_{\mathrm{QITE}}(t)
        &= \epsilon(0) \exp\left( -\eta \, \overline{K_{\mathrm{QITE}}} \, t \right).
\end{aligned}
\end{equation}

Therefore,
\begin{equation}
    \epsilon_{\mathrm{QITE}}(t)
    = \epsilon(0) \exp\left( -\eta \cdot \frac{N+1}{N} \cdot \overline{K_{\mathrm{GD}}} \cdot t \right).
\end{equation}

Dividing both error expressions, we obtain:
\begin{align}
    \frac{\epsilon_{\mathrm{QITE}}(t)}{\epsilon_{\mathrm{GD}}(t)}
    &= \exp\left( -\eta \, \overline{K_{\mathrm{GD}}} \, t \cdot \left( \frac{N+1}{N} - 1 \right) \right) \\
    &= \exp\left( -\frac{\eta \, t}{N} \, \overline{K_{\mathrm{GD}}} \right).
\end{align}

\ie,
\begin{equation}
    \epsilon_{\mathrm{QITE}}(t) \approx \epsilon_{\mathrm{GD}}(t) \cdot \exp\left( -\frac{\eta \, t}{N} \, \overline{K_{\mathrm{GD}}} \right).
\end{equation}

\section{Proof for Proposition \ref{Theorem_K_Linear}}
\label{sec:semi_proof_linear}

We present a simplified, semi-rigorous proof for Proposition~\ref{Theorem_K_Quadratic}. 

Firstly, we prove that

\begin{equation}
\overline{\lambda_{\text{QITE}} (t)} = \frac{N+1}{N} \overline{\lambda_{\text{GD}}(t)}
\end{equation}

holds under Assumption~\ref{assump:random}, without considering an explicit analysis of high-order fluctuations.

Recall the definitions:
\begin{equation}
\lambda_{\mathrm{QITE}}(t) = \frac{\mu_{\mathrm{QITE}}(t)}{K_{\mathrm{QITE}}(t)},
\qquad
\lambda_{\mathrm{GD}}(t) = \frac{\mu_{\mathrm{GD}}(t)}{K_{\mathrm{GD}}(t)},
\end{equation}
where
\begin{equation}
\mu_{\mathrm{QITE}} =
\sum_{\ell_1,\ell_2, \ell_3,\ell_4}
g^+_{\ell_1 \ell_3} \,
g^+_{\ell_2 \ell_4} \,
\frac{\partial^2 \epsilon}{\partial \bm \theta_{\ell_1} \partial \bm \theta_{\ell_2}} \,
\frac{\partial \epsilon}{\partial \bm \theta_{\ell_3}} \,
\frac{\partial \epsilon}{\partial \bm \theta_{\ell_4}},
\end{equation}
and
\begin{equation}
\mu_{\mathrm{GD}} =
\sum_{\ell_1,\ell_2}
\frac{\partial^2 \epsilon}{\partial \bm \theta_{\ell_1} \partial \bm \theta_{\ell_2}} \,
\frac{\partial \epsilon}{\partial \bm \theta_{\ell_1}} \,
\frac{\partial \epsilon}{\partial \bm \theta_{\ell_2}}.
\end{equation}

Under the Haar random initialization, we have:
\begin{equation}
\begin{aligned}
    &\overline{ \frac{\partial \epsilon}{\partial \bm \theta_\ell} } = 0, \\[4pt]
    &\overline{ \frac{\partial \epsilon}{\partial \bm \theta_{\ell_3}} \frac{\partial \epsilon}{\partial \bm \theta_{\ell_4}} }
    = \delta_{\ell_3 \ell_4} \cdot \overline{ \left( \frac{\partial \epsilon}{\partial \bm \theta_\ell} \right)^2 }, \\[4pt]
    &\overline{ \frac{\partial^2 \epsilon}{\partial \bm \theta_{\ell_1} \partial \bm \theta_{\ell_2}} }
    = \delta_{\ell_1 \ell_2} \cdot \overline{ \frac{\partial^2 \epsilon}{\partial \bm \theta_\ell^2} }.
\end{aligned}
\end{equation}

We similarly apply diagonal approximation to both metric tensors \(g^+_{\ell_1 \ell_3} \approx \delta_{\ell_1\ell_3} \, g^+_{\ell_1\ell_1}\) and \(g^+_{\ell_2 \ell_4} \approx \delta_{\ell_2\ell_4} \, g^+_{\ell_2\ell_2}\), yielding~(justified by Fig.~\ref{fig:Kt_diagonal_approx_linear} that denotes ``\textit{Diagonal Approx.}"):

\begin{figure*}[t]
    \centering
    \includegraphics[width=0.8\textwidth]{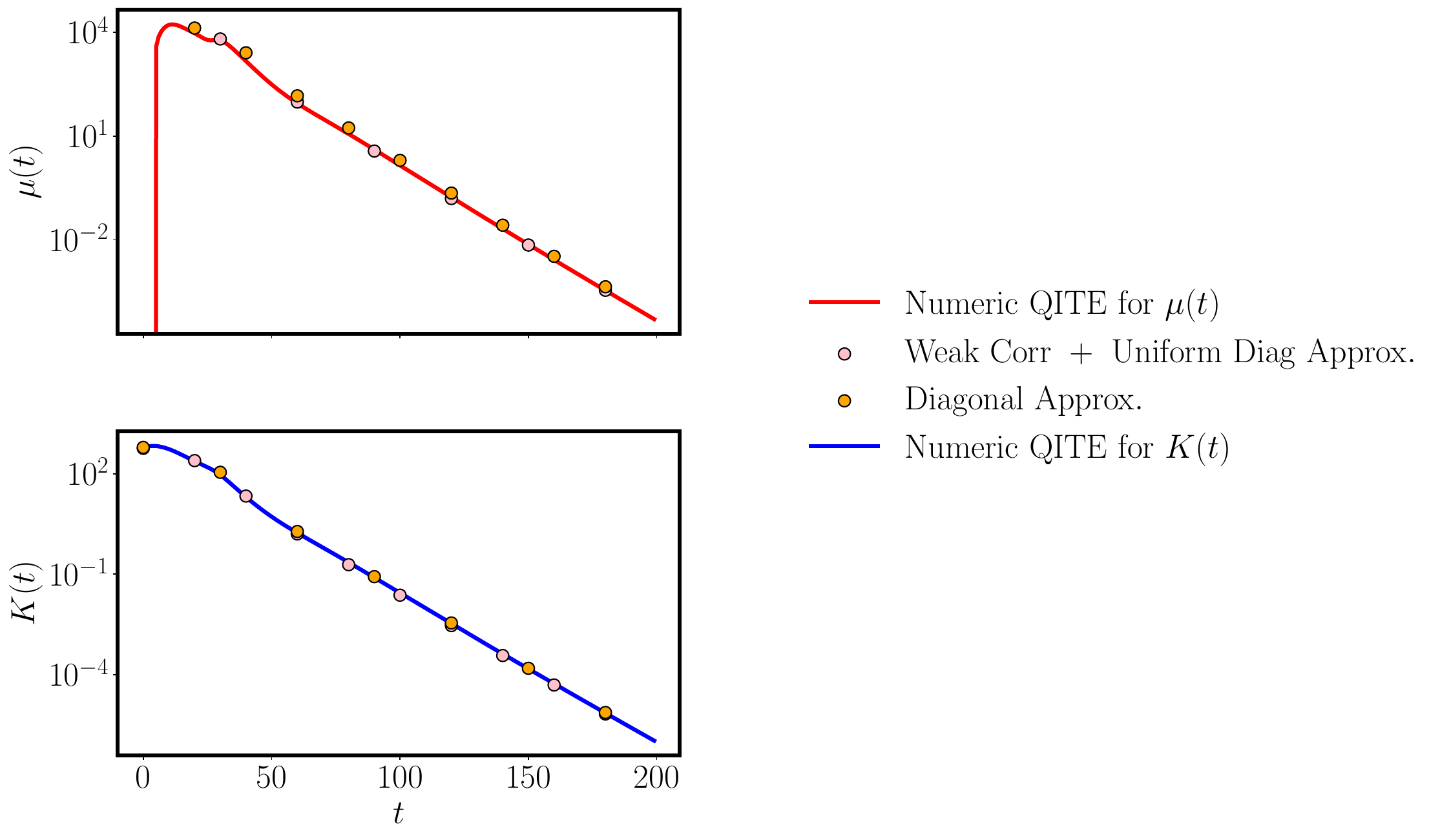}
    \caption{
       The diagonal approximation (yellow circles) is highlighted at selected steps to evaluate its agreement with the full \(K_\text{QITE}\). The agreement confirms that diagonal elements dominate the kernel structure, supporting the validity of the approximation in practice.
    }
    \label{fig:Kt_diagonal_approx_linear}
\end{figure*}

\begin{equation}
\begin{aligned}
        \overline{\mu_{\mathrm{QITE}}} &= \overline{\sum_{\ell_1,\ell_2, \ell_3,\ell_4}
g^+_{\ell_1 \ell_3} \,
g^+_{\ell_2 \ell_4} \,
\frac{\partial^2 \epsilon}{\partial \bm \theta_{\ell_1} \partial \bm \theta_{\ell_2}} \,
\frac{\partial \epsilon}{\partial \bm \theta_{\ell_3}} \,
\frac{\partial \epsilon}{\partial \bm \theta_{\ell_4}}} \\
&\approx \overline{ \sum_{\ell_1,\ell_2} g^+_{\ell_1\ell_1} g^+_{\ell_2\ell_2} \frac{\partial^2 \epsilon}{\partial \bm \theta_{\ell_1} \partial \bm \theta_{\ell_2}} \frac{\partial \epsilon}{\partial \bm \theta_{\ell_1}} \frac{\partial \epsilon}{\partial \bm \theta_{\ell_2}} }
\end{aligned}
\end{equation}


Assuming small fluctuations of \(g^+_{\ell\ell}\), we similarly approximate 

\begin{equation}
  \overline{(g^+_{\ell\ell})^2} \approx \left( \overline{g^+_{\ell\ell}} \right)^2 = \left( \frac{N+1}{N} \right)^2 
\end{equation}
(justified by Fig.~\ref{fig:Kt_diagonal_approx_linear} that denotes ``\textit{Weak Corr + Uniform Diag Approx.}''), leading to:
\begin{equation}
\overline{\mu_{\mathrm{QITE}}(t)} \approx \left( \frac{N+1}{N} \right)^2 \overline{ \sum_{\ell_1,\ell_2} \frac{\partial^2 \epsilon}{\partial \bm \theta_{\ell_1} \partial \bm \theta_{\ell_2}} \frac{\partial \epsilon}{\partial \bm \theta_{\ell_1}} \frac{\partial \epsilon}{\partial \bm \theta_{\ell_2}} }.
\end{equation}

Meanwhile, similar to \textit{Appendix \ref{sec:semi_proof_quadratic}}, we have~(justified by Fig.~\ref{fig:Kt_diagonal_approx_linear} that denotes ``\textit{Diagonal Approx.}" and ``\textit{Weak Corr + Uniform Diag Approx.}''):
\begin{equation}
\begin{aligned}
        \overline{K_{\mathrm{QITE}}}
&\approx \frac{N+1}{N} \left( \sum_{\ell} \overline{ \frac{\partial \epsilon}{\partial \bm \theta_\ell} \frac{\partial \epsilon}{\partial \bm \theta_\ell} }  \right)
\end{aligned}
\end{equation}

Combining numerator and denominator:
\begin{equation}
\overline{\lambda_{\mathrm{QITE}}(t)}
= \frac{\overline{\mu_{\mathrm{QITE}}(t)}}{\overline{K_{\mathrm{QITE}}(t)}}
\approx \frac{N+1}{N} \, \overline{\lambda_{\mathrm{GD}}(t)}.
\end{equation}

\qed

\textbf{Dynamics of \(K_{\text{QITE}}(t)\) and \(\epsilon_{\text{QITE}}(t).\)} We assume initial conditions \(K_{\text{QITE}}(0) = K_{\text{GD}}(0)\). Given that both GD and QITE satisfy:

\begin{equation}
    2\overline{\lambda} \, \epsilon(t) = K(t) \propto e^{-2\eta \overline{\lambda} t},
\end{equation}

For GD-based VQAs, the decay dynamics are:

\begin{equation}
    K_{\text{GD}}(t) = K_{\text{GD}}(0)  e^{-2\eta \overline{\lambda}_{\text{GD}} t}.
\end{equation}

For QITE, with decay rate \(\overline{\lambda}_{\text{QITE}} = \frac{N+1}{N}\overline{\lambda}_{\text{GD}}\):

\begin{equation}
\begin{aligned}
    K_{\text{QITE}}(t) &= K_{\text{QITE}}(0)  e^{-2\eta \overline{\lambda}_{\text{QITE}} t} \\
    &= \frac{K_{\text{GD}}(t)}{e^{-2\eta \overline{\lambda}_{\text{GD}} t}} \cdot e^{-2\eta \frac{N+1}{N} \overline{\lambda}_{\text{GD}} t} \\
    &= K_{\text{GD}}(t) \cdot e^{-2\eta \frac{1}{N} \overline{\lambda}_{\text{GD}} t}.
\end{aligned}
\end{equation}

In terms of \(\epsilon_{\text{QITE}}\), the error for QITE is:

\begin{equation}
    \epsilon_{\text{QITE}}(t) = \frac{K_{\text{QITE}}(t)}{2\overline{\lambda}_{\text{QITE}}}.
\end{equation}

Substitute \(\overline{\lambda}_{\text{QITE}} = \frac{N+1}{N}\overline{\lambda}_{\text{GD}}\) and the expression for \(K_{\text{QITE}}(t)\):

\begin{equation}
    \epsilon_{\text{QITE}}(t) = \frac{N}{2(N+1)\overline{\lambda}_{\text{GD}}} \left[ K_{\text{GD}}(t) \cdot e^{-2\eta \frac{1}{N} \overline{\lambda}_{\text{GD}} t} \right].
\end{equation}

\section{Additional Numerical Studies}

\subsection{Numerical Validation of Eq.~(\ref{eq:diag_inv})}
\label{subsec:validate_diag_inv}

\begin{figure*}[t]
    \centering
    \includegraphics[width=0.8\textwidth]{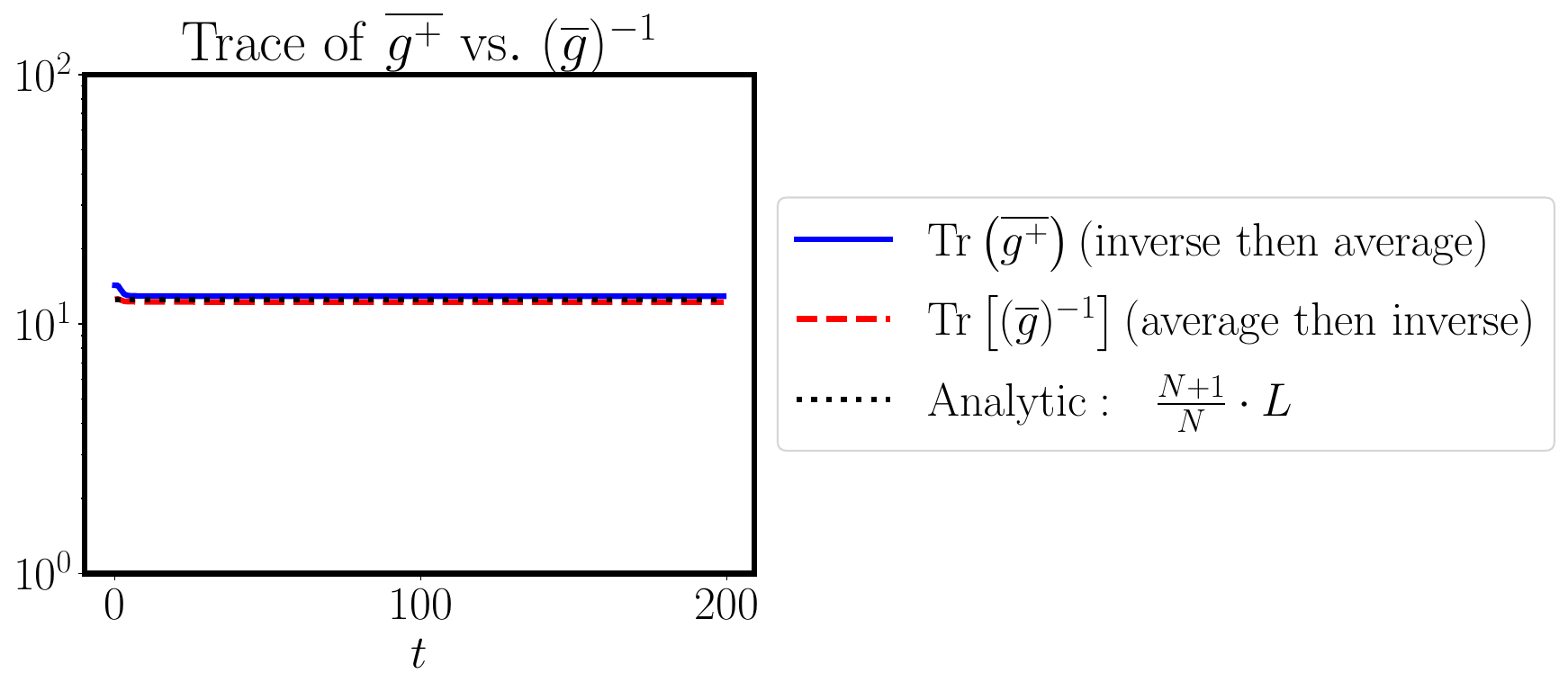}
    \caption{Comparison between the trace of the empirical average pseudoinverse, \(\mathrm{Tr}[\overline{g^+}]\), and the trace of the inverse of the average metric tensor, \(\mathrm{Tr}[(\overline{g})^{-1}]\), over training steps. Both quantities are with number of parameters \(L\), and averaged over random initializations. The black dashed line denotes the analytic prediction \(\frac{N+1}{N} \cdot L\) under the Haar-random assumption. The numerical agreement supports the validity of the diagonal approximation \(\overline{g^+_{\ell\ell}} \approx (\overline{g}^{-1})_{\ell\ell} \approx \frac{N+1}{N}\).}
    \label{fig:diag_inv_validation}
\end{figure*}

To validate the used approximation
\begin{equation}
     \overline{g^+_{\ell\ell}} \approx \left(\overline{g}^{-1}\right)_{\ell\ell}  = \frac{N+1}{N},
\end{equation}
we perform a numerical comparison of the trace of both sides across training steps. Here, \(g^+\) denotes \(g\)'s pseudoinverse. The approximation suggests that on average, the diagonal elements of \(g^+\) can be estimated using the diagonal of the inverse of the averaged matrix \(\overline{g}\). Specifically, we compare the trace of the average pseudoinverse, \(\mathrm{Tr}[\overline{g^+}]\), with the trace of the inverse of the average metric tensor, \(\mathrm{Tr}[(\overline{g})^{-1}]\). These two quantities are evaluated across training steps and averaged over random initializations. As shown in Fig.~\ref{fig:diag_inv_validation}, the two traces closely match throughout training and align well with the analytic prediction \(\mathrm{Tr} \approx \frac{N+1}{N} \cdot L\), thus supporting the validity of the diagonal approximation under Haar-random assumptions.


\subsection{Numerical Studies with Scaling Qubits and Layers}

\begin{figure*}[t]
    \centering
    \includegraphics[width=0.8\textwidth]{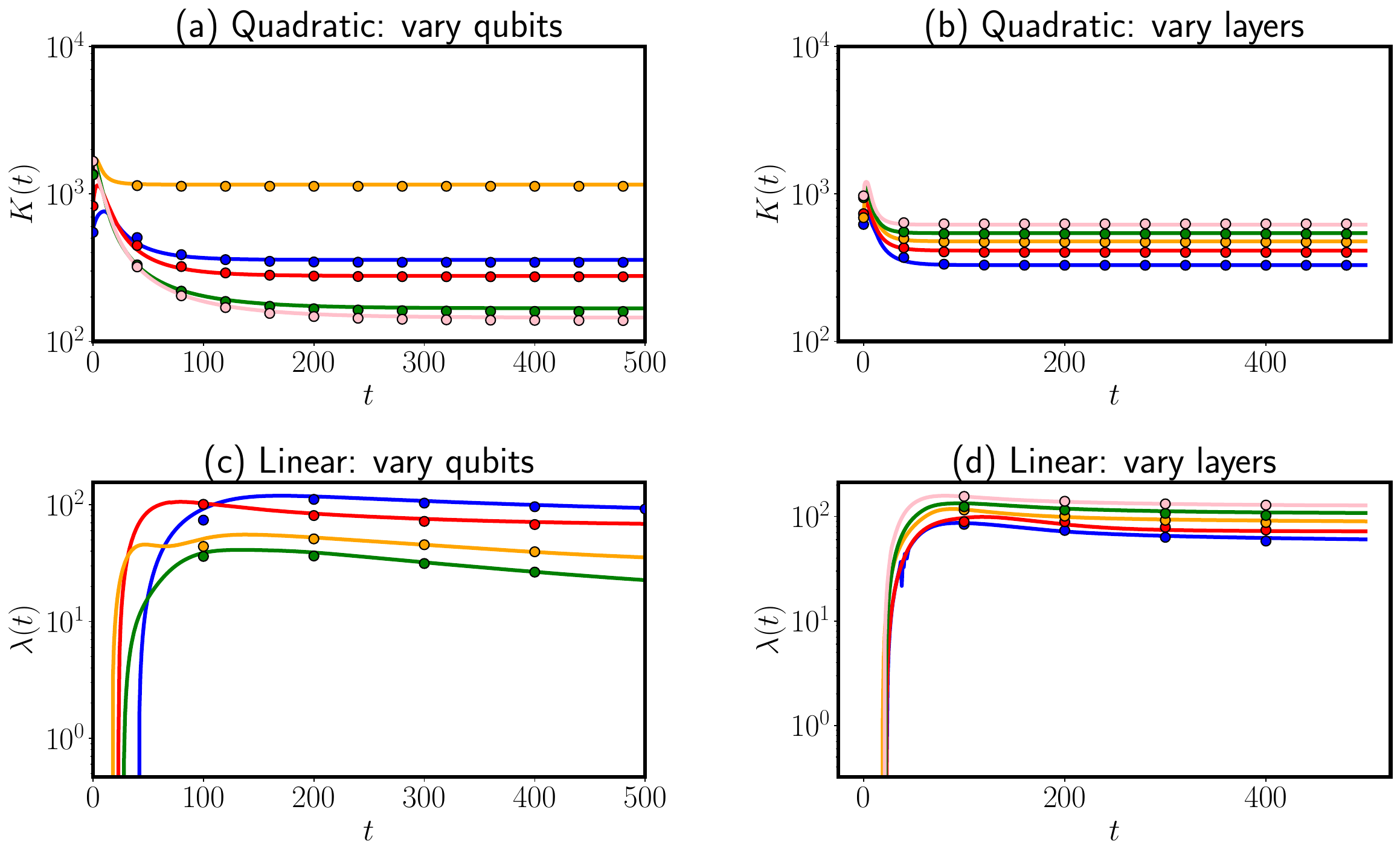}
    \caption{Scaling behavior of QNTK$K(t)$ and the relative dQNTK $\lambda(t)$ under quadratic and linear loss functions. Panels (a,b) show the scaling with the number of qubits $n=2,3,4,5,6$ and the number of layers $D=6,7,8,9,10$ for the quadratic loss, while panels (c,d) show the corresponding scaling for the linear loss. Solid lines denote numerical results averaged over 25 independent runs and circles indicate analytical predictions based on Haar random sampling. The color coding corresponds to system size: blue ($n=2$ qubits or $D=6$ layers), red ($n=3$ qubits or $D=7$ layers), orange ($n=4$ qubits or $D=8$ layers), green ($n=5$ qubits or $D=9$ layers), and pink ($n=6$ qubits or $D=10$ layers). Circles represent the analytic predictions, while solid lines represent the numerical results.}
    \label{fig:quadratic_scaling_L}
\end{figure*}

In this section, we examine the scaling behavior of the QNTK \(K\) and the relative QNTK \(\lambda\) under both quadratic and linear loss functions. 
Our analysis considers two complementary aspects: (i) scaling with circuit depth \(D\) at fixed qubit number \(n\) and ansatz architecture, and (ii) scaling with qubit number \(n\) under Haar-random circuit ensembles. 
For each loss function, we compare closed-form analytic predictions with numerical simulations. 
The analytic expressions for \(K(D)\), \(\lambda(D)\), and their \(n\)-dependence are derived under the assumption of sufficiently random circuits, modeled as unitary \(k\)-designs, yielding explicit scaling laws in the overparameterized regime. 
To validate these predictions, we sweep both depth and qubit number across broad ranges and extract the empirical scaling from simulations. We further report crossover behavior between shallow- and deep-depth regimes, where saturation of \(K(D)\) or changes in the decay rate \(\lambda(D)\) may emerge once the circuit approaches the effective mixing depth. 
The corresponding numerical results are summarized in Fig.~\ref{fig:quadratic_scaling_L}, which confirm that the observed scaling behaviors are in close agreement with the analytic predictions.

\section{Time Difference Equation for \(K_\text{QITE}(t)\)}
\label{sec:time_diff_K_QITE}

Below we provide the detailed derivation for the time difference equation for \(K_\text{QITE}(t)\):

\begin{equation}
\begin{aligned}
    &\delta K_\text{QITE} \equiv K_\text{QITE}(t+1) - K_\text{QITE}(t) = \delta \sum_{\ell_1,\ell_2} g_{\ell_1 \ell_2}^{+} \frac{\partial \epsilon }{ \partial \bm \theta_{\ell_1} }  \frac{\partial \epsilon}{\partial \bm \theta_{\ell_2} } \\
    &= \sum_{\ell_1,\ell_2} g_{\ell_1 \ell_2}^{+} (t + 1) \frac{\partial \epsilon }{ \partial \bm \theta_{\ell_1} } (t + 1)  \frac{\partial \epsilon}{\partial \bm \theta_{\ell_2} } (t + 1) - \sum_{\ell_1,\ell_2} g_{\ell_1 \ell_2}^{+} (t) \frac{\partial \epsilon }{ \partial \bm \theta_{\ell_1} } (t)  \frac{\partial \epsilon}{\partial \bm \theta_{\ell_2} } (t) \\
    &= \sum_{\ell_1,\ell_2} g_{\ell_1 \ell_2}^{+} (t + 1) \frac{\partial \epsilon }{ \partial \bm \theta_{\ell_1} } (t + 1)  \frac{\partial \epsilon}{\partial \bm \theta_{\ell_2} } (t + 1) - \sum_{\ell_1,\ell_2} g_{\ell_1 \ell_2}^{+} (t + 1) \frac{\partial \epsilon }{ \partial \bm \theta_{\ell_1} } (t + 1)  \frac{\partial \epsilon}{\partial \bm \theta_{\ell_2} } (t) + \sum_{\ell_1,\ell_2} g_{\ell_1 \ell_2}^{+} (t + 1) \frac{\partial \epsilon }{ \partial \bm \theta_{\ell_1} } (t + 1)  \frac{\partial \epsilon}{\partial \bm \theta_{\ell_2} } (t) \\
    &- \sum_{\ell_1,\ell_2} g_{\ell_1 \ell_2}^{+} (t) \frac{\partial \epsilon }{ \partial \bm \theta_{\ell_1} } (t + 1)  \frac{\partial \epsilon}{\partial \bm \theta_{\ell_2} } (t) + \sum_{\ell_1,\ell_2} g_{\ell_1 \ell_2}^{+} (t) \frac{\partial \epsilon }{ \partial \bm \theta_{\ell_1} } (t + 1)  \frac{\partial \epsilon}{\partial \bm \theta_{\ell_2} } (t) - \sum_{\ell_1,\ell_2} g_{\ell_1 \ell_2}^{+} (t) \frac{\partial \epsilon }{ \partial \bm \theta_{\ell_1} } (t)  \frac{\partial \epsilon}{\partial \bm \theta_{\ell_2} } (t) \\
    &= \sum_{\ell_1,\ell_2} g_{\ell_1 \ell_2}^{+} (t + 1) \frac{\partial \epsilon }{ \partial \bm \theta_{\ell_1} } (t + 1) \delta \left(\frac{\partial \epsilon}{\partial \bm \theta_{\ell_2} } (t) \right) + \sum_{\ell_1,\ell_2} \delta \left(  g_{\ell_1 \ell_2}^{+} (t) \right) \frac{\partial \epsilon }{ \partial \bm \theta_{\ell_1} } (t + 1)  \frac{\partial \epsilon}{\partial \bm \theta_{\ell_2} } (t) + \sum_{\ell_1,\ell_2} g_{\ell_1 \ell_2}^{+} (t) \delta \left( \frac{\partial \epsilon }{ \partial \bm \theta_{\ell_1} } (t) \right)  \frac{\partial \epsilon}{\partial \bm \theta_{\ell_2} } (t)\\
    &= \sum_{\ell_1,\ell_2} g_{\ell_1 \ell_2}^{+} (t + 1) \frac{\partial \epsilon }{ \partial \bm \theta_{\ell_1} } (t + 1) \delta \left(\frac{\partial \epsilon}{\partial \bm \theta_{\ell_2} } (t) \right) \\
    &- \sum_{\ell_1,\ell_2} g_{\ell_1 \ell_2}^{+} (t) \frac{\partial \epsilon }{ \partial \bm \theta_{\ell_1} } (t + 1) \delta \left(\frac{\partial \epsilon}{\partial \bm \theta_{\ell_2} } (t) \right) + \sum_{\ell_1,\ell_2} g_{\ell_1 \ell_2}^{+} (t) \frac{\partial \epsilon }{ \partial \bm \theta_{\ell_1} } (t + 1) \delta \left(\frac{\partial \epsilon}{\partial \bm \theta_{\ell_2} } (t) \right) \\
    &- \sum_{\ell_1,\ell_2} g_{\ell_1 \ell_2}^{+} (t) \frac{\partial \epsilon }{ \partial \bm \theta_{\ell_1} } (t) \delta \left(\frac{\partial \epsilon}{\partial \bm \theta_{\ell_2} } (t) \right) + \sum_{\ell_1,\ell_2} g_{\ell_1 \ell_2}^{+} (t) \frac{\partial \epsilon }{ \partial \bm \theta_{\ell_1} } (t) \delta \left(\frac{\partial \epsilon}{\partial \bm \theta_{\ell_2} } (t) \right) \\
    &+ \sum_{\ell_1,\ell_2} \delta \left(  g_{\ell_1 \ell_2}^{+} (t) \right) \frac{\partial \epsilon }{ \partial \bm \theta_{\ell_1} } (t + 1)  \frac{\partial \epsilon}{\partial \bm \theta_{\ell_2} } (t) - \sum_{\ell_1,\ell_2} \delta \left(  g_{\ell_1 \ell_2}^{+} (t) \right) \frac{\partial \epsilon }{ \partial \bm \theta_{\ell_1} } (t)  \frac{\partial \epsilon}{\partial \bm \theta_{\ell_2} } (t) + \sum_{\ell_1,\ell_2} \delta \left(  g_{\ell_1 \ell_2}^{+} (t) \right) \frac{\partial \epsilon }{ \partial \bm \theta_{\ell_1} } (t)  \frac{\partial \epsilon}{\partial \bm \theta_{\ell_2} } (t) \\
    &+ \sum_{\ell_1,\ell_2} g_{\ell_1 \ell_2}^{+} (t) \delta \left( \frac{\partial \epsilon }{ \partial \bm \theta_{\ell_1} } (t) \right)  \frac{\partial \epsilon}{\partial \bm \theta_{\ell_2} } (t)\\
    &=\sum_{\ell_1,\ell_2} \delta \left( g_{\ell_1 \ell_2}^{+} (t) \right) \frac{\partial \epsilon }{ \partial \bm \theta_{\ell_1} } (t + 1) \delta \left(\frac{\partial \epsilon}{\partial \bm \theta_{\ell_2} } (t) \right) + \sum_{\ell_1,\ell_2} g_{\ell_1 \ell_2}^{+} (t) \delta \left( \frac{\partial \epsilon }{ \partial \bm \theta_{\ell_1} } (t) \right) \delta \left(\frac{\partial \epsilon}{\partial \bm \theta_{\ell_2} } (t) \right) + \sum_{\ell_1,\ell_2} g_{\ell_1 \ell_2}^{+} (t) \frac{\partial \epsilon }{ \partial \bm \theta_{\ell_1} } (t) \delta \left(\frac{\partial \epsilon}{\partial \bm \theta_{\ell_2} } (t) \right)\\
    &+ \sum_{\ell_1,\ell_2} \delta \left(  g_{\ell_1 \ell_2}^{+} (t) \right) \delta \left( \frac{\partial \epsilon }{ \partial \bm \theta_{\ell_1} } (t) \right) \frac{\partial \epsilon}{\partial \bm \theta_{\ell_2} } (t) + \sum_{\ell_1,\ell_2} \delta \left(  g_{\ell_1 \ell_2}^{+} (t) \right) \frac{\partial \epsilon }{ \partial \bm \theta_{\ell_1} } (t)  \frac{\partial \epsilon}{\partial \bm \theta_{\ell_2} } (t) + \sum_{\ell_1,\ell_2} g_{\ell_1 \ell_2}^{+} (t) \delta \left( \frac{\partial \epsilon }{ \partial \bm \theta_{\ell_1} } (t) \right)  \frac{\partial \epsilon}{\partial \bm \theta_{\ell_2} } (t)
\end{aligned}
\end{equation}

We can ignore each term with two \(\delta\) in higher orders in \(\eta\), then we have the final formula for \(\delta K\) as follows:

\begin{equation}
    \delta K = \sum_{\ell_1,\ell_2} g_{\ell_1 \ell_2}^{+} (t) \frac{\partial \epsilon }{ \partial \bm \theta_{\ell_1} } (t) \delta \left(\frac{\partial \epsilon}{\partial \bm \theta_{\ell_2} } (t) \right) + \sum_{\ell_1,\ell_2} \delta \left(  g_{\ell_1 \ell_2}^{+} (t) \right) \frac{\partial \epsilon }{ \partial \bm \theta_{\ell_1} } (t)  \frac{\partial \epsilon}{\partial \bm \theta_{\ell_2} } (t) + \sum_{\ell_1,\ell_2} g_{\ell_1 \ell_2}^{+} (t) \delta \left( \frac{\partial \epsilon }{ \partial \bm \theta_{\ell_1} } (t) \right)  \frac{\partial \epsilon}{\partial \bm \theta_{\ell_2} } (t)
\end{equation}

Because \(g\) is a symmetry matrix, we have:

\begin{equation}
    \sum_{\ell_1,\ell_2} g_{\ell_1 \ell_2}^{+} (t) \delta \left( \frac{\partial \epsilon }{ \partial \bm \theta_{\ell_1} } (t) \right)  \frac{\partial \epsilon}{\partial \bm \theta_{\ell_2} } (t) = \sum_{\ell_2,\ell_1} g_{\ell_1 \ell_2}^{+} (t) \delta \left( \frac{\partial \epsilon }{ \partial \bm \theta_{\ell_2} } (t) \right)  \frac{\partial \epsilon}{\partial \bm \theta_{\ell_1} } (t)
\end{equation}

Thus, \(\delta K\) can be reduced to the following formula:

\begin{equation}
\begin{aligned}
    \delta K &= 2\sum_{\ell_1,\ell_2} g_{\ell_1 \ell_2}^{+} (t) \delta \left( \frac{\partial \epsilon }{ \partial \bm \theta_{\ell_1} } (t) \right)  \frac{\partial \epsilon}{\partial \bm \theta_{\ell_2} } (t) + \sum_{\ell_1,\ell_2} \delta \left(  g_{\ell_1 \ell_2}^{+} (t) \right) \frac{\partial \epsilon }{ \partial \bm \theta_{\ell_1} } (t)  \frac{\partial \epsilon}{\partial \bm \theta_{\ell_2} } (t)\\
\end{aligned}
\end{equation}

With quadratic loss function, we utilize the leading order Talor expansion on \(\delta (\partial \epsilon / \partial \bm \theta_l)\):

\begin{equation}
    \delta \left( \frac{\partial \epsilon}{\partial \bm \theta_{\ell_1}}(t) \right) = 
\sum_{\ell_2} \frac{\partial^2 \epsilon}{\partial \bm \theta_{\ell_2} \partial \bm \theta_{\ell_1}} \delta \bm \theta_{\ell_2} + \mathcal{O}(\eta^2) 
= - \eta \epsilon \sum_{\ell_2, \ell_3} g_{\ell_2 \ell_3}^{+} \frac{\partial^2 \epsilon}{\partial \bm \theta_{\ell_2} \partial \bm \theta_{\ell_1}} \frac{\partial \epsilon}{\partial \bm \theta_{\ell_3}} 
+ \mathcal{O}(\eta^2).
\end{equation}

Similarly, with linear loss function, the result should be:

\begin{equation}
    - \eta \sum_{\ell_2, \ell_3} g_{\ell_2 \ell_3}^{+} \frac{\partial^2 \epsilon}{\partial \bm \theta_{\ell_2} \partial \bm \theta_{\ell_1}} \frac{\partial \epsilon}{\partial \bm \theta_{\ell_3}} 
+ \mathcal{O}(\eta^2).
\end{equation}

Thus, the first term in \(\delta K\) can be rewritten using \(\mu\). With quadratic loss function, this is derived as follows:

\begin{equation}
\begin{aligned}
    2\sum_{\ell_1,\ell_2} g_{\ell_1 \ell_2}^{+} (t) \delta \left( \frac{\partial \epsilon }{ \partial \bm \theta_{\ell_1} } (t) \right)  \frac{\partial \epsilon}{\partial \bm \theta_{\ell_2} } (t) &= - 2\eta \epsilon \sum_{\ell_1, \ell_2, \ell_3, \ell_4} g_{\ell_1 \ell_2}^{+} g_{\ell_3 \ell_4}^{+} \frac{\partial^2 \epsilon}{\partial \bm \theta_{\ell_3} \partial \bm \theta_{\ell_1}} \frac{\partial \epsilon}{\partial \bm \theta_{\ell_4}} \frac{\partial \epsilon}{\partial \bm \theta_{\ell_2}} + \mathcal{O}(\eta^2) \\
    &=- 2\eta \epsilon \mu + \mathcal{O}(\eta^2).
\end{aligned} 
\end{equation}

Similarly, with linear loss function:

\begin{equation}
    - 2\eta \mu + \mathcal{O}(\eta^2).
\end{equation}

Therefore,  

\begin{equation}
\delta K_{\mathrm{QITE}} =
\begin{cases}
    -2\eta\, \epsilon(t)\, \mu(t)
    + \displaystyle\sum_{\ell_1,\ell_2} \delta \left( g^+_{\ell_1 \ell_2}(t) \right)
    \frac{\partial \epsilon}{\partial \bm \theta_{\ell_1}}(t)
    \frac{\partial \epsilon}{\partial \bm \theta_{\ell_2}}(t)
    + \mathcal{O}(\eta^2), & \text{(Quadratic loss)} \\[12pt]
    
    -2\eta\, \mu(t)
    + \displaystyle\sum_{\ell_1,\ell_2} \delta \left( g^+_{\ell_1 \ell_2}(t) \right)
    \frac{\partial \epsilon}{\partial \bm \theta_{\ell_1}}(t)
    \frac{\partial \epsilon}{\partial \bm \theta_{\ell_2}}(t)
    + \mathcal{O}(\eta^2), & \text{(Linear loss)}
\end{cases}
\label{eq:delta_KQITE_casewise}
\end{equation}

\begin{figure*}[t]
    \centering
    \includegraphics[width=0.8\textwidth]{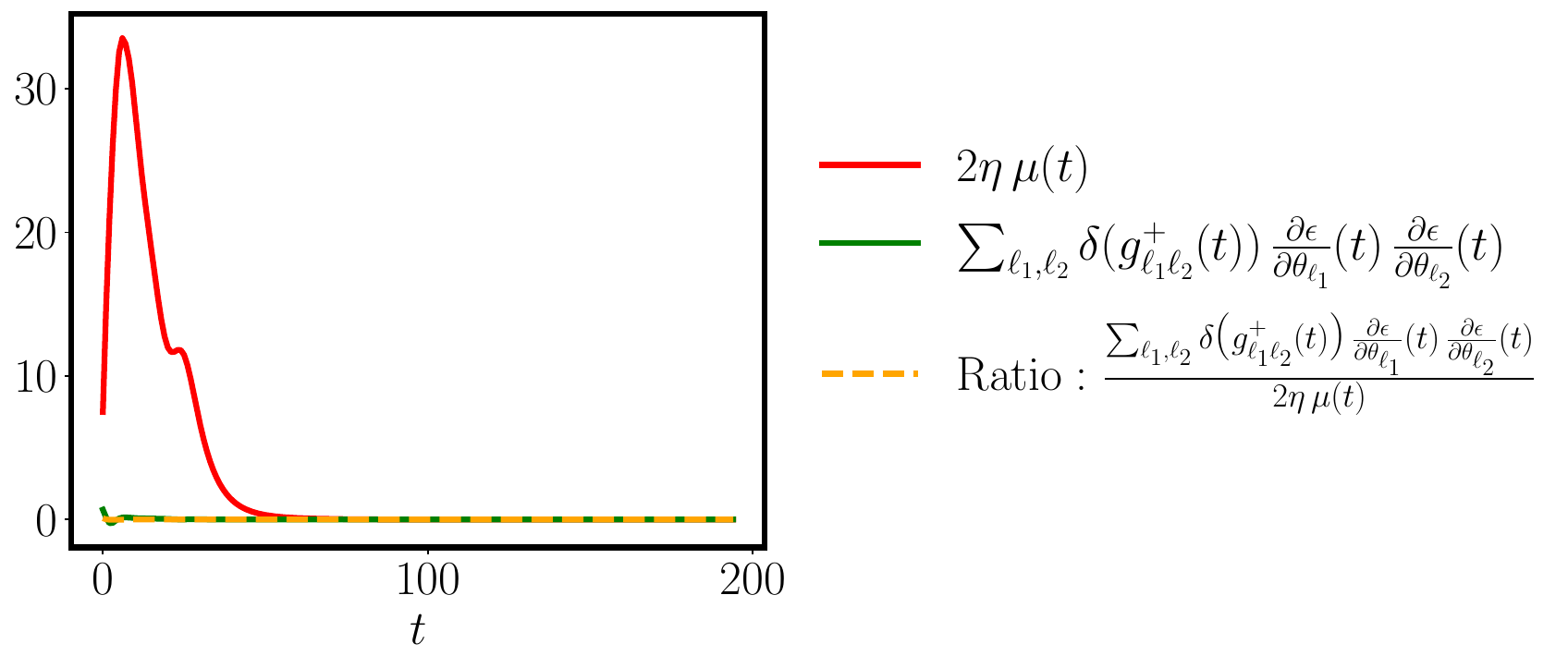}
    \caption{Numerical verification of the relative importance between the two leading-order terms in the time difference equation of \(K_{\mathrm{QITE}}\). The figure plots the ratio\(\frac{\sum_{\ell_1,\ell_2} \delta \left( g^+_{\ell_1\ell_2}(t) \right) \frac{\partial \epsilon}{\partial \bm \theta_{\ell_1}}(t) \frac{\partial \epsilon}{\partial \bm \theta_{\ell_2}}(t)}{2\eta \, \mu(t)}\), confirming that the term \(\sum_{\ell_1,\ell_2} \delta \left( g^+_{\ell_1\ell_2}(t) \right) \frac{\partial \epsilon}{\partial \bm \theta_{\ell_1}}(t) \frac{\partial \epsilon}{\partial \bm \theta_{\ell_2}}(t)\) remains significantly smaller throughout training and can be neglected to leading order in \(\eta\).}
    \label{fig:two_term_ratio}
\end{figure*}

Here, we focus on linear loss function because we assume \(K\) to be constant at late time with quadratic loss function. We examine if the term \(\sum_{\ell_1,\ell_2} \delta \left( g^+_{\ell_1 \ell_2}(t) \right)
    \frac{\partial \epsilon}{\partial \bm \theta_{\ell_1}}(t)
    \frac{\partial \epsilon}{\partial \bm \theta_{\ell_2}}(t)\) can be ignored. This is verified through a numerical study by calculating the ratio of \(2\eta \mu\) and \(\sum_{\ell_1,\ell_2} \delta \left(  g_{\ell_1 \ell_2}^{+} (t) \right) \frac{\partial \epsilon }{ \partial \bm \theta_{\ell_1} } (t)  \frac{\partial \epsilon}{\partial \bm \theta_{\ell_2} } (t)\). See Fig.\ref{fig:two_term_ratio} for the details.

By examining the numerical study, we make the simplification below:

\begin{equation}
\begin{aligned}
    \delta K_\text{QITE} &= -2\eta \mu  + \mathcal{O}(\eta^2)\\
\end{aligned}
\end{equation}

\section{Results with Haar Random Ensemble}

As shown in the main text, the parameterized unitary \({U}(\boldsymbol{\bm \theta})\) we consider is defined below: 
\begin{align}
{U}(\boldsymbol{\bm \theta})=\prod_{k=1}^L {W}_{\ell} {V}_{\ell}\left(\bm \theta_{\ell}\right)
\end{align}

We follow the notations in \citet{zhang2024dynamical}. Notably, we omit the constant factor \(1/2\) in the exponent and instead define \(\hat{V}_\ell(\bm \theta_\ell) = e^{-i\bm \theta_\ell \hat{X}_\ell}\) as below:

\begin{equation}
    \hat{U}(\boldsymbol{\bm \theta}) = \prod_{\ell=1}^{L} \hat{W}_\ell \hat{V}_\ell(\bm \theta_\ell),
\end{equation}
with:

\begin{equation}
    \hat{V}_\ell(\bm \theta_\ell) = e^{-i\bm \theta_\ell \hat{X}_\ell}
\end{equation}

This choice does not affect the generality of our results, as it amounts to a simple rescaling of the parameter range.
 For \(\ell\)-th parameter and specific \(\ell_1\), \(\ell_2\)-th parameters, we split \({U}(\boldsymbol{\bm \theta})\):

\begin{equation}
\begin{aligned}
    {U}(\boldsymbol{\bm \theta}) = U_{\ell^{-}} U_{\ell^{+}} = U_{\ell_1^{-}} U_{\ell_1 \to \ell_2} U_{\ell_2^{+}},
\end{aligned}
\end{equation}
where we define:

\begin{equation}
\begin{aligned}
&U_{\ell^{-}}=\prod_{k=1}^{\ell-1} W_k V_k\left(\bm \theta_k\right)\nonumber\\ &U_{\ell^{+}}=\prod_{k=\ell}^L W_k V_k\left(\bm \theta_k\right)\\
&U_{\ell_1 \to \ell_2}=\prod_{k=\ell_1}^{\ell_2-1} W_k V_k\left(\bm \theta_k\right)
\end{aligned}
\end{equation}

Indicating: 

\begin{equation}
    U_{\ell_1^{-}} U_{\ell_1^{+}} = U_{\ell_2^{-}} U_{\ell_2^{+}} = U_{\ell_1^{-}} U_{\ell_1 \to \ell_2} U_{\ell_2^{+}}
\end{equation}

\subsection{Average Fubini-Study Metric Tensor \(g\) Result under Haar Random Ensemble}
\label{Average Fisher information haar}

The variational output state \(\ket{\psi(\bm \theta)}\) can be given as:

\begin{equation}
   \ket{\psi(\bm \theta)} = U_{\ell^+} U_{\ell^-} | \psi_0 \rangle 
\end{equation}

For calculating \(g\), we firstly calculate the quantum geometric tensor \(G\):

\begin{equation}
    G_{\ell_1\ell_2}(\bm \theta) = \left\langle \frac{\partial \psi(\bm \theta)}{\partial \bm \theta_{\ell_1}} \bigg| \frac{\partial \psi(\bm \theta)}{\partial \bm \theta_{\ell_2}} \right\rangle 
- \left\langle \frac{\partial \psi(\bm \theta)}{\partial \bm \theta_{\ell_1}}\bigg| \psi(\bm \theta) \right\rangle \left\langle \psi(\bm \theta) \bigg| \frac{\partial \psi(\bm \theta)}{\partial \bm \theta_{\ell_2}} \right\rangle.
\end{equation}

Below we show the Haar ensemble average results for \(\left\langle \frac{\partial \psi(\bm \theta)}{\partial \bm \theta_{\ell_1}} \bigg| \frac{\partial \psi(\bm \theta)}{\partial \bm \theta_{\ell_2}} \right\rangle\) and \(\left\langle \frac{\partial \psi(\bm \theta)}{\partial \bm \theta_{\ell_1}}\bigg| \psi(\bm \theta) \right\rangle \left\langle \psi(\bm \theta) \bigg| \frac{\partial \psi(\bm \theta)}{\partial \bm \theta_{\ell_2}} \right\rangle\).

We can categorize the calculations into the diagonal case and off-diagonal case for \(G_{\ell_1 \ell_2}\). 

Firstly, we focus on the diagonal case, where \(\ell = \ell_1 = \ell_2\). For this, the analytic formula for \(\left\langle \frac{\partial \psi(\bm \theta)}{\partial \bm \theta_{\ell}} \bigg| \frac{\partial \psi(\bm \theta)}{\partial \bm \theta_{\ell}} \right\rangle\) as follows:

\begin{equation}
\begin{aligned}
&\left| \frac{\partial \psi(\bm \theta)}{\partial \bm \theta_\ell} \right\rangle = i U_{\ell^+} X_\ell U_{\ell^-} | \psi_0 \rangle \\
&\overline{\left\langle \frac{\partial \psi(\bm \theta)}{\partial \bm \theta_{\ell}} \bigg| \frac{\partial \psi(\bm \theta)}{\partial \bm \theta_{\ell}} \right\rangle} = 1
\end{aligned}
\end{equation}

For the calculation of \(\left\langle \frac{\partial \psi(\bm \theta)}{\partial \bm \theta_{\ell}}\bigg| \psi(\bm \theta) \right\rangle \left\langle \psi(\bm \theta) \bigg| \frac{\partial \psi(\bm \theta)}{\partial \bm \theta_{\ell}} \right\rangle\):

\begin{equation}
\begin{aligned}
\overline{\left\langle \frac{\partial \psi(\bm \theta)}{\partial \bm \theta_{\ell}} \middle| \psi(\bm \theta) \right\rangle \cdot \left\langle \psi(\bm \theta) \middle| \frac{\partial \psi(\bm \theta)}{\partial \bm \theta_{\ell}} \right\rangle}
&= \int dU_{\ell^-} dU_{\ell^+} (- i) \left\langle \psi_0 \right| U_{\ell^-}^\dagger X_{\ell} U_{\ell^+}^\dagger U_{\ell^+} U_{\ell^-} \left| \psi_0 \right\rangle (i) \left\langle \psi_0 \right| U_{\ell^-} U_{\ell^+} U_{\ell^+}^\dagger X_{\ell} U_{\ell^+} \left| \psi_0 \right\rangle \\
&= \int dU_{\ell^-} (- i) \left\langle \psi_0 \right| U_{\ell^-}^\dagger X_{\ell} U_{\ell^-} \left| \psi_0 \right\rangle (i) \left\langle \psi_0 \right| U_{\ell^-}^\dagger X_{\ell} U_{\ell^-} \left| \psi_0 \right\rangle \\
&= \frac{1}{N+1},
\end{aligned}
\end{equation}
where we leverage RTNI package~\cite{fukuda2019rtni} to calculate the last equation. Therefore

\begin{equation}
\overline{G_{\ell \ell}} = \overline{\left\langle \frac{\partial \psi(\bm \theta)}{\partial \bm \theta_{\ell}} \bigg| \frac{\partial \psi(\bm \theta)}{\partial \bm \theta_{\ell}} \right\rangle} - \overline{\left\langle \frac{\partial \psi(\bm \theta)}{\partial \bm \theta_{\ell}} \middle| \psi(\bm \theta) \right\rangle \cdot \left\langle \psi(\bm \theta) \middle| \frac{\partial \psi(\bm \theta)}{\partial \bm \theta_{\ell}} \right\rangle} = 1 - \frac{1}{N+1}
\end{equation}

For off-diagonal case where \(\ell_1 \neq \ell_2\). Because \({U}(\boldsymbol{\bm \theta}) = U_{\ell_1^{-}} U_{\ell_1^{+}} = U_{\ell_1^{-}} U_{\ell_1 \to \ell_2} U_{\ell_2^{+}}\), we have:

\begin{equation}
\begin{aligned}
    &U_{\ell_1^{+}} = U_{\ell_1 \to \ell_2} U_{\ell_2^{+}} \\
    &U_{\ell_{1}^{+}}^\dagger =  U_{\ell_2^{+}}^\dagger U_{\ell_1 \to \ell_2}^\dagger 
\end{aligned}
\end{equation}

Accordingly, we can calculate the analytic formula for \(\left\langle \frac{\partial \psi(\bm \theta)}{\partial \bm \theta_{\ell_1}} \bigg| \frac{\partial \psi(\bm \theta)}{\partial \bm \theta_{\ell_2}} \right\rangle\) and \(\left\langle \frac{\partial \psi(\bm \theta)}{\partial \bm \theta_{\ell_1}}\bigg| \psi(\bm \theta) \right\rangle \left\langle \psi(\bm \theta) \bigg| \frac{\partial \psi(\bm \theta)}{\partial \bm \theta_{\ell_2}} \right\rangle\) respectively, where 


\begin{equation}
\begin{aligned}
\left\langle \frac{\partial \psi(\bm \theta)}{\partial \bm \theta_{\ell_1}} \bigg| \frac{\partial \psi(\bm \theta)}{\partial \bm \theta_{\ell_2}} \right\rangle &= \left\langle \psi_0 \right| U_{\ell_1^-}^\dagger X U_{\ell_1^+}^\dagger U_{\ell_2^+} X U_{\ell_2^-} \left| \psi_0 \right\rangle \\
&= \left\langle \psi_0 \right| U_{\ell_1^-}^\dagger X U_{\ell_1 \to \ell_2}^\dagger U_{\ell_2^{+}}^\dagger U_{\ell_2^+} X U_{\ell_2^-} \left| \psi_0 \right\rangle \\
&= \left\langle \psi_0 \right| U_{\ell_1^-}^\dagger X U_{\ell_1 \to \ell_2}^\dagger X U_{\ell_2^-} \left| \psi_0 \right\rangle \\
&= \Tr{\rho_0 U_{\ell_1^-}^\dagger X U_{\ell_1 \to \ell_2}^\dagger X U_{\ell_2^-} } \\
\end{aligned}
\end{equation}

Therefore, the result for \(\left\langle \frac{\partial \psi(\bm \theta)}{\partial \bm \theta_{\ell_1}} \bigg| \frac{\partial \psi(\bm \theta)}{\partial \bm \theta_{\ell_2}} \right\rangle\) is as follows:

\begin{equation}
\begin{aligned}
    \overline{\left\langle \frac{\partial \psi(\bm \theta)}{\partial \bm \theta_{\ell_1}} \bigg| \frac{\partial \psi(\bm \theta)}{\partial \bm \theta_{\ell_2}} \right\rangle} &= \int dU_{\ell_1^-} U_{\ell_1 \to \ell_2} dU_{\ell_2^-} \left\langle \psi_0 \right| U_{\ell_1^-}^\dagger X U_{\ell_1 \to \ell_2}^\dagger X U_{\ell_2^-} \left| \psi_0 \right\rangle\\
    &= \int dU_{\ell_1^-} U_{\ell_1 \to \ell_2} dU_{\ell_2^-} \Tr{\rho_0  U_{\ell_1^-}^\dagger X U_{\ell_1 \to \ell_2}^\dagger X U_{\ell_2^-} } \\
    &= 0
\end{aligned}
\end{equation}

Similarly, we can derive the result for \(\left\langle \frac{\partial \psi(\bm \theta)}{\partial \bm \theta_{\ell_1}} \middle| \psi(\bm \theta) \right\rangle \cdot \left\langle \psi(\bm \theta) \middle| \frac{\partial \psi(\bm \theta)}{\partial \bm \theta_{\ell_2}} \right\rangle\) as below:

\begin{equation}
\begin{aligned}
    \overline{\left\langle \frac{\partial \psi(\bm \theta)}{\partial \bm \theta_{\ell_1}} \middle| \psi(\bm \theta) \right\rangle \cdot \left\langle \psi(\bm \theta) \middle| \frac{\partial \psi(\bm \theta)}{\partial \bm \theta_{\ell_2}} \right\rangle} 
&= \int dU_{\ell_1^-} dU_{\ell_2^+} (- i) \left\langle \psi_0 \right| U_{\ell_1^-}^\dagger X_{\ell_1} U_{\ell_1^+}^\dagger U_{\ell_1^+} U_{\ell_1^-} \left| \psi_0 \right\rangle (i) \left\langle \psi_0 \right| U_{\ell_2^-}^\dagger
U_{\ell_2^+}^\dagger U_{\ell_2^+} X_{\ell_2} U_{\ell_2^-}  \left| \psi_0 \right\rangle \\
&= \int dU_{\ell_1^-} dU_{\ell_2^+} \left\langle \psi_0 \right| U_{\ell_1^-}^\dagger X_{\ell_1} U_{\ell_1^-} \left| \psi_0 \right\rangle \left\langle \psi_0 \right| U_{\ell_2^-}^\dagger
X_{\ell_2} U_{\ell_2^-} \left| \psi_0 \right\rangle \\
&= \int dU_{\ell_1^-} dU_{\ell_2^+} \Tr{\rho_0 U_{\ell_1^-}^\dagger X_{\ell_1} U_{\ell_1^-} \rho_0 U_{\ell_2^-}^\dagger
X_{\ell_2} U_{\ell_2^-} \left| \psi_0 \right\rangle}  \\
&= 0
\end{aligned}
\end{equation}

Therfore, we can have a general formula for \(\overline{G_{\ell_1 \ell_2}}\):

\begin{equation}
    \overline{G_{\ell_1 \ell_2}} = (1 - \frac{1}{N+1}) \delta_{\ell_1 \ell_2},
\end{equation}
where \(\delta_{\ell_1 \ell_2}\) represents the Kronecker delta defined as:

\begin{equation}
    \delta_{ij} = 
\begin{cases} 
0 & \text{if } i \neq j, \\
1 & \text{if } i = j.
\end{cases}
\end{equation}

\subsection{Fluctuations of Fubini-Study Metric Tensor \(g\) Under Haar Random Ensemble}
\label{sec:fluctuations_g}

In this section, we analyze the fluctuations in \(g\) around \(\overline{g}\), \ie, \(\Delta g_{\ell_1 \ell_2}^2 = \mathbb{E}(g_{\ell_1 \ell_2}^2) - \overline{g_{\ell_1 \ell_2}}^2 = \overline{g_{\ell_1 \ell_2}^2} - \overline{g_{\ell_1 \ell_2}}^2\)

For the off-diagonal case where \(\ell_1 \neq \ell_2\), we already have \(\overline{g_{\ell_1 \ell_2}}^2 = 0\). According to:

\begin{equation}
\begin{aligned}
&\left| \frac{\partial \psi(\bm \theta)}{\partial \bm \theta_\ell} \right\rangle = i U_{\ell^+} X_\ell U_{\ell^-} | \psi_0 \rangle \\
&   \ket{\psi(\bm \theta)} = U_{\ell^+} U_{\ell^-} | \psi_0 \rangle \\
\end{aligned}
\end{equation}

For \(G_{\ell_1 \ell_2}\):

\begin{equation}
\begin{aligned}
    G_{\ell_1\ell_2}(\bm \theta) &= \left\langle \frac{\partial \psi(\bm \theta)}{\partial \bm \theta_{\ell_1}} \bigg| \frac{\partial \psi(\bm \theta)}{\partial \bm \theta_{\ell_2}} \right\rangle 
- \left\langle \frac{\partial \psi(\bm \theta)}{\partial \bm \theta_{\ell_1}}\bigg| \psi(\bm \theta) \right\rangle \left\langle \psi(\bm \theta) \bigg| \frac{\partial \psi(\bm \theta)}{\partial \bm \theta_{\ell_2}} \right\rangle \\
&= \left\langle \psi_0 \big| U_{\ell_1^-}^\dagger X_{\ell_1} U_{\ell_1^+}^\dagger U_{\ell_2^+} X_{\ell_2} U_{\ell_2^-} \big| \psi_0 \right\rangle \\
&\quad - \left\langle \psi_0 \big| U_{\ell_1^-}^\dagger X_{\ell_1} U_{\ell_1^+}^\dagger U_{\ell_1^+} U_{\ell_1^-} \big| \psi_0 \right\rangle
\left\langle \psi_0 \big| U_{\ell_2^-}^\dagger U_{\ell_2^+}^\dagger U_{\ell_2^+}  X_{\ell_2} U_{\ell_2^-} \big| \psi_0 \right\rangle\\
&= \left\langle \psi_0 \big| U_{\ell_1^-}^\dagger X_{\ell_1} U_{\ell_1^+}^\dagger U_{\ell_2^+} X_{\ell_2} U_{\ell_2^-} \big| \psi_0 \right\rangle \\
&\quad - \left\langle \psi_0 \big| U_{\ell_1^-}^\dagger X_{\ell_1} U_{\ell_1^-} \big| \psi_0 \right\rangle
\left\langle \psi_0 \big| U_{\ell_2^-}^\dagger  X_{\ell_2} U_{\ell_2^-} \big| \psi_0 \right\rangle
\end{aligned}
\end{equation}

Therefore,

\begin{equation}
\begin{aligned}
G_{\ell_1\ell_2}^*(\bm \theta)
&=\left\langle \psi_0 \Big|
      U_{\ell_2^-}^\dagger X_{\ell_2} U_{\ell_2^+}^\dagger
      U_{\ell_1^+} X_{\ell_1} U_{\ell_1^-}
   \Big| \psi_0 \right\rangle \\[2pt]
&\quad-
  \left\langle \psi_0 \Big|
      U_{\ell_1^-}^\dagger X_{\ell_1} U_{\ell_1^-}
   \Big| \psi_0 \right\rangle
  \left\langle \psi_0 \Big|
      U_{\ell_2^-}^\dagger X_{\ell_2} U_{\ell_2^-}
   \Big| \psi_0 \right\rangle.
\end{aligned}
\end{equation}

For \(g = \Re{G} = \frac{G + G^*}{2}\), we have:

\begin{equation}
\begin{aligned}
g_{\ell_1\ell_2}(\bm \theta)
&=\frac{G_{\ell_1\ell_2}(\bm \theta)+G_{\ell_1\ell_2}^*(\bm \theta)}{2}\\[4pt]
&=\frac{1}{2}\!
  \Big(
     \left\langle \psi_0 \Big|
        U_{\ell_1^-}^\dagger X_{\ell_1} U_{\ell_1^+}^\dagger
        U_{\ell_2^+} X_{\ell_2} U_{\ell_2^-}
     \Big| \psi_0 \right\rangle
     +
     \left\langle \psi_0 \Big|
        U_{\ell_2^-}^\dagger X_{\ell_2} U_{\ell_2^+}^\dagger
        U_{\ell_1^+} X_{\ell_1} U_{\ell_1^-}
     \Big| \psi_0 \right\rangle
  \Big)\\[4pt]
&\quad-
  \left\langle \psi_0 \Big|
      U_{\ell_1^-}^\dagger X_{\ell_1} U_{\ell_1^-}
   \Big| \psi_0 \right\rangle
  \left\langle \psi_0 \Big|
      U_{\ell_2^-}^\dagger X_{\ell_2} U_{\ell_2^-}
   \Big| \psi_0 \right\rangle.
\end{aligned}
\end{equation}

Therefore, \(g_{\ell_1 \ell_2}^2\) can be derived:

\begin{equation}
\begin{aligned}
g_{\ell_1\ell_2}^2(\bm \theta)
&=\left[
      \frac{
        \left\langle \psi_0 \Big|
           U_{\ell_1^-}^\dagger X_{\ell_1} U_{\ell_1^+}^\dagger
           U_{\ell_2^+} X_{\ell_2} U_{\ell_2^-}
        \Big| \psi_0 \right\rangle
        +
        \left\langle \psi_0 \Big|
           U_{\ell_2^-}^\dagger X_{\ell_2} U_{\ell_2^+}^\dagger
           U_{\ell_1^+} X_{\ell_1} U_{\ell_1^-}
        \Big| \psi_0 \right\rangle
      }{2}
      -
      \left\langle \psi_0 \Big|
         U_{\ell_1^-}^\dagger X_{\ell_1} U_{\ell_1^-}
      \Big| \psi_0 \right\rangle
      \left\langle \psi_0 \Big|
         U_{\ell_2^-}^\dagger X_{\ell_2} U_{\ell_2^-}
      \Big| \psi_0 \right\rangle
   \right]^2 \\[6pt]
&=\frac14 \!
   \left(
      \left\langle \psi_0 \Big|
         U_{\ell_1^-}^\dagger X_{\ell_1} U_{\ell_1^+}^\dagger
         U_{\ell_2^+} X_{\ell_2} U_{\ell_2^-}
      \Big| \psi_0 \right\rangle
      +
      \left\langle \psi_0 \Big|
         U_{\ell_2^-}^\dagger X_{\ell_2} U_{\ell_2^+}^\dagger
         U_{\ell_1^+} X_{\ell_1} U_{\ell_1^-}
      \Big| \psi_0 \right\rangle
   \right)^2 \\[6pt]
&\quad-
   \left(
      \left\langle \psi_0 \Big|
         U_{\ell_1^-}^\dagger X_{\ell_1} U_{\ell_1^+}^\dagger
         U_{\ell_2^+} X_{\ell_2} U_{\ell_2^-}
      \Big| \psi_0 \right\rangle
      +
      \left\langle \psi_0 \Big|
         U_{\ell_2^-}^\dagger X_{\ell_2} U_{\ell_2^+}^\dagger
         U_{\ell_1^+} X_{\ell_1} U_{\ell_1^-}
      \Big| \psi_0 \right\rangle
   \right)
   \left\langle \psi_0 \Big|
      U_{\ell_1^-}^\dagger X_{\ell_1} U_{\ell_1^-}
   \Big| \psi_0 \right\rangle
   \left\langle \psi_0 \Big|
      U_{\ell_2^-}^\dagger X_{\ell_2} U_{\ell_2^-}
   \Big| \psi_0 \right\rangle \\[6pt]
&\quad+
   \left(
      \left\langle \psi_0 \Big|
         U_{\ell_1^-}^\dagger X_{\ell_1} U_{\ell_1^-}
      \Big| \psi_0 \right\rangle
      \left\langle \psi_0 \Big|
         U_{\ell_2^-}^\dagger X_{\ell_2} U_{\ell_2^-}
      \Big| \psi_0 \right\rangle
   \right)^2.\\
&=\frac14
  \left\langle \psi_0 \Big|
     U_{\ell_1^-}^\dagger X_{\ell_1} U_{\ell_1^+}^\dagger
     U_{\ell_2^+} X_{\ell_2} U_{\ell_2^-}
   \Big| \psi_0 \right\rangle^{\!2} \\[4pt]
&\quad+\frac14
  \left\langle \psi_0 \Big|
     U_{\ell_2^-}^\dagger X_{\ell_2} U_{\ell_2^+}^\dagger
     U_{\ell_1^+} X_{\ell_1} U_{\ell_1^-}
   \Big| \psi_0 \right\rangle^{\!2} \\[4pt]
&\quad+\frac12
  \left\langle \psi_0 \Big|
     U_{\ell_1^-}^\dagger X_{\ell_1} U_{\ell_1^+}^\dagger
     U_{\ell_2^+} X_{\ell_2} U_{\ell_2^-}
   \Big| \psi_0 \right\rangle
  \left\langle \psi_0 \Big|
     U_{\ell_2^-}^\dagger X_{\ell_2} U_{\ell_2^+}^\dagger
     U_{\ell_1^+} X_{\ell_1} U_{\ell_1^-}
   \Big| \psi_0 \right\rangle \\[6pt]
&\quad-
  \left\langle \psi_0 \Big|
     U_{\ell_1^-}^\dagger X_{\ell_1} U_{\ell_1^+}^\dagger
     U_{\ell_2^+} X_{\ell_2} U_{\ell_2^-}
   \Big| \psi_0 \right\rangle
  \left\langle \psi_0 \Big|
     U_{\ell_1^-}^\dagger X_{\ell_1} U_{\ell_1^-}
   \Big| \psi_0 \right\rangle
  \left\langle \psi_0 \Big|
     U_{\ell_2^-}^\dagger X_{\ell_2} U_{\ell_2^-}
   \Big| \psi_0 \right\rangle \\[4pt]
&\quad-
  \left\langle \psi_0 \Big|
     U_{\ell_2^-}^\dagger X_{\ell_2} U_{\ell_2^+}^\dagger
     U_{\ell_1^+} X_{\ell_1} U_{\ell_1^-}
   \Big| \psi_0 \right\rangle
  \left\langle \psi_0 \Big|
     U_{\ell_1^-}^\dagger X_{\ell_1} U_{\ell_1^-}
   \Big| \psi_0 \right\rangle
  \left\langle \psi_0 \Big|
     U_{\ell_2^-}^\dagger X_{\ell_2} U_{\ell_2^-}
   \Big| \psi_0 \right\rangle \\[4pt]
&\quad+
  \left\langle \psi_0 \Big|
     U_{\ell_1^-}^\dagger X_{\ell_1} U_{\ell_1^-}
   \Big| \psi_0 \right\rangle^{\!2}
  \left\langle \psi_0 \Big|
     U_{\ell_2^-}^\dagger X_{\ell_2} U_{\ell_2^-}
   \Big| \psi_0 \right\rangle^{\!2}.
\end{aligned}
\end{equation}

For \(\overline{\left\langle \psi_0 \Big|
     U_{\ell_1^-}^\dagger X_{\ell_1} U_{\ell_1^+}^\dagger
     U_{\ell_2^+} X_{\ell_2} U_{\ell_2^-}
   \Big| \psi_0 \right\rangle^{\!2}}\),

\begin{equation}
\begin{aligned}
    \overline{\left\langle \psi_0 \Big|
     U_{\ell_1^-}^\dagger X_{\ell_1} U_{\ell_1^+}^\dagger
     U_{\ell_2^+} X_{\ell_2} U_{\ell_2^-}
   \Big| \psi_0 \right\rangle^{\!2}} = &\int dU_{\ell_1^-} dU_{\ell_1^+} dU_{\ell_2^-} dU_{\ell_2^+} \left\langle \psi_0 \Big|
     U_{\ell_1^-}^\dagger X_{\ell_1} U_{\ell_1^+}^\dagger
     U_{\ell_2^+} X_{\ell_2} U_{\ell_2^-}
   \Big| \psi_0 \right\rangle \left\langle \psi_0 \Big|
     U_{\ell_1^-}^\dagger X_{\ell_1} U_{\ell_1^+}^\dagger
     U_{\ell_2^+} X_{\ell_2} U_{\ell_2^-}
   \Big| \psi_0 \right\rangle \\
   &= \int dU_{\ell_1^-} dU_{\ell_1^+} dU_{\ell_2^-} dU_{\ell_2^+} \Tr{\rho_0
     U_{\ell_1^-}^\dagger X_{\ell_1} U_{\ell_1^+}^\dagger
     U_{\ell_2^+} X_{\ell_2} U_{\ell_2^-}
   \rho_0
     U_{\ell_1^-}^\dagger X_{\ell_1} U_{\ell_1^+}^\dagger
     U_{\ell_2^+} X_{\ell_2} U_{\ell_2^-}
   }\\
   &= 0
\end{aligned}
\end{equation}

For \(\overline{\left\langle \psi_0 \Big|
   U_{\ell_2^-}^\dagger X_{\ell_2} U_{\ell_2^+}^\dagger
   U_{\ell_1^+} X_{\ell_1} U_{\ell_1^-}
 \Big| \psi_0 \right\rangle^{\!2}}\),

\begin{equation}
\begin{aligned}
\overline{\left\langle \psi_0 \Big|
   U_{\ell_2^-}^\dagger X_{\ell_2} U_{\ell_2^+}^\dagger
   U_{\ell_1^+} X_{\ell_1} U_{\ell_1^-}
 \Big| \psi_0 \right\rangle^{\!2}}
&= \!\!\int dU_{\ell_1^-}dU_{\ell_1^+}dU_{\ell_2^-}dU_{\ell_2^+}\;
   \Tr{\rho_0
       U_{\ell_2^-}^\dagger X_{\ell_2} U_{\ell_2^+}^\dagger
       U_{\ell_1^+} X_{\ell_1} U_{\ell_1^-}\,
       \rho_0\,
       U_{\ell_2^-}^\dagger X_{\ell_2} U_{\ell_2^+}^\dagger
       U_{\ell_1^+} X_{\ell_1} U_{\ell_1^-}}
\; \\
&=0
\end{aligned}
\end{equation}

For \(\overline{\left\langle \psi_0 \Big|
   U_{\ell_1^-}^\dagger X_{\ell_1} U_{\ell_1^+}^\dagger
   U_{\ell_2^+} X_{\ell_2} U_{\ell_2^-}
 \Big| \psi_0 \right\rangle\!
 \left\langle \psi_0 \Big|
   U_{\ell_2^-}^\dagger X_{\ell_2} U_{\ell_2^+}^\dagger
   U_{\ell_1^+} X_{\ell_1} U_{\ell_1^-}
 \Big| \psi_0 \right\rangle}\),

\begin{equation}
\begin{aligned}
&\overline{\left\langle \psi_0 \Big|
   U_{\ell_1^-}^\dagger X_{\ell_1} U_{\ell_1^+}^\dagger
   U_{\ell_2^+} X_{\ell_2} U_{\ell_2^-}
 \Big| \psi_0 \right\rangle\!
 \left\langle \psi_0 \Big|
   U_{\ell_2^-}^\dagger X_{\ell_2} U_{\ell_2^+}^\dagger
   U_{\ell_1^+} X_{\ell_1} U_{\ell_1^-}
 \Big| \psi_0 \right\rangle}\\
&= \!\!\int dU_{\ell_1^-}dU_{\ell_1^+}dU_{\ell_2^-}dU_{\ell_2^+}\;
   \Tr{\rho_0
       U_{\ell_1^-}^\dagger X_{\ell_1} U_{\ell_1^+}^\dagger
       U_{\ell_2^+} X_{\ell_2} U_{\ell_2^-}\,
       \rho_0\,
       U_{\ell_2^-}^\dagger X_{\ell_2} U_{\ell_2^+}^\dagger
       U_{\ell_1^+} X_{\ell_1} U_{\ell_1^-}}
\; \\
&=\frac{1}{N}
\end{aligned}
\end{equation}

For \(\overline{\left\langle \psi_0 \Big|
   U_{\ell_1^-}^\dagger X_{\ell_1} U_{\ell_1^+}^\dagger
   U_{\ell_2^+} X_{\ell_2} U_{\ell_2^-}
 \Big| \psi_0 \right\rangle
 \left\langle \psi_0 \Big|
   U_{\ell_1^-}^\dagger X_{\ell_1} U_{\ell_1^-}
 \Big| \psi_0 \right\rangle
 \left\langle \psi_0 \Big|
   U_{\ell_2^-}^\dagger X_{\ell_2} U_{\ell_2^-}
 \Big| \psi_0 \right\rangle}\),

\begin{equation}
\begin{aligned}
&\overline{\left\langle \psi_0 \Big|
   U_{\ell_1^-}^\dagger X_{\ell_1} U_{\ell_1^+}^\dagger
   U_{\ell_2^+} X_{\ell_2} U_{\ell_2^-}
 \Big| \psi_0 \right\rangle
 \left\langle \psi_0 \Big|
   U_{\ell_1^-}^\dagger X_{\ell_1} U_{\ell_1^-}
 \Big| \psi_0 \right\rangle
 \left\langle \psi_0 \Big|
   U_{\ell_2^-}^\dagger X_{\ell_2} U_{\ell_2^-}
 \Big| \psi_0 \right\rangle} \\
&= \!\!\int dU_{\ell_1^-}dU_{\ell_1^+}dU_{\ell_2^-}dU_{\ell_2^+}\;
   \Tr{\rho_0
       U_{\ell_1^-}^\dagger X_{\ell_1} U_{\ell_1^+}^\dagger
       U_{\ell_2^+} X_{\ell_2} U_{\ell_2^-}\,
       \rho_0\,
       U_{\ell_1^-}^\dagger X_{\ell_1} U_{\ell_1^-}\,
       \rho_0\,
       U_{\ell_2^-}^\dagger X_{\ell_2} U_{\ell_2^-}}
\; \\
&=0.
\end{aligned}
\end{equation}

For \(\overline{\left\langle \psi_0 \Big|
   U_{\ell_2^-}^\dagger X_{\ell_2} U_{\ell_2^+}^\dagger
   U_{\ell_1^+} X_{\ell_1} U_{\ell_1^-}
 \Big| \psi_0 \right\rangle
 \left\langle \psi_0 \Big|
   U_{\ell_1^-}^\dagger X_{\ell_1} U_{\ell_1^-}
 \Big| \psi_0 \right\rangle
 \left\langle \psi_0 \Big|
   U_{\ell_2^-}^\dagger X_{\ell_2} U_{\ell_2^-}
 \Big| \psi_0 \right\rangle}\),

\begin{equation}
\begin{aligned}
&\overline{\left\langle \psi_0 \Big|
   U_{\ell_2^-}^\dagger X_{\ell_2} U_{\ell_2^+}^\dagger
   U_{\ell_1^+} X_{\ell_1} U_{\ell_1^-}
 \Big| \psi_0 \right\rangle
 \left\langle \psi_0 \Big|
   U_{\ell_1^-}^\dagger X_{\ell_1} U_{\ell_1^-}
 \Big| \psi_0 \right\rangle
 \left\langle \psi_0 \Big|
   U_{\ell_2^-}^\dagger X_{\ell_2} U_{\ell_2^-}
 \Big| \psi_0 \right\rangle} \\
&= \!\!\int dU_{\ell_1^-}dU_{\ell_1^+}dU_{\ell_2^-}dU_{\ell_2^+}\;
   \Tr{\rho_0
       U_{\ell_2^-}^\dagger X_{\ell_2} U_{\ell_2^+}^\dagger
       U_{\ell_1^+} X_{\ell_1} U_{\ell_1^-}\,
       \rho_0\,
       U_{\ell_1^-}^\dagger X_{\ell_1} U_{\ell_1^-}\,
       \rho_0\,
       U_{\ell_2^-}^\dagger X_{\ell_2} U_{\ell_2^-}}
\;\\
&=0.
\end{aligned}
\end{equation}

For \(\overline{\left\langle \psi_0 \Big|
   U_{\ell_1^-}^\dagger X_{\ell_1} U_{\ell_1^-}
 \Big| \psi_0 \right\rangle^{\!2}
 \left\langle \psi_0 \Big|
   U_{\ell_2^-}^\dagger X_{\ell_2} U_{\ell_2^-}
 \Big| \psi_0 \right\rangle^{\!2}}\),

\begin{equation}
\begin{aligned}
&\overline{\bigl\langle \psi_0 \big|
   U_{\ell_1^-}^\dagger X_{\ell_1} U_{\ell_1^-}
 \big| \psi_0 \bigr\rangle^{\!2}
   \bigl\langle \psi_0 \big|
   U_{\ell_2^-}^\dagger X_{\ell_2} U_{\ell_2^-}
 \big| \psi_0 \bigr\rangle^{\!2}} \\
&= \!\!\int dU_{\ell_1^-}dU_{\ell_2^-}\;
   \Tr{\rho_0
       U_{\ell_1^-}^\dagger X_{\ell_1} U_{\ell_1^-}\,
       \rho_0\,
       U_{\ell_1^-}^\dagger X_{\ell_1} U_{\ell_1^-}\,
       \rho_0\,
       U_{\ell_2^-}^\dagger X_{\ell_2} U_{\ell_2^-}\,
       \rho_0\,
       U_{\ell_2^-}^\dagger X_{\ell_2} U_{\ell_2^-}}
\; \\
&=\frac{1}{(N+1)^2}.
\end{aligned}
\end{equation}

Therefore, 

\begin{equation}
\begin{aligned}
    &\overline{g_{\ell_1\ell_2}^2(\bm \theta)} = \frac{1}{2N} + \frac{1}{(N+1)^2},\\
    &\Delta g_{\ell_1 \ell_2}^2 = \mathbb{E}(g_{\ell_1 \ell_2}^2) - \overline{g_{\ell_1 \ell_2}}^2 = \overline{g_{\ell_1 \ell_2}^2} - \overline{g_{\ell_1 \ell_2}}^2 =\frac{1}{2N} + \frac{1}{(N+1)^2}.
\end{aligned}
\end{equation}

In large-N limit, 

\begin{equation}
    \Delta g_{\ell_1 \ell_2}^2 \approx \frac{1}{2N}
\end{equation}

When \(\ell_1 = \ell_2\),

\begin{equation}
\begin{aligned}
    &\Bigl[g_{\ell_1\ell_2}^2(\bm \theta)\Bigr]_{\ell_1=\ell_2=\ell} = g_{\ell \ell}^2(\bm \theta)
\\[2pt]
&= \frac14
   \Bigl\langle\psi_0\Big|
     U_{\ell^-}^{\dagger} X_{\ell} U_{\ell^+}^{\dagger}
     U_{\ell^+} X_{\ell} U_{\ell^-}
   \Big|\psi_0\Bigr\rangle^{2}
 + \frac14
   \Bigl\langle\psi_0\Big|
     U_{\ell^-}^{\dagger} X_{\ell} U_{\ell^+}^{\dagger}
     U_{\ell^+} X_{\ell} U_{\ell^-}
   \Big|\psi_0\Bigr\rangle^{2}
\\
&\quad +\frac12
   \Bigl\langle\psi_0\Big|
     U_{\ell^-}^{\dagger} X_{\ell} U_{\ell^+}^{\dagger}
     U_{\ell^+} X_{\ell} U_{\ell^-}
   \Big|\psi_0\Bigr\rangle
   \Bigl\langle\psi_0\Big|
     U_{\ell^-}^{\dagger} X_{\ell} U_{\ell^+}^{\dagger}
     U_{\ell^+} X_{\ell} U_{\ell^-}
   \Big|\psi_0\Bigr\rangle
\\
&\quad - 
   \Bigl\langle\psi_0\Big|
     U_{\ell^-}^{\dagger} X_{\ell} U_{\ell^+}^{\dagger}
     U_{\ell^+} X_{\ell} U_{\ell^-}
   \Big|\psi_0\Bigr\rangle
   \Bigl\langle\psi_0\Big|
     U_{\ell^-}^{\dagger} X_{\ell} U_{\ell^-}
   \Big|\psi_0\Bigr\rangle^{2}
\\
&\quad -
   \Bigl\langle\psi_0\Big|
     U_{\ell^-}^{\dagger} X_{\ell} U_{\ell^+}^{\dagger}
     U_{\ell^+} X_{\ell} U_{\ell^-}
   \Big|\psi_0\Bigr\rangle
   \Bigl\langle\psi_0\Big|
     U_{\ell^-}^{\dagger} X_{\ell} U_{\ell^-}
   \Big|\psi_0\Bigr\rangle^{2}
\\
&\quad +
   \Bigl\langle\psi_0\Big|
     U_{\ell^-}^{\dagger} X_{\ell} U_{\ell^-}
   \Big|\psi_0\Bigr\rangle^{4}
\\[6pt]
&=
   \frac14
   \Bigl\langle\psi_0\Big|
     U_{\ell^-}^{\dagger} X_{\ell}^{2} U_{\ell^-}
   \Big|\psi_0\Bigr\rangle^{2}
 + \frac14
   \Bigl\langle\psi_0\Big|
     U_{\ell^-}^{\dagger} X_{\ell}^{2} U_{\ell^-}
   \Big|\psi_0\Bigr\rangle^{2}
\\
&\quad +\frac12
   \Bigl\langle\psi_0\Big|
     U_{\ell^-}^{\dagger} X_{\ell}^{2} U_{\ell^-}
   \Big|\psi_0\Bigr\rangle^{2}
 - 2\,
   \Bigl\langle\psi_0\Big|
     U_{\ell^-}^{\dagger} X_{\ell}^{2} U_{\ell^-}
   \Big|\psi_0\Bigr\rangle
   \Bigl\langle\psi_0\Big|
     U_{\ell^-}^{\dagger} X_{\ell} U_{\ell^-}
   \Big|\psi_0\Bigr\rangle^{2}
\\
&\quad +
   \Bigl\langle\psi_0\Big|
     U_{\ell^-}^{\dagger} X_{\ell} U_{\ell^-}
   \Big|\psi_0\Bigr\rangle^{4}
\\[6pt]
&= 1 - 2
   \Bigl\langle\psi_0\Big|
     U_{\ell^-}^{\dagger} X_{\ell} U_{\ell^-}
   \Big|\psi_0\Bigr\rangle^{2} +
   \Bigl\langle\psi_0\Big|
     U_{\ell^-}^{\dagger} X_{\ell} U_{\ell^-}
   \Big|\psi_0\Bigr\rangle^{4}
\\[6pt]
\end{aligned}
\end{equation}

For \(\overline{\bigl\langle\psi_0\big|U_{\ell^-}^{\dagger}X_{\ell}U_{\ell^-}\big|\psi_0\bigr\rangle^{2}}\),

\begin{equation}
\begin{aligned}
&\overline{\bigl\langle\psi_0\big|U_{\ell^-}^{\dagger}X_{\ell}U_{\ell^-}\big|\psi_0\bigr\rangle^{2}}
\\
&=\;\int dU_{\ell^-}\;
\Bigl[\Tr{\rho_0\,U_{\ell^-}^{\dagger}X_{\ell}U_{\ell^-}}\Bigr]^{2}\; \\
&=\frac{1}{N+1}.
\end{aligned}
\end{equation}

For \(\overline{\bigl\langle\psi_0\big|U_{\ell^-}^{\dagger}X_{\ell}U_{\ell^-}\big|\psi_0\bigr\rangle^{4}}\), we adopt Wick contraction technique as it provides an efficient method for computing high-order moments of Haar-random unitaries in the large-$N$ limit, where matrix elements become approximately Gaussian \cite{collins2006integration}. This makes the $2k$-point correlators decompose into $(2k-1)!!$ pairwise contractions, each contributing $1/N$, simplifying calculations compared to exact Weingarten methods \cite{cotler2017chaos}.

\begin{equation}
\begin{aligned}
    \overline{\left\langle \psi_0 \left| U_{\ell^-}^{\dagger} X_{\ell} U_{\ell^-} \right| \psi_0 \right\rangle^4}
&= \int dU_{\ell^-}  \left( \langle \psi_0 | U_{\ell^-}^{\dagger} X_{\ell} U_{\ell^-} | \psi_0 \rangle \right)^4  \\[4pt]
&= \int dU_{\ell^-}  \prod_{a=1}^4 \langle \psi_0 | U_{\ell^-}^{\dagger} X_{\ell} U_{\ell^-} | \psi_0 \rangle  \\[4pt]
&= \int dU_{\ell^-}  \langle \psi_0 | U_{\ell^-}^{\dagger} X_{\ell} U_{\ell^-} | \psi_0 \rangle \langle \psi_0 | U_{\ell^-}^{\dagger} X_{\ell} U_{\ell^-} | \psi_0 \rangle \left( \langle \psi_0 | U_{\ell^-}^{\dagger} X_{\ell} U_{\ell^-} | \psi_0 \rangle \right)^2  \\[4pt]
&= 3 \cdot \frac{1}{N} \int dU_{\ell^-}  \langle \psi_0 | \psi_0 \rangle \langle \psi_0 | U_{\ell^-}^{\dagger} X_{\ell} X_{\ell} U_{\ell^-} | \psi_0 \rangle \left( \langle \psi_0 | U_{\ell^-}^{\dagger} X_{\ell} U_{\ell^-} | \psi_0 \rangle \right)^2  \\[4pt]
&= 3 \cdot \frac{1}{N} \int dU_{\ell^-}  1 \cdot \langle \psi_0 | U_{\ell^-}^{\dagger} I  U_{\ell^-} | \psi_0 \rangle \left( \langle \psi_0 | U_{\ell^-}^{\dagger} X_{\ell} U_{\ell^-} | \psi_0 \rangle \right)^2 \\[4pt]
&= 3 \cdot \frac{1}{N} \int dU_{\ell^-}  \langle \psi_0 | \psi_0 \rangle \left( \langle \psi_0 | U_{\ell^-}^{\dagger} X_{\ell} U_{\ell^-} | \psi_0 \rangle \right)^2  \\[4pt]
&= 3 \cdot \frac{1}{N} \int dU_{\ell^-}  \left( \langle \psi_0 | U_{\ell^-}^{\dagger} X_{\ell} U_{\ell^-} | \psi_0 \rangle \right)^2  \\[4pt]
&= 3 \cdot \frac{1}{N} \cdot \frac{1}{N+1}  \\
&= \frac{3}{N(N+1)}.
\end{aligned}
\end{equation}

Therefore, 
\begin{equation}
    \overline{g_{\ell \ell}^2(\bm \theta)} = 1 - \frac{2}{N+1} + \frac{3}{N(N+1)}
\end{equation}

Given that \(\overline{g_{\ell \ell}(\bm \theta)}^2 = (1 - \frac{1}{N+1})^2\), we have \(\Delta g_{\ell \ell}^2\):

\begin{equation}
\begin{aligned}
        \Delta g_{\ell \ell}^2 &= 1 - \frac{2}{N+1} + \frac{3}{N(N+1)} - \frac{N^2}{(N+1)^2}\\
    &= \frac{2N+3}{N^3+2N^2+N}
\end{aligned}
\end{equation}

In the large-N limit, we get:

\begin{equation}
    \Delta g_{\ell \ell}^2 \approx \frac{2}{N^2}
\end{equation}

In sum:

\begin{equation}
    \Delta g_{\ell_1 \ell_2}^2 \approx
    \begin{cases}
        \dfrac{2}{N^2}, & \text{if } \ell_1 = \ell_2 = \ell, \\[6pt]
        \dfrac{1}{2N}, & \text{if } \ell_1 \ne \ell_2.
    \end{cases}
\end{equation}

Therefore, the fluctuation of the elements in \(g\) is negligible when \(N \to \infty\).

\end{document}